\documentclass[12pt,twoside, a4paper]{article}

\def\mc{\mathcal}

\usepackage[dvips]{graphicx}
\usepackage{amssymb}
\usepackage{amssymb,amsmath}
\usepackage{graphicx}
\usepackage{dsfont}
\usepackage{caption}
\usepackage{subcaption}
\input{epsf.sty}  
\begin{document}
\begin{center}
\LARGE{\textbf{Two-dimensional SCFTs from matter-coupled $7D$ $N=2$ gauged supergravity}}
\end{center}
\vspace{1 cm}
\begin{center}
\large{\textbf{Parinya Karndumri}$^a$ and \textbf{Patharadanai Nuchino}$^b$}
\end{center}
\begin{center}
String Theory and Supergravity Group, Department
of Physics, Faculty of Science, Chulalongkorn University, 254 Phayathai Road, Pathumwan, Bangkok 10330, Thailand
\end{center}
E-mail: $^a$parinya.ka@hotmail.com \\
E-mail: $^b$danai.nuchino@hotmail.com \vspace{1 cm}\\
\begin{abstract}
We study supersymmetric $AdS_3\times M^4$ solutions of $N=2$ gauged supergravity in seven dimensions coupled to three vector multiplets with $SO(4)\sim SO(3)\times SO(3)$ gauge group and $M^4$ being a four-manifold with constant curvature. The gauged supergravity admits two supersymmetric $AdS_7$ critical points with $SO(4)$ and $SO(3)$ symmetries corresponding to $N=(1,0)$ superconformal field theories (SCFTs) in six dimensions. For $M^4=\Sigma^2\times\Sigma^2$ with $\Sigma^2$ being a Riemann surface, we obtain a large class of supersymmetric $AdS_3\times \Sigma^2\times \Sigma^2$ solutions preserving four supercharges and $SO(2)\times SO(2)$ symmetry for one of the $\Sigma^2$ being a hyperbolic space $H^2$, and the solutions are dual to $N=(2,0)$ SCFTs in two dimensions. For a smaller symmetry $SO(2)$, only $AdS_3\times H^2\times H^2$ solutions exist. Some of these are also solutions of pure $N=2$ gauged supergravity with $SU(2)\sim SO(3)$ gauge group. We numerically study domain walls interpolating between the two supersymmetric $AdS_7$ vacua and these geometries. The solutions describe holographic RG flows across dimensions from $N=(1,0)$ SCFTs in six dimensions to $N=(2,0)$ two-dimensional SCFTs in the IR. Similar solutions for $M^4$ being a Kahler four-cycle with negative curvature are also given. In addition, unlike $M^4=\Sigma^2\times \Sigma^2$ case, it is possible to twist by $SO(3)_{\textrm{diag}}$ gauge fields resulting in two-dimensional $N=(1,0)$ SCFTs. Some of the solutions can be uplifted to eleven dimensions and provide a new class of $AdS_3\times M^4\times S^4$ solutions in M-theory.
\end{abstract}
\newpage
\section{Introduction}
One of the most interesting implications of the AdS/CFT correspondence \cite{maldacena} is the study of holographic RG flows. These solutions take the form of  a domain wall interpolating between $AdS$ vacua and holographically describe deformations of a conformal field theory (CFT) in the UV to another CFT in the IR or in some cases to a non-conformal field theory dual to a singular geometry, see \cite{non-CFT_flow_Zaffaroni,Gubser_non_CFT,Freedman_lecture} for example. Of particular interest are RG flows across dimensions in which a higher dimensional CFT flows to a lower dimensional CFT. This type of RG flows allows us to investigate the structure and dynamics of less known CFTs in higher, especially five and six, dimensions using the well-understood lower dimensional CFTs. In this paper, we will consider this type of RG flows in six-dimensional CFTs to two dimensions. Furthermore, the study along this direction is much more fruitful and controllable in the presence of supersymmetry. We are then mainly interested in RG flows within superconformal field theories (SCFTs).  
\\
\indent Supersymmetric solutions of gauged supergravities play an important role in studying the aforementioned RG flows. In general, RG flows across dimensions from a $d$-dimensional SCFT to a $(d-n)$-dimensional SCFT are obtained by twisted compactification of the former on an $n$-dimensional manifold $M^n$. The twist is needed for the compactification to preserve some amount of supersymmetry. This is achieved by turning on some gauge fields to cancel the spin connection on $M^n$. In the supergravity dual, these RG flows are described by domain walls interpolating between an $AdS_{d+1}$ vacuum to an $AdS_{d+1-n}\times M^n$ geometry. Solutions of this type have been studied in various dimensions, see \cite{Maldacena_Nunez_nogo,Kim_AdS_factor,Cucu_AdSD-2,BB,Wraped_D3,3D_CFT_from_LS_point,flow_acrossD_bobev,BBC,
4D_SCFT_from_M5,Wraped_M5,N3_AdS2,flow_across_Betti,AdS2_trisasakian,7D_twist,6D_twist,5D_N4_flow,5Dtwist,BH_microstate_6D2,
Minwoo_6D_BH2,Calos_6D_flow1,flow_across_Gauntlett1,flow_across_Gauntlett2} for an incomplete list.
\\
\indent In this paper, we are interested in supersymmetric $AdS_3\times M^4$ solutions of $N=2$ gauged supergravity in seven dimensions with $SO(4)\sim SO(3)\times SO(3)$ gauge group. This gauged supergravity is obtained by coupling three vector multiplets to pure $N=2$ gauged supergravity with $SU(2)$ gauge group constructed in \cite{Pure_N2_7D1,Pure_N2_7D2}. The matter-coupled gauged supergravity has been constructed in \cite{Eric_N2_7D,Park_7D,Salam_7DN2} with an extension to include a topological mass term for the three-form field, dual to the two-form in the $N=2$ supergravity multiplet, given in \cite{Eric_N2_7Dmassive}. This massive gauged supergravity admits supersymmetric $AdS_7$ vacua which has been extensively studied in \cite{7D_flow,7D_noncompact,AdS_7_N2_Jan}. These vacua are dual to $N=(1,0)$ SCFTs in six dimensions, and a number of RG flows of various types have already been studied \cite{7D_twist,7D_flow,7D_defect}. However, holographic RG flows from $N=(1,0)$ six-dimensional SCFTs to two-dimensional SCFTs in the framework of matter-coupled $N=2$ gauged supergravity have not appeared so far. To fill this gap, we will give a large class of $AdS_3\times M^4$ fixed points and the corresponding RG flows across dimensions within six-dimensional $N=(1,0)$ SCFTs.   
\\
\indent We will consider a four-manifold $M^4$ with constant curvature of two types, a product of two Riemann surfaces $\Sigma^2\times \Sigma^2$ and a Kahler four-cycle $M^4_k$. In the first case, the twists can be performed by using $SO(2)_R\subset SO(3)_R$ with $SO(3)_R$ being the R-symmetry. We will look for solutions with $SO(2)\times SO(2)$, $SO(2)_{\textrm{diag}}$ and $SO(2)_R$ symmetries. In the second case, $M^4_k$ has a $U(2)\sim SU(2)\times U(1)$ spin connection. Therefore, we can perform the twists by turning on either $SO(2)_R\subset SO(3)_R$ or the full $SO(3)_R$ to cancel the $U(1)$ or the $SU(2)$ parts of the spin connection, respectively. It should also be noted that a twist by cancelling the full $U(2)$ spin connection is not possible since the R-symmetry of $N=2$ gauged supergravity is not large enough.  
\\
\indent In general, the two $SO(3)\sim SU(2)$ factors in the $SO(4)$ gauge group can have different coupling constants. However, for a particular case of equal $SU(2)$ coupling constants, the resulting gauged supergravity can be embedded in eleven-dimensional supergravity via a truncation on $S^4$ \cite{7D_from_11D}. The seven-dimensional solutions can accordingly be uplifted to eleven dimensions giving rise to new $AdS_3\times M^4\times S^4$ solutions of eleven-dimensional supergravity. Therefore, these solutions provide a number of new two-dimensional SCFTs with known M-theory dual. We also consider the uplifted solutions in this case.
\\
\indent The paper is organized as follow. In section \ref{N2_7D_SUGRA}, we give a short review of the
matter coupled $N=2$ seven-dimensional gauged supergravity and supersymmetric $AdS_7$ vacua. In sections \ref{AdS3_Sigma_sigma} and \ref{AdS3_M4k}, we look for supersymmetric $AdS_3\times \Sigma^2\times \Sigma^2$ and $AdS_3\times M^4_k$ solutions and numerically study interpolating solutions between these geometries and the $AdS_7$ fixed points. We finally give some conclusions and comments in section \ref{conclusion}. Relevant formulae for the truncation of eleven-dimensional supergravity on $S^4$ giving rise to $N=2$ gauged supergravity with $SO(4)$ gauge group are reviewed in the appendix.

\section{Seven-dimensional $N=2,\ SO(4)$ gauged supergravity and supersymmetric $AdS_7$ vacua}\label{N2_7D_SUGRA}
We firstly review $N=2$ gauged supergravity in seven dimensions coupled to three vector multiplets with $SO(4)$ gauge group. Only relevant formulae involving bosonic Lagrangian and supersymmetry transformations of fermions will be presented. The detailed construction of general $N=2$ seven-dimensional gauged supergravity can be found in \cite{Eric_N2_7Dmassive}, see also \cite{Dibietto_7D_embedding_tensor} for gaugings in the embedding tensor formalism.

\subsection{Seven-dimensional $N=2,\ SO(4)$ gauged supergravity}
The seven-dimensional $N=2,\ SO(4)$ gauged supergravity is obtained by coupling the minimal $N=2$ supergravity to three vector multiplets. The supergravity multiplet consists of the graviton $e^{\hat{\mu}}_\mu$, two gravitini $\psi^a_\mu$, three vectors $A^i_\mu$, two spin-$\frac{1}{2}$ fields $\chi^a$, a two-form field $B_{\mu\nu}$ and the dilaton $\sigma$. Each vector multiplet contains a vector field $A_\mu$, two gaugini $\lambda^a$, and three scalars $\phi^i$. We will use the convention that curved and flat space-time indices are denoted by $\mu$, $\nu$ and $\hat{\mu}$, $\hat{\nu}$ respectively. Indices $i, j=1, 2, 3$ and $a,b=1,2$ label triplet and doublet of $SO(3)_R\sim SU(2)_R$ R-symmetry with the latter being suppressed throughout this work. The three vector multiplets will be labeled by indices $r,s=1,2,3$ which in turn describe the triplet of the matter symmetry $SO(3)$ under which the three vector multiplets transform.
\\
\indent From both supergravity and vector multiplets, there are in total six vector fields denoted collectively by $A^I=(A^i,A^r)$. Indices $I,J,\ldots=1,2,\ldots,6$ describe fundamental representation of the global symmetry $SO(3,3)$ and are lowered and raised by the $SO(3,3)$ invariant tensor $\eta_{IJ}=\text{diag}(-1,-1,-1,1,1,1)$ and its inverse $\eta^{IJ}$. The two-form field will be dualized to a three-form $C_{\mu\nu\rho}$, which admits a topological mass term required by the existence of $AdS_7$ vacua.
\\
\indent The nine scalar fields $\phi^{ir}$ parametrize $SO(3,3)/SO(3)\times SO(3)$ coset manifold. They can be described by the coset representative
\begin{equation}\label{DefL}
{L_I}^A=({L_I}^i,{L_I}^r)
\end{equation}
with an index $A=(i,r)$ corresponding to representations of the compact $SO(3)\times SO(3)$ local symmetry. The inverse of ${L_I}^A$ will be denoted by
\begin{equation}\label{DefLi}
{L_A}^{I}=({L_i}^I,{L_r}^I)
\end{equation}
with the relation
\begin{equation}
{L_j}^I{L_I}^i=\delta^i_j,\qquad {L_s}^I{L_I}^r=\delta^r_s\, .
\end{equation}
Being an element of $SO(3,3)$, the coset representative also satisfies the relation
\begin{equation}
\eta_{IJ}= -{L_I}^i{L_J}^i+{L_I}^r{L_J}^r\, .
\end{equation}
\indent The bosonic Lagrangian of the $N=2,\ SO(4)$ gauged supergravity in form language can be written as
\begin{eqnarray}\label{Lag}
\mathcal{L}&=&\ \frac{1}{2}R\ast\mathbf{1}-\frac{1}{2}e^{\sigma}a_{IJ}\ast F^I_{(2)}\wedge F^J_{(2)}-\frac{1}{2}e^{-2\sigma}\ast H_{(4)}\wedge H_{(4)}-\frac{5}{8}\ast d\sigma\wedge d\sigma\nonumber \\
& &-\frac{1}{2}\ast P^{ir}\wedge P^{ir}+\frac{1}{\sqrt{2}}H_{(4)}\wedge\omega_{(3)}-4hH_{(4)}\wedge C_{(3)}-\mathbf{V}\ast\mathbf{1}\, .
\end{eqnarray}
The constant $h$ describes the topological mass term for the three-form $C_{(3)}$ with the field strength $H_{(4)}=dC_{(3)}$. The gauge field strength is defined by
\begin{equation}
F^I_{(2)}=dA^I_{(1)}+\frac{1}{2}{f_{JK}}^IA^J_{(1)}\wedge A^K_{(1)}\, .
\end{equation}
The definition of the $SO(4)$ structure constants ${f_{IJ}}^K$ includes the gauge coupling constants
\begin{equation}
f_{IJK}=(g_1\epsilon_{ijk},-g_2\varepsilon_{rst})
\end{equation}
where $g_1$ and $g_2$ are coupling constants of $SO(3)_R$ and $SO(3)$, respectively.
\\
\indent The scalar matrix $a_{IJ}$ appearing in the kinetic term of vector fields is given in term of the coset representative as follow
\begin{equation}\label{aIJ}
a_{IJ}={L_I}^i{L_J}^i+{L_I}^r{L_J}^r\, .
\end{equation}
The Chern-Simons three-form satisfying $d\omega_{(3)}=F^I_{(2)}\wedge F^I_{(2)}$ is defined by
\begin{equation}
\omega_{(3)}=F^I_{(2)}\wedge A^I_{(1)}-\frac{1}{6}{f_{IJ}}^KA^I_{(1)}\wedge A^J_{(1)}\wedge A_{(1)K}\, .
\end{equation}
\indent The scalar potential is given by
\begin{equation}\label{Pot}
\mathbf{V}=\frac{1}{4}e^{-\sigma}\left(C^{ir}C_{ir}-\frac{1}{9}C^2\right)+16h^2e^{4\sigma}-\frac{4\sqrt{2}}{3}he^{\frac{3\sigma}{2}}C,
\end{equation}
where $C$-functions, or fermion-shift matrices, are defined as
\begin{eqnarray}\label{CFn}
C&=&-\frac{1}{\sqrt{2}}{f_{IJ}}^K{L_i}^I{L_j}^JL_{Kk}\varepsilon^{ijk},\\
C^{ir}&=& \frac{1}{\sqrt{2}}{f_{IJ}}^K{L_j}^I{L_k}^J{L_K}^r\varepsilon^{ijk},\\[5pt]
C_{rsi}&=& {f_{IJ}}^K{L_r}^I{L_s}^JL_{Ki}.
\end{eqnarray}
It should also be noted that indices $i,j$ and $r,s$ are raised and lowered by $\delta_{ij}$ and $\delta_{rs}$, respectively. Finally, the scalar kinetic term is defined in term of the vielbein on the $SO(3,3)/SO(3)\times SO(3)$ coset as
\begin{equation}\label{P^ir}
P^{ir}_\mu=L^{rI}\left(\delta^K_I\partial_\mu+{f_{IJ}}^KA^J_\mu\right){L_K}^i.
\end{equation}
\indent To find supersymmetric solutions, we need supersymmetry transformations of fermionic fields $\psi_\mu$, $\chi$ and $\lambda^r$. With all fermionic fields vanishing, these transformations read
\begin{eqnarray}\label{SUSY}
\delta\psi_\mu&=& 2D_\mu\epsilon-\frac{\sqrt{2}}{30}e^{-\frac{\sigma}{2}}C\gamma_\mu\epsilon-\frac{4}{5}he^{2\sigma}\gamma_\mu\epsilon-\frac{i}{20}e^{\frac{\sigma}{2}}F^i_{\rho\sigma}\sigma^i(3\gamma_\mu\gamma^{\rho\sigma}-5\gamma^{\rho\sigma}\gamma_\mu)\epsilon\nonumber\\&&-\frac{1}{240\sqrt{2}}e^{-\sigma}H_{\rho\sigma\lambda\tau}(\gamma_\mu\gamma^{\rho\sigma\lambda\tau}+5\gamma^{\rho\sigma\lambda\tau}\gamma_\mu)\epsilon,\\
\delta\chi&=&-\frac{1}{2}\gamma^\mu\partial_\mu\sigma\epsilon+\frac{\sqrt{2}}{30}e^{-\frac{\sigma}{2}}C\epsilon-\frac{16}{5}e^{2\sigma}h\epsilon-\frac{i}{10}e^{\frac{\sigma}{2}}F^i_{\mu\nu}\sigma^i\gamma^{\mu\nu}\epsilon\nonumber\\&&-\frac{1}{60\sqrt{2}}e^{-\sigma}H_{\mu\nu\rho\sigma}\gamma^{\mu\nu\rho\sigma}\epsilon,\\
\delta\lambda^r&=&i\gamma^\mu P^{ir}_\mu \sigma^i\epsilon-\frac{1}{2}e^{\frac{\sigma}{2}}F^r_{\mu\nu}\gamma^{\mu\nu}\epsilon-\frac{i}{\sqrt{2}}e^{-\frac{\sigma}{2}}C^{ir}\sigma^i\epsilon\label{dlambda}
\end{eqnarray}
where $\sigma^i$ are the usual Pauli matrices.
\\
\indent The dressed field strengths $F^i$ and $F^r$ are defined by the relations
\begin{equation}
F^i_{(2)}={L_I}^iF^I_{(2)}\qquad  \textrm{and}\qquad F^r_{(2)}={L_I}^rF^I_{(2)}\, .
\end{equation}
The covariant derivative of the supersymmetry parameter $\epsilon$ is given by
\begin{equation}\label{Depsilon}
D_\mu\epsilon=\partial_\mu\epsilon+\frac{1}{4}{\omega_\mu}^{\hat{\nu}\hat{\rho}}\gamma_{\hat{\nu}\hat{\rho}}\epsilon+\frac{1}{2\sqrt{2}}Q^i_\mu\sigma^i\epsilon
\end{equation}
where $Q^i_\mu$ is defined in term of the composite connection $Q^{ij}_\mu$ as
\begin{equation}
Q^i_\mu=\frac{i}{\sqrt{2}}\varepsilon^{ijk}Q^{jk}_\mu
\end{equation}
with
\begin{equation}\label{DefQ}
Q^{ij}_\mu=L^{jI}\left(\delta^K_I\partial_\mu+{f_{IJ}}^KA^J_\mu\right){L_K}^i\, .
\end{equation}
\indent For convenience, we also give the full bosonic field equations derived from the Lagrangian given in \eqref{Lag}
\begin{eqnarray}
d(e^{-2\sigma}\ast H_{(4)})+8hH_{(4)}-\frac{1}{\sqrt{2}}F^I_{(2)}\wedge F^I_{(2)} &=&0,\label{C3_eq}\\
D(e^\sigma a_{IJ}\ast F^I_{(2)})-\sqrt{2}H_{(4)}\wedge F^J_{(2)}+\ast P^{ir}{f_{IJ}}^K{L_r}^IL_{Ki} &=&0,\\
D(\ast P^{ir})-2e^\sigma{L_I}^i{L_J}^r\ast F^I_{(2)}\wedge F^J_{(2)} \qquad \qquad\qquad & &\nonumber \\-\left(\frac{1}{\sqrt{2}}e^{-\sigma}C^{js}C_{rsk}\varepsilon^{ijk}+4\sqrt{2}he^{\frac{3\sigma}{2}}C^{ir}\right)\varepsilon_{(7)}&=&0,\\
\frac{5}{4}d(\ast d\sigma)-\frac{1}{2}e^\sigma a_{IJ}\ast F^I_{(2)}\wedge F^J_{(2)}+e^{-2\sigma}\ast H_{(4)}\wedge H_{(4)}\qquad & &\nonumber\\ +\left[\frac{1}{4}e^{-\sigma}\left(C^{ir}C_{ir}-\frac{1}{9}C^2\right)+2\sqrt{2}he^{\frac{3\sigma}{2}}C-64h^2e^{4\sigma}\right]\varepsilon_{(7)} &=&0,\\
R_{\mu\nu}-\frac{5}{4}\partial_\mu \sigma\partial_\nu \sigma-a_{IJ}e^\sigma\left(F^I_{\mu\rho}{F^J_\nu}^\rho-\frac{1}{10}g_{\mu\nu}F^I_{\rho\sigma}F^{J\ \rho\sigma}\right)\qquad & & \nonumber \\ -P^{ir}_\mu P^{ir}_\nu-\frac{2}{5}g_{\mu\nu}\mathbf{V}-\frac{1}{6}e^{-2\sigma}\left(H_{\mu\rho\sigma\lambda}{H_\nu}^{\rho\sigma\lambda}-\frac{3}{20}g_{\mu\nu}H_{\rho\sigma\lambda\tau}H^{\rho\sigma\lambda\tau}\right)&=&0\, .\quad\label{EinsteinEQ}
\end{eqnarray}

\subsection{Supersymmetric $AdS_7$ critical points}
We now give a brief review of supersymmetric $AdS_7$ vacua found in \cite{7D_flow}. There are two supersymmetric $N=2$ $AdS_7$ critical points with $SO(4)\sim SO(3)\times SO(3)$ and $SO(3)_{\text{diag}}\subset SO(3)\times SO(3)$ symmetries. To compute the scalar potential, we need an explicit parametrization of $SO(3,3)$ $/SO(3)\times SO(3)$ coset. By defining the following $GL(6,\mathbb{R})$ matrices
\begin{equation}
(e_{IJ})_{KL}=\delta_{IK}\delta_{JL},
\end{equation}
we can write non-compact generators of $SO(3,3)$ as
\begin{equation}\label{nonComGen}
Y_{ir}=e_{i,r+3}+e_{r+3,i}\, .
\end{equation}
\indent Among the nine scalars from $SO(3,3)/SO(3)\times SO(3)$, there is one $SO(3)_{\textrm{diag}}$ singlet corresponding to the non-compact generator
\begin{equation}\label{SO(3)diagnonComGen}
Y_s=Y_{11}+Y_{22}+Y_{33}\, .
\end{equation}
The coset representative is then given by
\begin{equation}
L=e^{\phi Y_s}.\label{SO3d_coset}
\end{equation}
The scalar potential for the dilaton $\sigma$ and the $SO(3)_{\text{diag}}$ singlet scalar $\phi$ is readily computed to be
\begin{eqnarray}
\mathbf{V}&=&\frac{1}{32}e^{-\sigma}\left[(g_1^2+g_2^2)\left(\cosh(6\phi)-9\cosh(2\phi)\right)+8g_1g_2\sinh^3(2\phi)\right.\nonumber \\
& &\left.+8\left[g_2^2-g_1^2+64h^2e^{5\sigma}-32e^{\frac{5\sigma}{2}}h(g_1\cosh^3\phi+g_2\sinh^3\phi)\right]\right].
\end{eqnarray}
This potential admits two supersymmetric $AdS_7$ critical points
\begin{eqnarray}
\textrm{I}:\qquad  \sigma&=&\phi=0, \qquad\mathbf{V}_0=-240h^2, \label{SO(4)vacuum}\\
\textrm{II}:\qquad \sigma&=&\frac{1}{5}\ln\left[\frac{g_2^2}{g_2^2-256h^2}\right], \qquad \phi=\frac{1}{2}\ln\left[\frac{g_2-16h}{g_2+16h}\right], \nonumber \\
\mathbf{V}_0&=&-\frac{240g_2^{\frac{8}{5}}h^2}{(g^2-256h^2)^{\frac{4}{5}}}\, .\label{SO(3)diagvacuum}
\end{eqnarray}
Critical points I and II have $SO(4)$ and $SO(3)_{\textrm{diag}}$ symmetries, respectively. We have also chosen $g_1=16h$ to bring the $SO(4)$ critical point to the value $\sigma=0$. The cosmological constant is denoted by $\mathbf{V}_0$. According to the AdS/CFT correspondence, these critical points correspond to $N=(1,0)$ SCFTs in six dimensions with $SO(4)$ and $SO(3)$ symmetries, respectively. A holographic RG flow interpolating between these two critical points has already been studied in \cite{7D_flow}, see also \cite{6D_flow_Tomasiello} for more general solutions. In subsequent sections, we will find supersymmetric $AdS_3\times M^4$ solutions to this $N=2$ $SO(4)$ gauged supergravity and RG flow solutions from the above $AdS_7$ vacua to these geometries in the IR.

\section{Supersymmetric $AdS_3\times\Sigma^2\times\Sigma^2$ solutions and RG flows}\label{AdS3_Sigma_sigma}
In this section, we look for supersymmetric solutions of the form $AdS_3\times\Sigma^2_{k_1}\times\Sigma^2_{k_2}$ with $\Sigma^2_{k_i}$ for $i=1,2$ being two-dimensional Riemann surfaces. Constants $k_i$ describe the curvature of $\Sigma^2_{k_i}$ with values $k_i=1,0,-1$ corresponding to a two-dimensional sphere $S^2$, a flat space $\mathbb{R}^2$ or a hyperbolic space $H^2$, respectively.
\\
\indent We will choose the ansatz for the seven-dimensional metric of the form
\begin{equation}\label{AdS3xSig2xSig2metric}
ds_7^2=e^{2U(r)}dx^2_{1,1}+dr^2+e^{2V(r)}ds^2_{\Sigma^2_{k_1}}+e^{2W(r)}ds^2_{\Sigma^2_{k_2}},
\end{equation}
in which $dx^2_{1,1}=\eta_{\alpha\beta}dx^\alpha dx^\beta$, $\alpha,\beta=0,1$ is the flat metric on the two-dimensional spacetime. The explicit form of the metric on $\Sigma^2_{k_i}$ can be written as
\begin{equation}\label{Sig2metric}
ds^2_{\Sigma^2_{k_i}}=d\theta_i^2+f_{k_i}(\theta_i)^2d\varphi_i^2\, .
\end{equation}
The functions $f_{k_i}(\theta_i)$ are defined as
\begin{equation}\label{fFn}
f_{k_i}(\theta_i)=\begin{cases}
                        	\sin{\theta_i}, \qquad  k_i=1 \\
                       	\theta_i,\phantom{\sin{\theta}} \qquad   k_i=0 \\
			\sinh{\theta_i}, \qquad  k_i=-1
                    \end{cases}\, .
\end{equation}
\indent By using an obvious choice of vielbein
\begin{eqnarray}
e^{\hat{\alpha}}&=& e^Udx^\alpha, \qquad e^{\hat{r}}=dr,\qquad e^{\hat{\theta}_1}= e^Vd\theta_1,\nonumber \\ e^{\hat{\varphi}_1}&=&e^Vf_{k_1}(\theta_1)d\varphi_1,\qquad
e^{\hat{\theta}_2}= e^Wd\theta_2,\qquad e^{\hat{\varphi}_2}=e^Wf_{k_2}(\theta_2)d\varphi_2,
\end{eqnarray}
we can compute the following non-vanishing components of the spin connection
\begin{eqnarray}\label{AdS3xSig2xSig2spinCon}
{\omega^{\hat{\alpha}}}_{\hat{r}}&=&U'e^{\hat{\alpha}}, \qquad {\omega^{\hat{\theta}_1}}_{\hat{r}}=V'e^{\hat{\theta}_1}, \qquad {\omega^{\hat{\varphi}_1}}_{\hat{r}}= V'e^{\hat{\varphi}_1}, \qquad {\omega^{\hat{\theta}_2}}_{\hat{r}}=W'e^{\hat{\theta}_2}, \nonumber \\ {\omega^{\hat{\varphi}_2}}_{\hat{r}}&=&W'e^{\hat{\varphi}_2},\qquad {\omega^{\hat{\varphi}_1}}_{\hat{\theta}_1}=e^{-V}\frac{f'_{k_1}(\theta_1)}{f_{k_1}(\theta_1)}e^{\hat{\varphi}_1}, \qquad {\omega^{\hat{\varphi}_2}}_{\hat{\theta}_2}=e^{-W}\frac{f'_{k_2}(\theta_2)}{f_{k_2}(\theta_2)}e^{\hat{\varphi}_2}\, .\quad 
\end{eqnarray}
Throughout the paper, we will use primes to denote derivatives of a function with respect to its argument for example $U'=dU/dr$ and $f'_{k_i}(\theta_i)=df_{k_i}(\theta_i)/d\theta_i$.
\\
\indent To find supersymmetric $AdS_3\times \Sigma^2_{k_1}\times \Sigma^2_{k_2}$ solutions which admit non-vanishing Killing spinors, we perform a twist by turning on gauge fields along $\Sigma^2_{k_1}\times \Sigma^2_{k_2}$. In the following discussions, we will consider various possible twists with different unbroken symmetries.

\subsection{$AdS_3$ vacua with $SO(2)\times SO(2)$ symmetry}
\begin{figure}[h!]
  \centering
  \begin{subfigure}[b]{0.32\linewidth}
    \includegraphics[width=\linewidth]{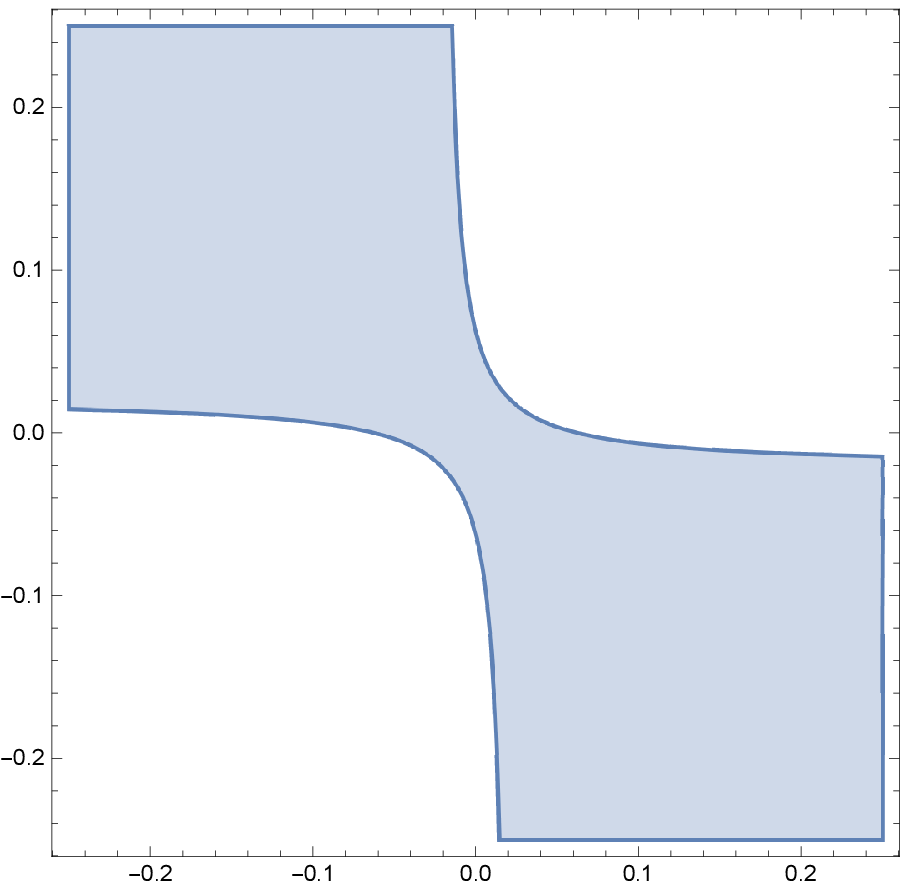}
  \end{subfigure}
  \begin{subfigure}[b]{0.32\linewidth}
    \includegraphics[width=\linewidth]{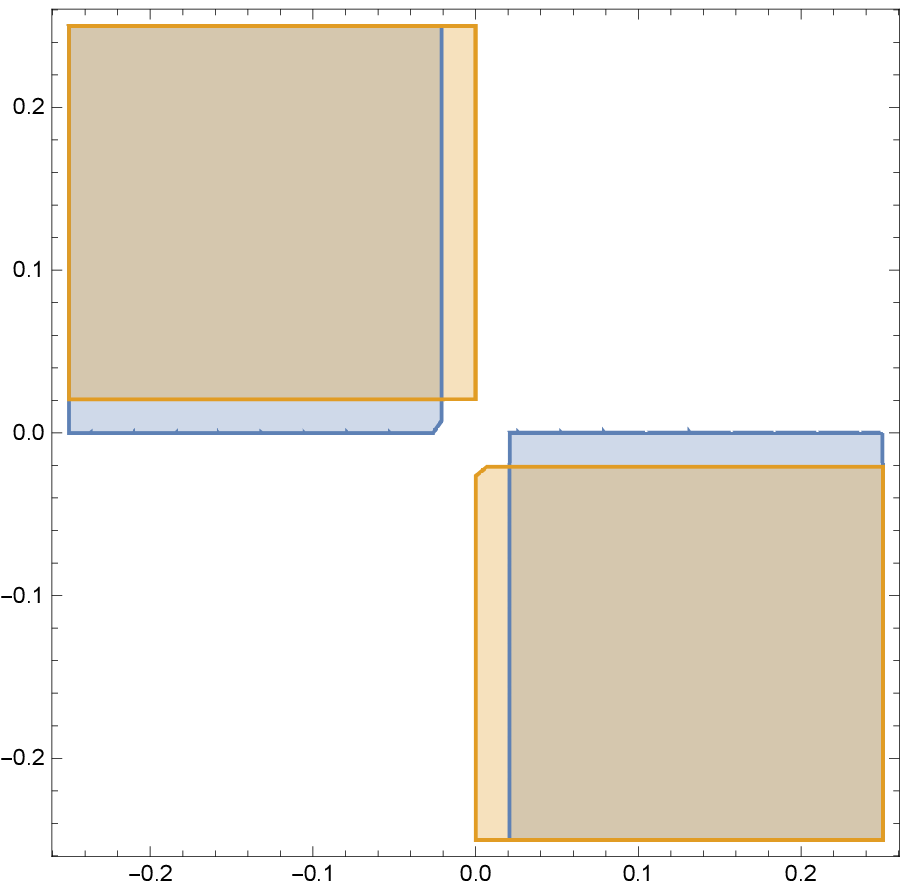}
  \end{subfigure}
  \begin{subfigure}[b]{0.32\linewidth}
    \includegraphics[width=\linewidth]{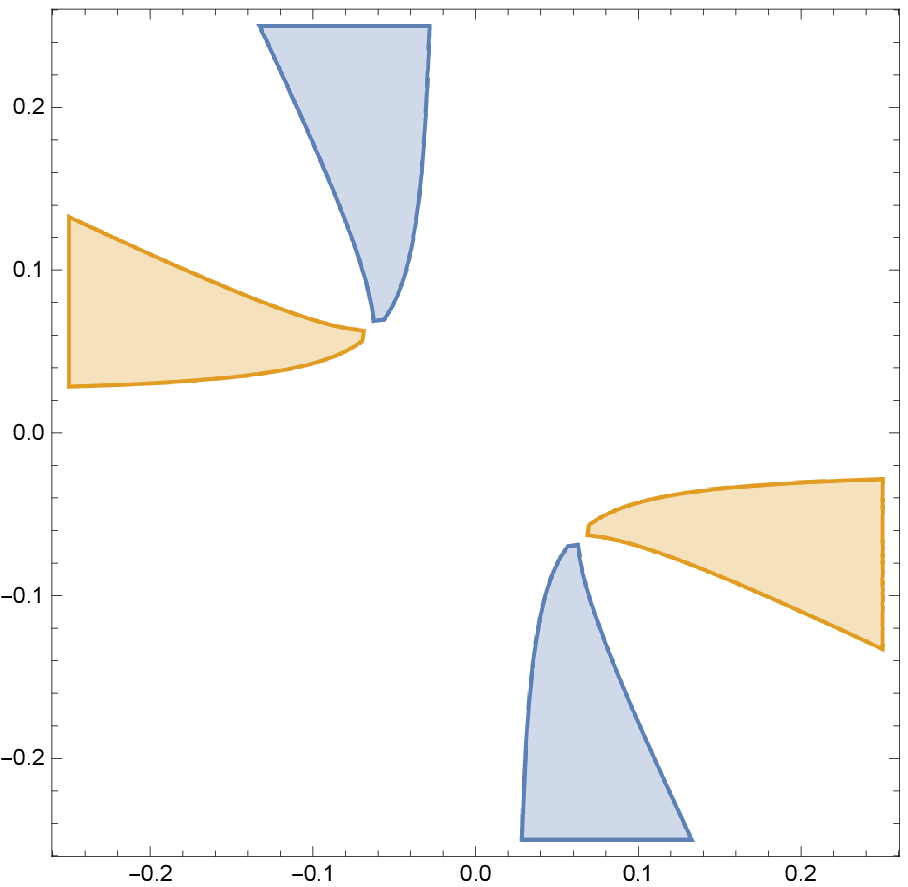}
  \end{subfigure}
  \caption{Regions (blue) in the parameter space ($p_{21}$, $p_{22}$) where good $AdS_3$ vacua exist. From left to right, these are the cases of $(k_1=k_2=-1)$, $(k_1=-1$, $k_2=0)$ and $(k_1=-k_2=-1)$, respectively. The orange regions correspond to interchanging $k_1$ and $k_2$.}
  \label{Sig2xSig2regions}
\end{figure}

We first consider solutions with $SO(2)\times SO(2)$ symmetry. To perform the twist, we turn on the following $SO(2)\times SO(2)$ gauge fields on $\Sigma^2_{k_1}\times \Sigma^2_{k_2}$
\begin{eqnarray}
A^3_{(1)}&=&-\frac{p_{11}}{k_1}e^{-V}\frac{f'_{k_1}(\theta_1)}{f_{k_1}(\theta_1)}e^{\hat{\varphi}_1}-\frac{p_{12}}{k_2}e^{-W}\frac{f'_{k_2}(\theta_2)}{f_{k_2}(\theta_2)}e^{\hat{\varphi}_2},\label{A3_1}\\
A^6_{(1)}&=&-\frac{p_{21}}{k_1}e^{-V}\frac{f'_{k_1}(\theta_1)}{f_{k_1}(\theta_1)}e^{\hat{\varphi}_1}-\frac{p_{22}}{k_2}e^{-W}\frac{f'_{k_2}(\theta_2)}{f_{k_2}(\theta_2)}e^{\hat{\varphi}_2},\label{A6_1}
\end{eqnarray}
where $p_{ij}$ are constants magnetic charges.
\\
\indent There is one $SO(2)\times SO(2)$ singlet scalar from $SO(3,3)/SO(3)\times SO(3)$ coset corresponding to the non-compact generator $Y_{33}$. We then parametrize the coset representative by
\begin{equation}
L=e^{\phi Y_{33}}\label{SO2_SO2_scalar}
\end{equation}
with $\phi$ depending only on the radial coordinate $r$. By computing the composite connection $Q^{ij}_\mu$ along $\Sigma^2_{k_1}\times \Sigma^2_{k_2}$, we can cancel the spin connections by imposing the following twist conditions
\begin{equation}
g_1p_{11}=k_1\qquad \textrm{and} \qquad g_1p_{12}=k_2 \label{Sig2xSig2QYM}
\end{equation}
together with the projection conditions
\begin{equation} \gamma_{\hat{\theta}_1\hat{\varphi}_1}\epsilon=\gamma_{\hat{\theta}_2\hat{\varphi}_2}\epsilon=i\sigma^3\epsilon\, .\label{SO2_SO2_projection1}
\end{equation}
Note that only the gauge field $A^3_{(1)}$ enters the twist procedure since $A^3_{(1)}$ is the gauge field of $SO(2)_R\subset SO(3)_R$ under which the gravitini and supersymmetry parameters are charged.
\\
\indent From the gauge fields given in \eqref{A3_1} and \eqref{A6_1}, we can straightforwardly compute the corresponding two-form field strengths
\begin{eqnarray}\label{Sig2xSig2F}
F^3_{(2)}&=& e^{-2V}p_{11}e^{\hat{\theta}_1}\wedge e^{\hat{\varphi}_1}+e^{-2W}p_{12}e^{\hat{\theta}_2}\wedge e^{\hat{\varphi}_2},\\
F^6_{(2)}&=& e^{-2V}p_{21}e^{\hat{\theta}_1}\wedge e^{\hat{\varphi}_1}+e^{-2W}p_{22}e^{\hat{\theta}_2}\wedge e^{\hat{\varphi}_2}\, .
\end{eqnarray}
It should also be noted that these field strengths give non-vanishing $F^I_{(2)}\wedge F^I_{(2)}$ term. This term is present in the field equation of the three-form fied $C_{(3)}$ as can be seen from equation \eqref{C3_eq}. Therefore, we need to turn on the three-form field with the corresponding four-form field strength given by
\begin{equation}\label{SO(2)xSO(2)Sig2xSig24form}
H_{(4)}=\frac{1}{8\sqrt{2}h}e^{-2(V+W)}(p_{21}p_{22}-p_{11}p_{12}) e^{\hat{\theta}_1}\wedge e^{\hat{\varphi}_1}\wedge e^{\hat{\theta}_2}\wedge e^{\hat{\varphi}_2}.
\end{equation}
This is very similar to the solutions of maximal $SO(5)$ gauged supergravity considered in \cite{BB}.
\\
\indent By imposing an additional projector
\begin{equation}
\gamma_r\epsilon=\epsilon\label{gamma_r_projection}
\end{equation}
required by $\delta \chi=0$ and $\delta\lambda^r=0$ conditions, we find the following BPS equations
\begin{eqnarray}
U'&=&\frac{1}{5}e^{\frac{\sigma}{2}}\left[\left(g_1e^{-\sigma}\cosh{\phi}+4he^{\frac{3\sigma}{2}}\right)+\frac{3}{8h}e^{-\frac{3\sigma}{2}-2(V+W)}(p_{11}p_{12}-p_{21}p_{22})\right.\quad \nonumber \\
& &\left.\phantom{\frac{1}{4}}-e^{-2V}(p_{11}\cosh{\phi}+p_{21}\sinh{\phi})-e^{-2W}(p_{12}\cosh{\phi}+p_{22}\sinh{\phi})\right],\qquad \\
V'&=&\frac{1}{5}e^{\frac{\sigma}{2}}\left[\left(g_1e^{-\sigma}\cosh{\phi}+4he^{\frac{3\sigma}{2}}\right)-\frac{1}{4h}e^{-\frac{3\sigma}{2}-2(V+W)}(p_{11}p_{12}-p_{21}p_{22})\right.\quad \nonumber\\
& &\left.\phantom{\frac{1}{4}}+4e^{-2V}(p_{11}\cosh{\phi}+p_{21}\sinh{\phi})-e^{-2W}(p_{12}\cosh{\phi}+p_{22}\sinh{\phi})\right],\qquad \\
W'&=&\frac{1}{5}e^{\frac{\sigma}{2}}\left[\left(g_1e^{-\sigma}\cosh{\phi}+4he^{\frac{3\sigma}{2}}\right)-\frac{1}{4h}e^{-\frac{3\sigma}{2}-2(V+W)}(p_{11}p_{12}-p_{21}p_{22})\right.\quad\nonumber \\
& &\left.\phantom{\frac{1}{4}}-e^{-2V}(p_{11}\cosh{\phi}+p_{21}\sinh{\phi})+4e^{-2W}(p_{12}\cosh{\phi}+p_{22}\sinh{\phi})\right],\qquad \\
\sigma'&=&\frac{2}{5}e^{\frac{\sigma}{2}}\left[\left(g_1e^{-\sigma}\cosh{\phi}-16he^{\frac{3\sigma}{2}}\right)-\frac{1}{4h}e^{-\frac{3\sigma}{2}-2(V+W)}(p_{11}p_{12}-p_{21}p_{22})\right.\quad \nonumber \\
& &\left.\phantom{\frac{1}{4}}-e^{-2V}(p_{11}\cosh{\phi}+p_{21}\sinh{\phi})-e^{-2W}(p_{12}\cosh{\phi}+p_{22}\sinh{\phi})\right],\qquad\\
\phi'&=&-e^{\frac{\sigma}{2}}\left[e^{-2V}(p_{11}\sinh{\phi}+p_{21}\cosh{\phi})+e^{-2W}(p_{12}\sinh{\phi}+p_{22}\cosh{\phi})\right]\nonumber \\ 
& &-g_1e^{-\frac{\sigma}{2}}\sinh{\phi}\, .
\end{eqnarray}
It can be verified that these BPS equations satisfy all the field equations. At large $r$, we have $U\sim V\sim W\sim r$ and $\phi\sim \sigma\sim e^{-\frac{4r}{L}}$ with the $AdS_7$ radius given by $L=\frac{1}{4h}$, and the terms involving gauge fields and the three-form field are highly suppressed. We find the $SO(4)$ $AdS_7$ fixed point from these BPS equations in this limit. The solutions are then asymptotically locally $AdS_7$ as $r\rightarrow \infty$.
\\
\indent We now look for supersymmetric $AdS_3$ solutions satisfying $V'=W'=\sigma'=\phi'=0$ and $U'=\frac{1}{L_{AdS_3}}$ in the limit $r\rightarrow -\infty$. We find a class of $AdS_3$ fixed point solutions
\begin{eqnarray}\label{SO(2)xSO(2)fixedpoint}
e^{\frac{5}{2}\sigma}&=&\frac{g_1Ze^\phi}{4h(p_{21}(p_{12}-3p_{22})+p_{11}(p_{12}+p_{22}))},\\
e^\phi&=&\sqrt{\frac{p_{21}(p_{12}-3p_{22})+p_{11}(p_{12}+p_{22})}{p_{11}(p_{12}-p_{22})-p_{21}(p_{12}+3p_{22})}},\\
e^{2V}&=&\frac{p_{21}-p_{11}-(p_{11}+p_{21})e^{2\phi}}{8he^{\phi+\frac{3}{2}\sigma}},\\
e^{2W}&=&\frac{p_{22}-p_{12}-(p_{12}+p_{22})e^{2\phi}}{8he^{\phi+\frac{3}{2}\sigma}},\\
L_{AdS_3}&=&\frac{8he^{\sigma+2V+2W}}{p_{11}p_{12}-p_{21}p_{22}+32h^2e^{2V+2W+3\sigma}}
\end{eqnarray}
where
\begin{equation}
Z=\frac{(p_{12}(p_{11}^2+p_{21}^2)-2p_{11}p_{21}p_{22})(-2p_{12}p_{21}p_{22}+p_{11}(p_{12}^2+p_{22}^2))}{(p_{11}^2(3p_{12}^2+p_{22}^2)+p_{21}^2(p_{12}^2+3p_{22}^2)-8p_{11}p_{12}p_{21}p_{22})}.
\end{equation}
Note that the coupling constant $g_2$ does not appear in the above equations, so the solutions can be uplifted to eleven dimensions by setting $g_2=g_1$.
\\
\indent To obtain real solutions, we require that $e^{2V}>0$, $e^{2W}>0$, $e^\sigma>0$, and $e^\phi>0$. It turns out that $AdS_3$ solutions are possible only for one of the two $k_i$ is equal to $-1$ with the seven-dimensional spacetime given by $AdS_3\times H^2\times H^2$, $AdS_3\times H^2\times \mathbb{R}^2$ and $AdS_3\times H^2\times S^2$. Since the charges $p_{11}$ and $p_{12}$ are fixed by the twist conditions \eqref{Sig2xSig2QYM}, there are only two parameters $p_{21}$ and $p_{22}$ characterizing the solutions. For $g_1=16h$ and $h=1$, regions in the parameter space ($p_{21}$, $p_{22}$) for good AdS$_3$ vacua to exist are shown in figure \ref{Sig2xSig2regions}. Note that these regions are precisely the same as supersymmetric $AdS_3\times \Sigma^2\times \Sigma^2$ solutions of maximal seven-dimensional $SO(5)$ gauged supergravity in \cite{BB}.

\begin{figure}[h!]
  \centering
  \begin{subfigure}[b]{0.45\linewidth}
    \includegraphics[width=\linewidth]{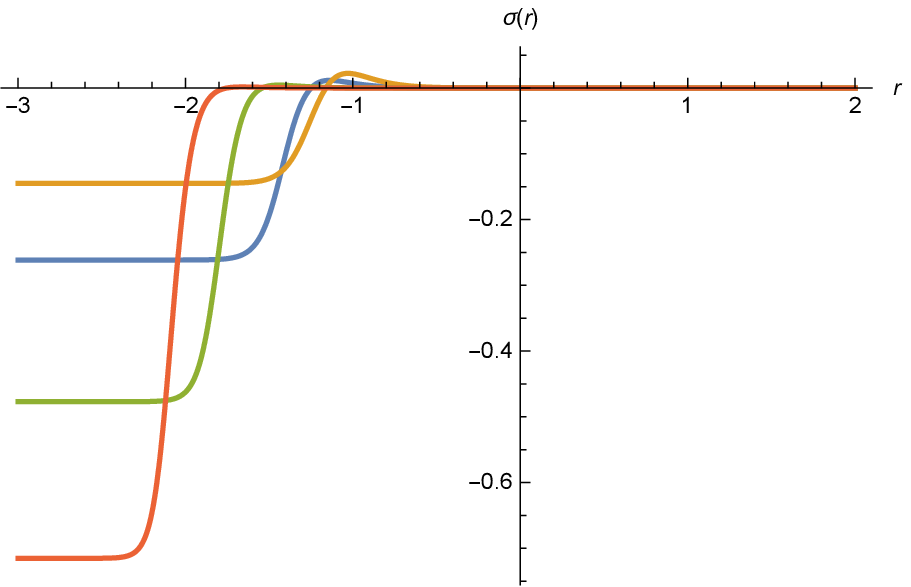}
  \caption{$\sigma$ solution}
  \end{subfigure}
  \begin{subfigure}[b]{0.45\linewidth}
    \includegraphics[width=\linewidth]{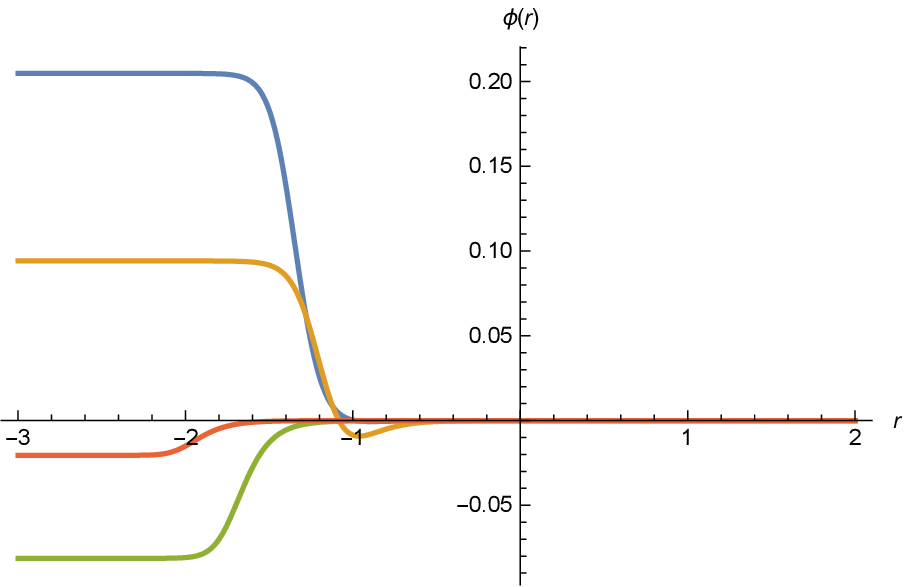}
  \caption{$\phi$ solution}
  \end{subfigure}\\
  \begin{subfigure}[b]{0.45\linewidth}
    \includegraphics[width=\linewidth]{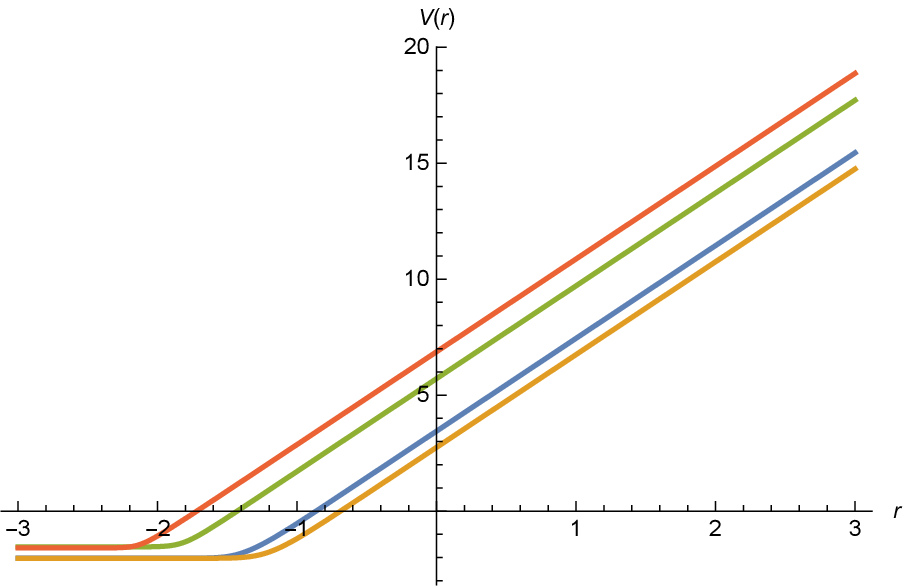}
  \caption{$V$ solution}
  \end{subfigure}
  \begin{subfigure}[b]{0.45\linewidth}
    \includegraphics[width=\linewidth]{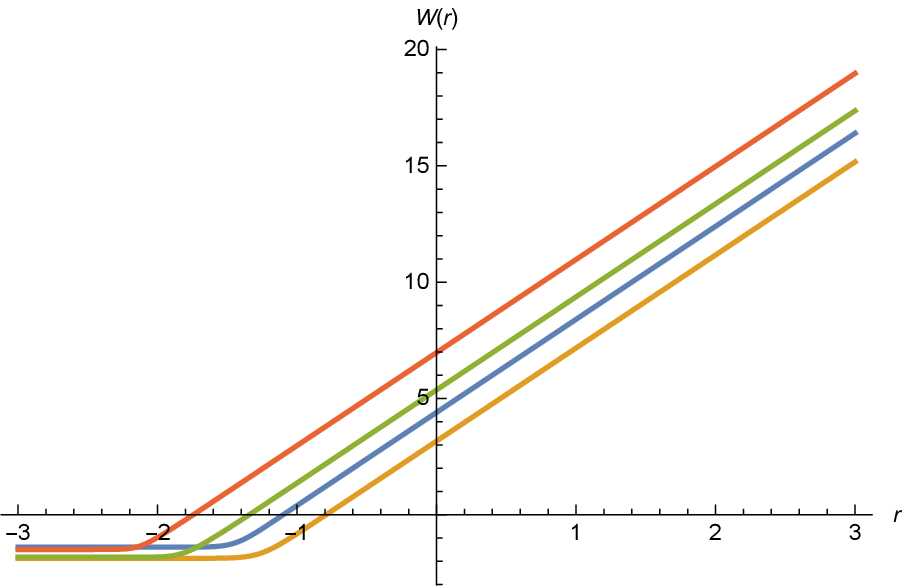}
  \caption{$W$ solution}
  \end{subfigure}
  \caption{RG flows from $SO(4)$ $N=(1, 0)$ SCFT in six dimensions to two-dimensional $N=(2,0)$ SCFTs with $SO(2)\times SO(2)$ symmetry dual to $AdS_3\times H^2\times H^2$ solutions for $(p_{21}, p_{22})=(\frac{1}{12},-\frac{1}{2}), (\frac{1}{12},-\frac{1}{7}), (\frac{1}{3},-\frac{1}{7}), (-\frac{1}{4},\frac{1}{3})$ (blue, yellow, green, red).}
  \label{H2xH2flow}
\end{figure}

\begin{figure}[h!]
  \centering
  \begin{subfigure}[b]{0.45\linewidth}
    \includegraphics[width=\linewidth]{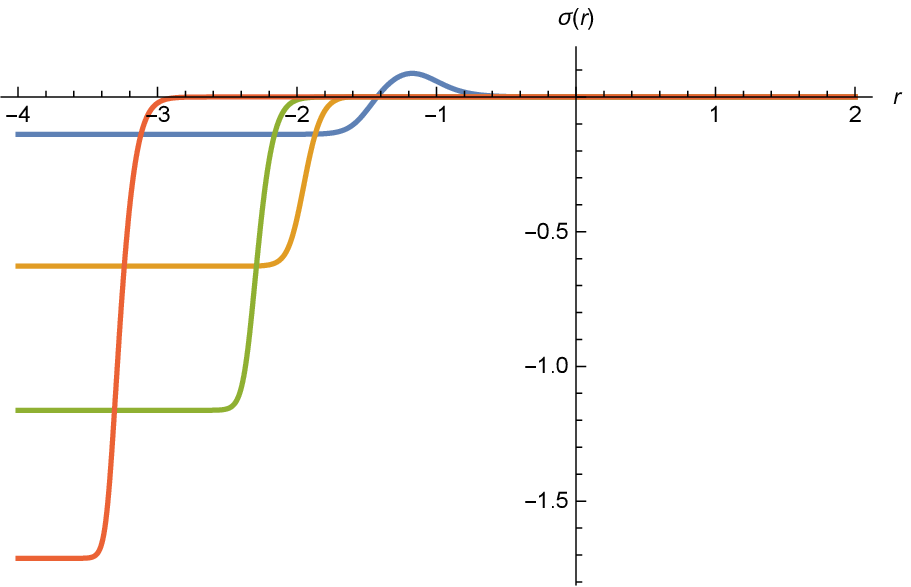}
  \caption{$\sigma$ solution}
  \end{subfigure}
  \begin{subfigure}[b]{0.45\linewidth}
    \includegraphics[width=\linewidth]{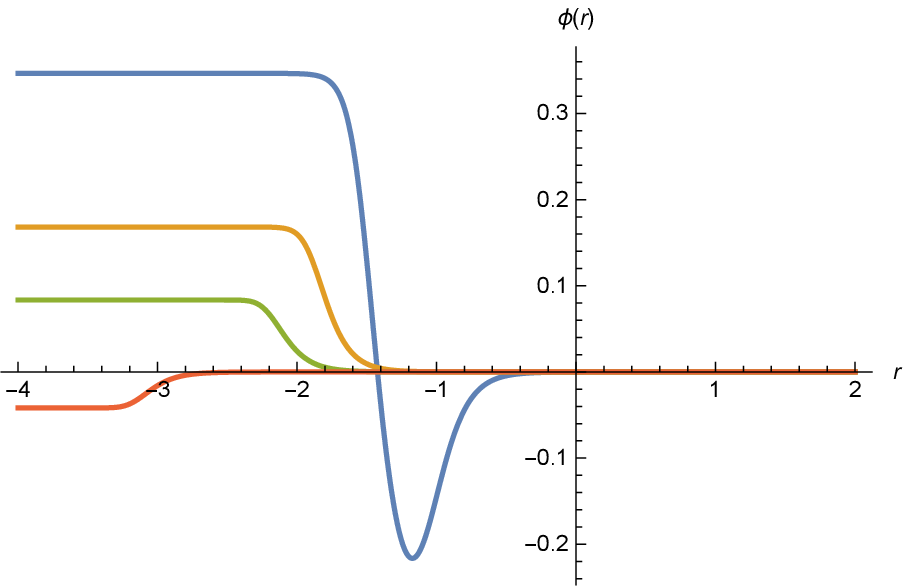}
  \caption{$\phi$ solution}
  \end{subfigure}\\
  \begin{subfigure}[b]{0.45\linewidth}
    \includegraphics[width=\linewidth]{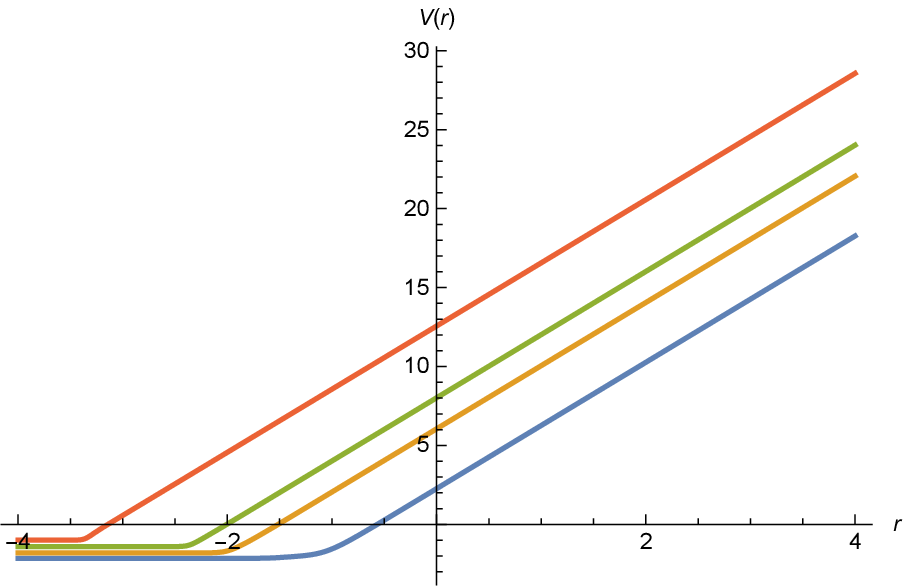}
  \caption{$V$ solution}
  \end{subfigure}
  \begin{subfigure}[b]{0.45\linewidth}
    \includegraphics[width=\linewidth]{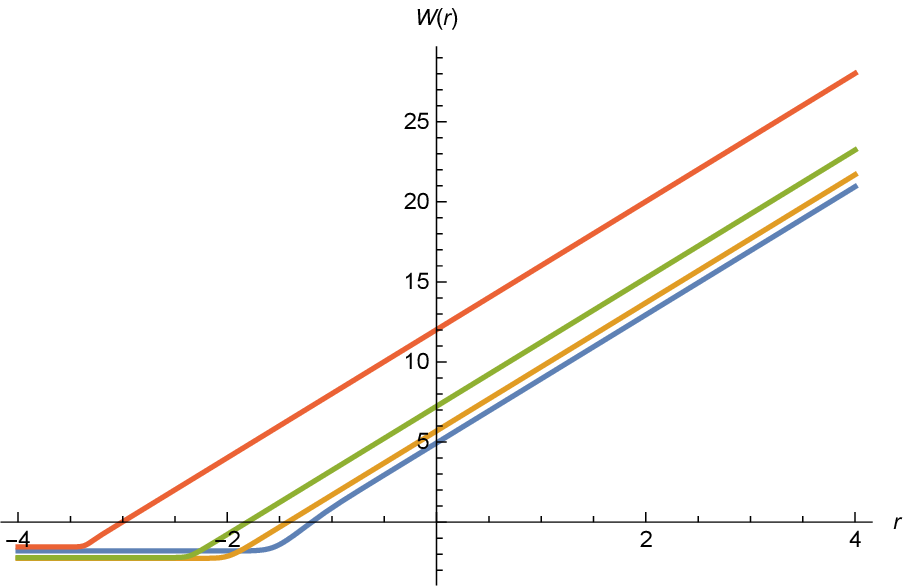}
  \caption{$W$ solution}
  \end{subfigure}
  \caption{RG flows from $SO(4)$ $N=(1, 0)$ SCFT in six dimensions to two-dimensional $N=(2,0)$ SCFTs with $SO(2)\times SO(2)$ symmetry dual to $AdS_3\times H^2\times \mathbb{R}^2$ solutions for $(p_{21}, p_{22})=(\frac{1}{16},-\frac{1}{4}), (\frac{1}{8},-\frac{1}{10}), (\frac{1}{4},-\frac{1}{10}), (-\frac{1}{2},\frac{1}{3})$ (blue, yellow, green, red).}
  \label{H2xR2flow}
\end{figure}

\begin{figure}[h!]
  \centering
  \begin{subfigure}[b]{0.45\linewidth}
    \includegraphics[width=\linewidth]{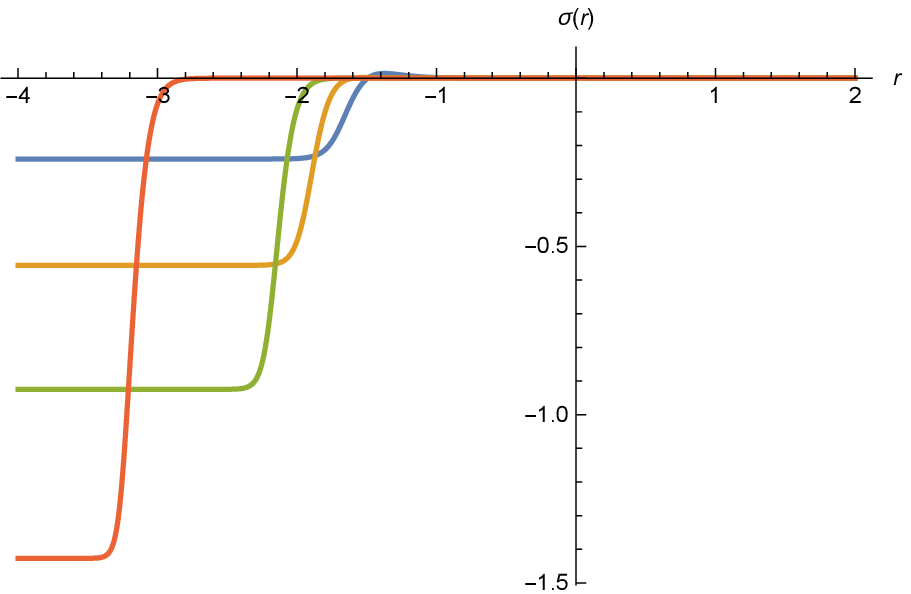}
  \caption{$\sigma$ solution}
  \end{subfigure}
  \begin{subfigure}[b]{0.45\linewidth}
    \includegraphics[width=\linewidth]{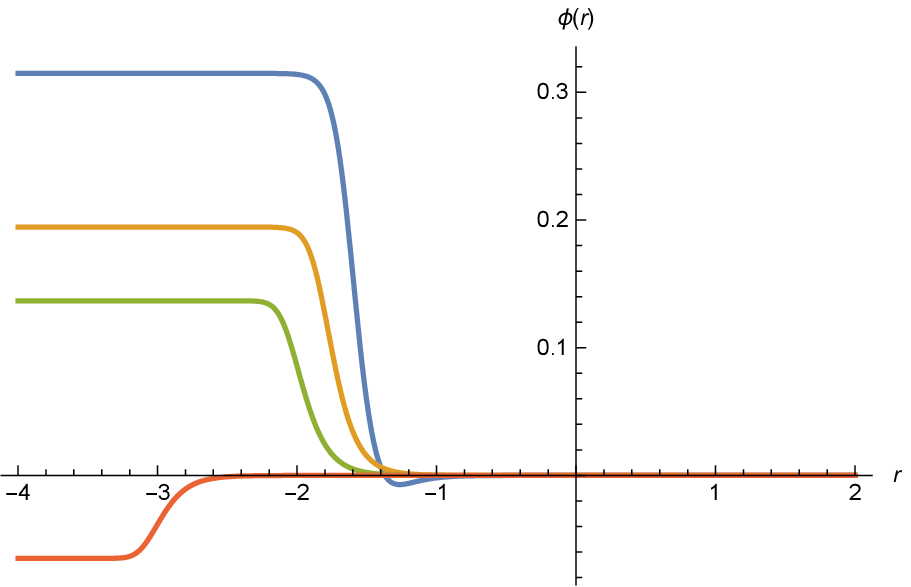}
  \caption{$\phi$ solution}
  \end{subfigure}\\
  \begin{subfigure}[b]{0.45\linewidth}
    \includegraphics[width=\linewidth]{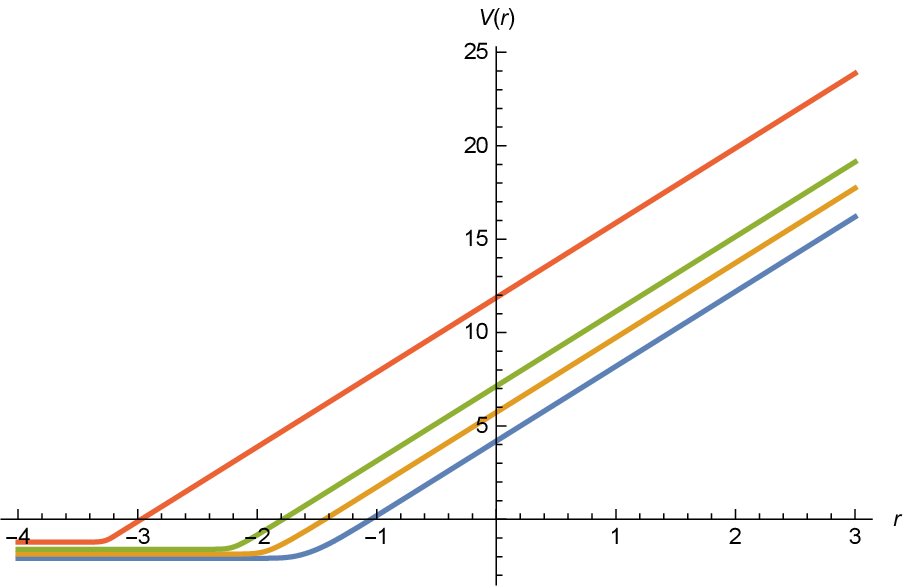}
  \caption{$V$ solution}
  \end{subfigure}
  \begin{subfigure}[b]{0.45\linewidth}
    \includegraphics[width=\linewidth]{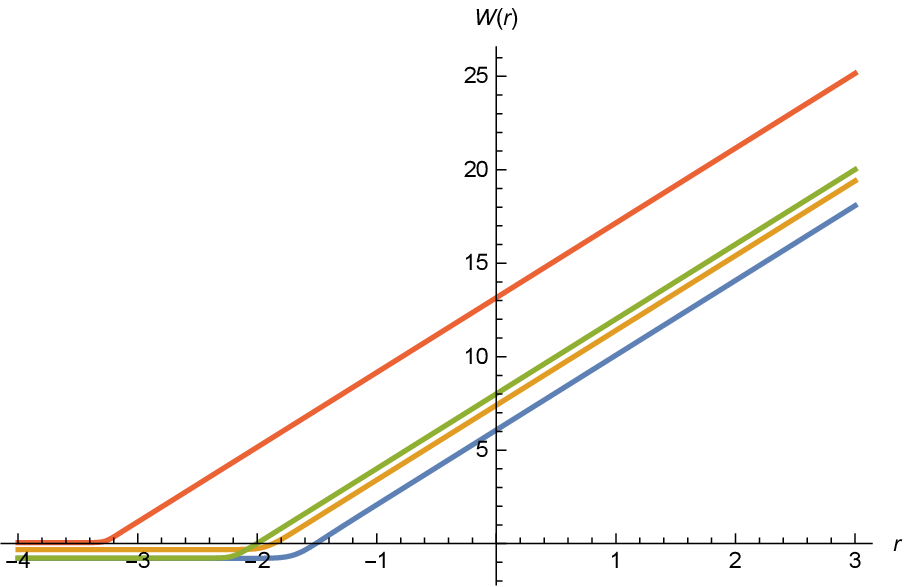}
  \caption{$W$ solution}
  \end{subfigure}
  \caption{RG flows from $SO(4)$ $N=(1, 0)$ SCFT in six dimensions to two-dimensional $N=(2,0)$ SCFTs with $SO(2)\times SO(2)$ symmetry dual to $AdS_3\times H^2\times S^2$ solutions for $(p_{21}, p_{22})=(\frac{1}{14},-2), (\frac{1}{9},-5), (\frac{1}{6},-2), (-\frac{1}{3},9)$ (blue, yellow, green, red).}
  \label{H2xS2flow}
\end{figure}

These $AdS_3$ fixed points preserve four supercharges due to the two projectors in \eqref{SO2_SO2_projection1} and correspond to $N=(2,0)$ SCFTs in two dimensions with $SO(2)\times SO(2)$ symmetry. On the other hand, the entire RG flow solutions interpolating between the $AdS_7$ fixed point and these $AdS_3$ geometries preserve only two supercharges due to an extra projector in \eqref{gamma_r_projection}. Examples of these RG flows from the $AdS_7$ fixed point to $AdS_3\times H^2\times H^2$, $AdS_3\times H^2\times \mathbb{R}^2$ and $AdS_3\times H^2\times S^2$ with $h=1$ and different values of $p_{21}$ and $p_{22}$ are shown in figures \ref{H2xH2flow}, \ref{H2xR2flow} and \ref{H2xS2flow}, respectively. 


These solutions can be uplifted to eleven dimensions using the truncation ansatz given in \cite{7D_from_11D}. By using the formulae reviewed in the appendix together with the $S^3$ coordinates
\begin{equation} 
\mu^\alpha=(\cos\psi\cos\alpha,\cos\psi\sin\alpha,\sin\psi\cos\beta,\sin\psi\sin\beta)
\end{equation}
and the $SL(4,\mathbb{R})/SO(4)$ matrix
\begin{equation}
\tilde{T}^{-1}_{\alpha\beta}=\textrm{diag}(e^\phi,e^\phi,e^{-\phi},e^{-\phi}),
\end{equation}
we find the eleven-dimensional metric
\begin{eqnarray}
d\hat{s}^2_{11}&=&\Delta^{\frac{1}{3}}\left[e^{2U}dx^2_{1,1}+dr^2+e^{2V}ds^2_{\Sigma^2_{k_1}}+e^{2W}ds^2_{\Sigma^2_{k_2}}\right]\nonumber \\
& &+\frac{2}{g^2}\Delta^{-\frac{2}{3}}\left[e^{-2\sigma}\cos^2\xi+e^{\frac{\sigma}{2}}\sin^2\xi(e^\phi\cos^2\psi+e^{-\phi}\sin^2\psi)\right]d\xi^2\nonumber \\
& &+\frac{1}{2g^2}\Delta^{-\frac{2}{3}}e^{\frac{\sigma}{2}}\cos^2\xi\left[(e^{\phi}\sin^2\psi+e^{-\phi}\cos^2\psi)d\psi^2\right. \nonumber \\
& &\left. +e^\phi\cos^2\psi(d\alpha-gA^{12})^2+e^{-\phi}\sin^2\psi(d\beta-gA^{34})^2\right]\label{11D_metric_SO2_SO2}
\end{eqnarray}
with $A^{12}=A^3_{(1)}+A^6_{(1)}$, $A^{34}=A^3_{(1)}-A^6_{(1)}$ and
\begin{equation}
\Delta=e^{2\sigma}\sin^2\xi+e^{-\frac{\sigma}{2}}\cos^2\xi\left(e^{-\phi}\cos^2\psi+e^\phi \sin^2\psi\right).
\end{equation}
From the metric, we see that the $SO(2)\times SO(2)$ symmetry corresponds to the isometry along the $\alpha$ and $\beta$ directions.

\subsection{$AdS_3$ vacua with $SO(2)_{\textrm{diag}}$ symmetry}
\begin{figure}[h!]
  \centering
  \begin{subfigure}[b]{0.45\linewidth}
    \includegraphics[width=\linewidth]{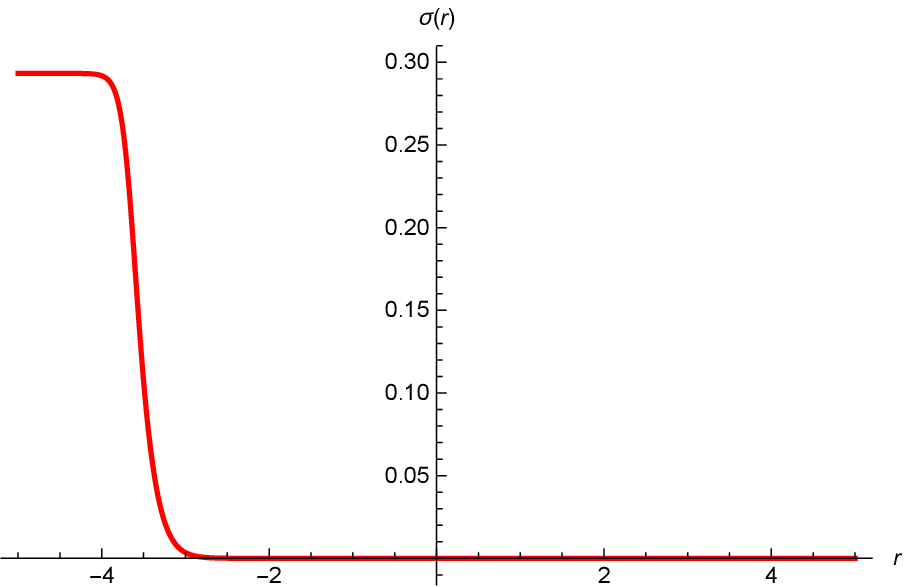}
  \caption{$\sigma$ solution}
  \end{subfigure}
  \begin{subfigure}[b]{0.45\linewidth}
    \includegraphics[width=\linewidth]{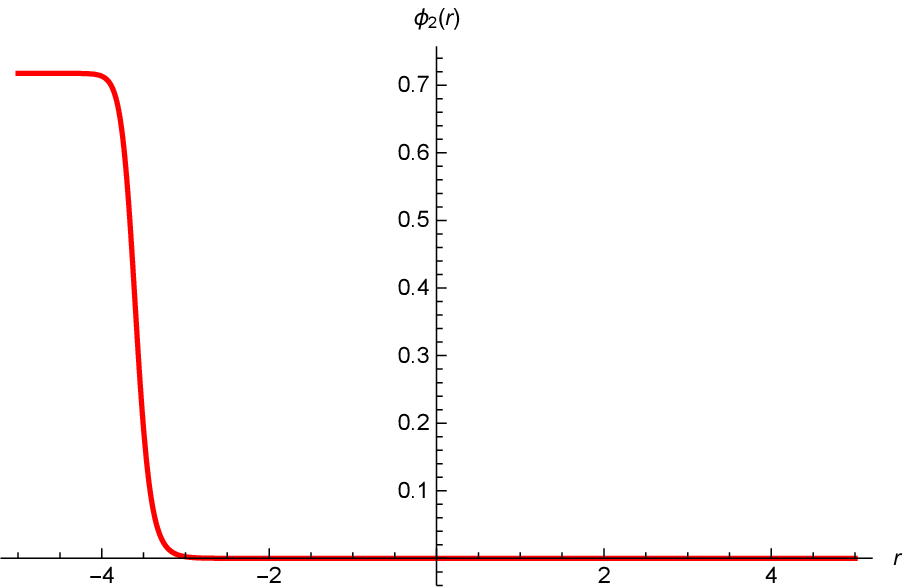}
  \caption{$\phi_2$ solution}
  \end{subfigure}
  \begin{subfigure}[b]{0.45\linewidth}
    \includegraphics[width=\linewidth]{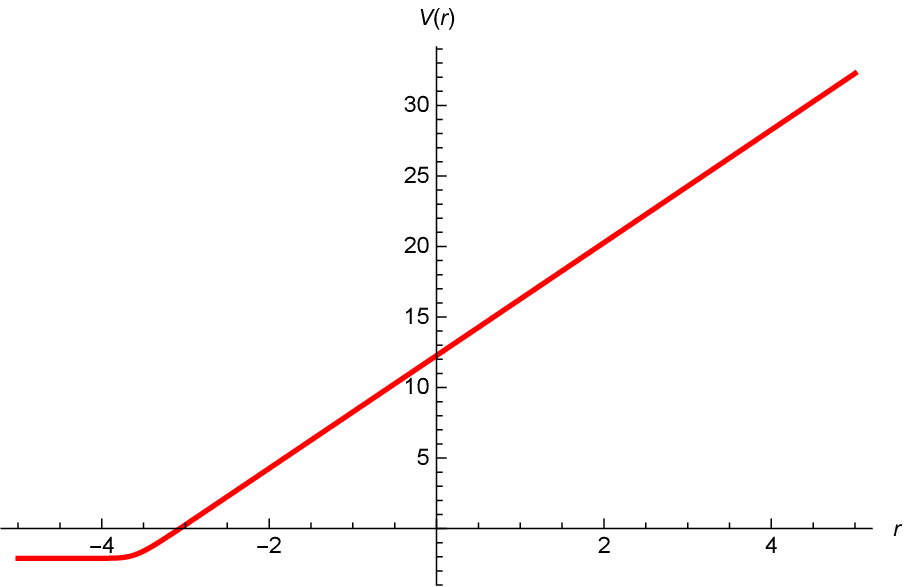}
  \caption{$V$ solution}
  \end{subfigure}
  \begin{subfigure}[b]{0.45\linewidth}
    \includegraphics[width=\linewidth]{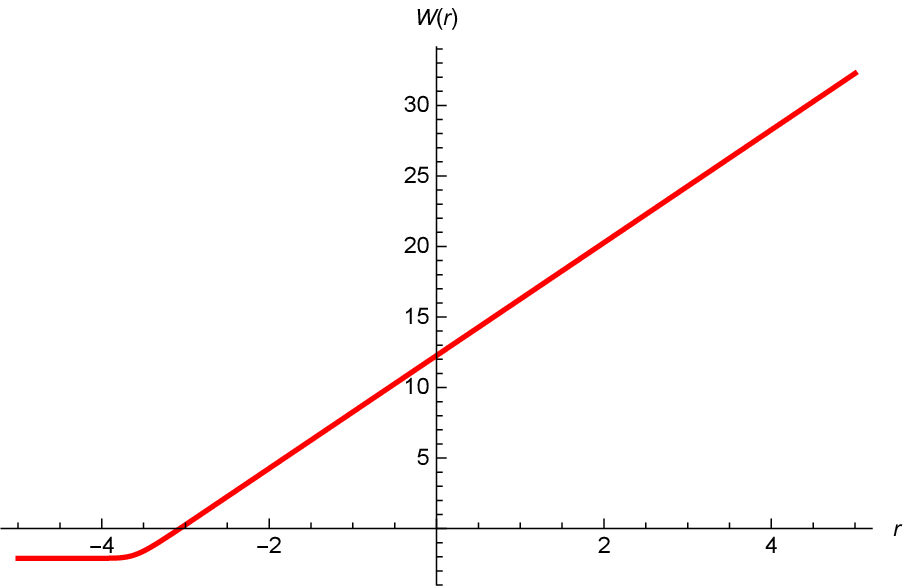}
  \caption{$W$ solution}
  \end{subfigure}
  \caption{An RG flow from $SO(4)$ $N=(1,0)$ SCFT in six dimensions to two-dimensional $N=(2,0)$ SCFT with $SO(2)_{\text{diag}}$ symmetry dual to $AdS_3\times H^2\times H^2$ solution.}
  \label{SO(2)diagH2xH2flow0}
\end{figure}

We now consider $AdS_3$ solutions with $SO(2)_{\text{diag}}\subset SO(2)\times SO(2)\subset SO(3)\times SO(3)$ symmetry. In this case, there are three $SO(2)_{\text{diag}}$ singlets from the nine scalars in $SO(3,3)/SO(3)\times SO(3)$ coset. These correspond to non-compact generators
\begin{equation}
\hat{Y}_1=Y_{11}+Y_{22},\qquad \hat{Y}_2=Y_{33}, \qquad \hat{Y}_3=Y_{12}-Y_{21}\, .
\end{equation}
The coset representative takes the form of
\begin{equation}
L=e^{\phi_1\hat{Y}_1}e^{\phi_2\hat{Y}_2}e^{\phi_3\hat{Y}_3}\, .\label{SO2d_scalar}
\end{equation}
The ansatz for $SO(2)_{\textrm{diag}}$ gauge fields is obtained from that of $SO(2)\times SO(2)$ given in \eqref{A3_1} and \eqref{A6_1} by setting $g_2A^6=g_1 A^3$ or, equivalently,
\begin{equation}
g_2p_{21}=g_1p_{11}\qquad \textrm{and}\qquad g_2p_{22}=g_1p_{12}\, .
\end{equation}
We will also simplify the notation by redefining the charges $p_1=p_{11}$ and $p_2=p_{12}$. In this case, the four-form field strength is given by
\begin{equation}\label{SO(2)diagSig2xSig24form}
H_{(4)}=\frac{p_{1}p_{2}}{8\sqrt{2}hg_2^2}e^{-2(V+W)}(g_1^2-g_2^2) e^{\hat{\theta}_1}\wedge e^{\hat{\varphi}_1}\wedge e^{\hat{\theta}_2}\wedge e^{\hat{\varphi}_2},
\end{equation}
and the twist conditions read
\begin{equation}
g_1p_{1}=k_1\qquad \textrm{and} \qquad g_1p_{2}=k_2\, .
\end{equation}
\indent Using the projection conditions \eqref{SO2_SO2_projection1} and \eqref{gamma_r_projection}, we obtain the corresponding BPS equations. It turns out that compatibility between these BPS equations and field equations requires either $\phi_1=0$ or $\phi_3=0$. Furthermore, setting $\phi_3=0$ gives the same BPS equations as setting $\phi_1=0$ with $\phi_3$ and $\phi_1$ interchanged. We will then consider only the $\phi_3=0$ case with the following BPS equations
\begin{eqnarray}
U'&=&\frac{1}{10}e^{\frac{\sigma}{2}}\left[\cosh{2\phi_1}(g_1e^{-\sigma}\cosh{\phi_2}+g_2e^{-\sigma}\sinh{\phi_2})+8he^{\frac{3\sigma}{2}}\phantom{\frac{g_1}{g_2}}\right. \nonumber \\
& &-2p_1e^{-2V}\left(\cosh{\phi_2}+\frac{g_1}{g_2}\sinh{\phi_2}\right)-2p_2e^{-2W}\left(\cosh{\phi_2}+\frac{g_1}{g_2}\sinh{\phi_2}\right)
\nonumber \\
& &\left. +g_1e^{-\sigma}\cosh{\phi_2}-g_2e^{-\sigma}\sinh{\phi_2}-\frac{3}{4hg_2^2}e^{-\frac{3\sigma}{2}-2(V+W)}(g_1^2-g_2^2)p_{1}p_{2}\right],\\
V'&=&\frac{1}{10}e^{\frac{\sigma}{2}}\left[\cosh{2\phi_1}(g_1e^{-\sigma}\cosh{\phi_2}+g_2e^{-\sigma}\sinh{\phi_2})+8he^{\frac{3\sigma}{2}}\phantom{\frac{g_1}{g_2}}\right. \nonumber \\
& &+8p_1e^{-2V}\left(\cosh{\phi_2}+\frac{g_1}{g_2}\sinh{\phi_2}\right)-2p_2e^{-2W}\left(\cosh{\phi_2}+\frac{g_1}{g_2}\sinh{\phi_2}\right)
\nonumber \\
& &\left. +g_1e^{-\sigma}\cosh{\phi_2}-g_2e^{-\sigma}\sinh{\phi_2}+\frac{1}{2hg_2^2}e^{-\frac{3\sigma}{2}-2(V+W)}(g_1^2-g_2^2)p_{1}p_{2}\right],\\
W'&=&\frac{1}{10}e^{\frac{\sigma}{2}}\left[\cosh{2\phi_1}(g_1e^{-\sigma}\cosh{\phi_2}+g_2e^{-\sigma}\sinh{\phi_2})+8he^{\frac{3\sigma}{2}}\phantom{\frac{g_1}{g_2}}\right.\nonumber \\ & &-2p_1e^{-2V}\left(\cosh{\phi_2}+\frac{g_1}{g_2}\sinh{\phi_2}\right)+8p_2e^{-2W}\left(\cosh{\phi_2}+\frac{g_1}{g_2}\sinh{\phi_2}\right)
\nonumber \\
& &\left. +g_1e^{-\sigma}\cosh{\phi_2}-g_2e^{-\sigma}\sinh{\phi_2}+\frac{1}{2hg_2^2}e^{-\frac{3\sigma}{2}-2(V+W)}(g_1^2-g_2^2)p_{1}p_{2}\right],\quad\\
\sigma'&=&\frac{1}{5}e^{\frac{\sigma}{2}}\left[\cosh{2\phi_1}(g_1e^{-\sigma}\cosh{\phi_2}+g_2e^{-\sigma}\sinh{\phi_2})-32he^{\frac{3\sigma}{2}}\phantom{\frac{g_1}{g_2}}\right.\nonumber \\ & &
-2p_1e^{-2V}\left(\cosh{\phi_2}+\frac{g_1}{g_2}\sinh{\phi_2}\right)-2p_2e^{-2W}\left(\cosh{\phi_2}+\frac{g_1}{g_2}\sinh{\phi_2}\right)\nonumber \\
& &\left.+g_1e^{-\sigma}\cosh{\phi_2}-g_2e^{-\sigma}\sinh{\phi_2}+\frac{1}{2hg_2^2}e^{-\frac{3\sigma}{2}-2(V+W)}(g_1^2-g_2^2)p_{1}p_{2} \right],\\
\phi'_1&=&-\frac{1}{2}e^{-\frac{\sigma}{2}}\sinh{2\phi_1}(g_1\cosh{\phi_2}+g_2\sinh{\phi_2}),\\
\phi'_2&=&\frac{1}{2}e^{\frac{\sigma}{2}}\left[e^{-\sigma}\left[g_2\cosh{\phi_2}-g_1\sinh{\phi_2}-\cosh{2\phi_1}(g_2\cosh{\phi_2}+g_1\sinh{\phi_2})\right]\phantom{\frac{g_1}{g_2}}\right. \nonumber \\
& &\left. -2p_1e^{-2V}\left(\sinh{\phi_2}+\frac{g_1}{g_2}\cosh{\phi_2}\right)-2p_2e^{-2W}\left(\sinh{\phi_2}+\frac{g_1}{g_2}\cosh{\phi_2}\right)\right].\nonumber \\
& &
\end{eqnarray}
In this case, solutions to the BPS equations are asymptotic to the two supersymmetric $AdS_7$ vacua with $SO(4)$ and $SO(3)_{\textrm{diag}}$ symmetries at large $r$. Furthermore, unlike the previous case, all charge parameters are fixed by the twist conditions, and there exist only $AdS_3\times H^2\times H^2$ solutions.
\\
\indent We now look for $AdS_3$ fixed points. The solutions also preserve four supercharges and correspond to $N=(2,0)$ SCFTs in two dimensions as in the previous case. We begin with a class of $AdS_3$ fixed points for $\phi_1=0$
\begin{eqnarray}\label{1SO(2)diagfixedpoint}
\sigma&=&\frac{2}{5}\phi_2+\frac{2}{5}\ln\left[\frac{g_1g_2^2}{12h(g_2^2+2g_1g_2-3g_1^2)}\right],\\
\phi_2&=&\frac{1}{2}\ln\left[\frac{3g_1^2-2g_1g_2-g_2^2}{3g_1^2+2g_1g_2-g_2^2}\right],\\
V&=&W=\frac{1}{10}\ln\left[\frac{27(g_1-g_2)^4(g_1+g_2)^4}{16h^2g_1^8g_2^6(g_2^2-9g_1^2)}\right],\\
L_{AdS_3}&=&\left[\frac{8(9g_1^4 g_2 - 10 g_1^2 g_2^3 + g_2^5)^2}{3hg_1^4(g_2^2-3g_1^2)^5}\right]^{\frac{1}{5}}
\end{eqnarray}
with $g_2>3g_1$ or $g_2<-3g_1$ for $AdS_3$ vacua to exist. An example of RG flows from the $SO(4)$ $AdS_7$ critical point to this $AdS_3\times H^2\times H^2$ fixed point for $g_2=4g_1$ and $h=1$ is shown in figure \ref{SO(2)diagH2xH2flow0} with $\phi_1$ set to zero along the flow.

\begin{figure}[h!]
  \centering
  \begin{subfigure}[b]{0.32\linewidth}
    \includegraphics[width=\linewidth]{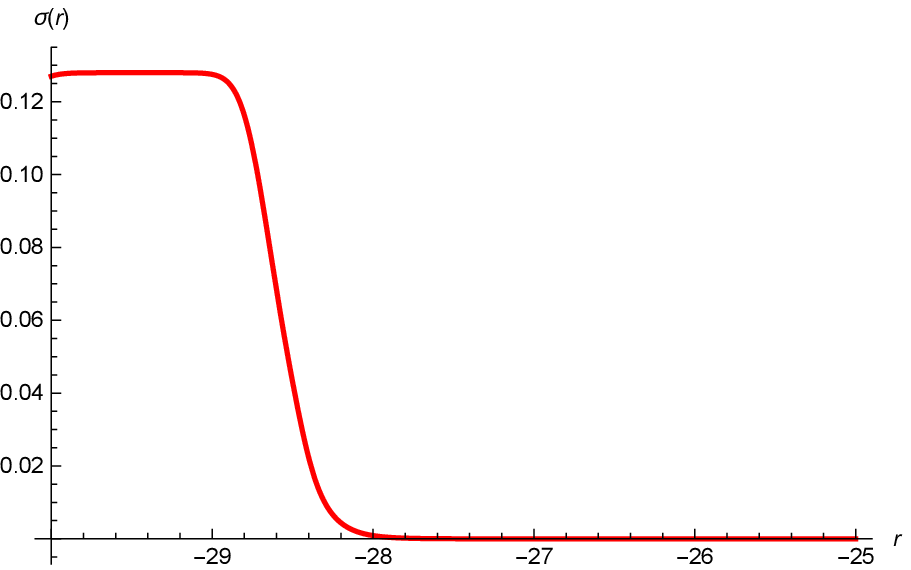}
  \caption{$\sigma$ solution}
  \end{subfigure}
  \begin{subfigure}[b]{0.32\linewidth}
    \includegraphics[width=\linewidth]{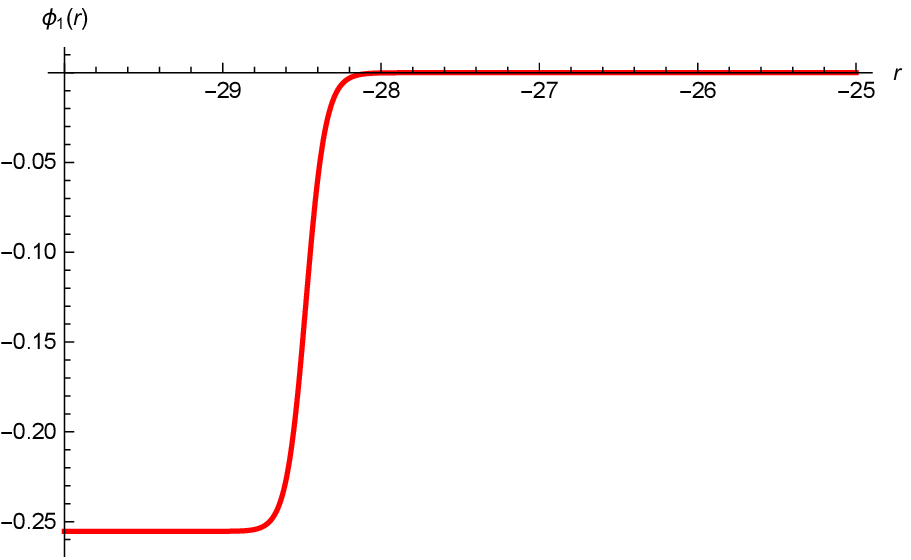}
  \caption{$\phi_1$ solution}
  \end{subfigure}
  \begin{subfigure}[b]{0.32\linewidth}
    \includegraphics[width=\linewidth]{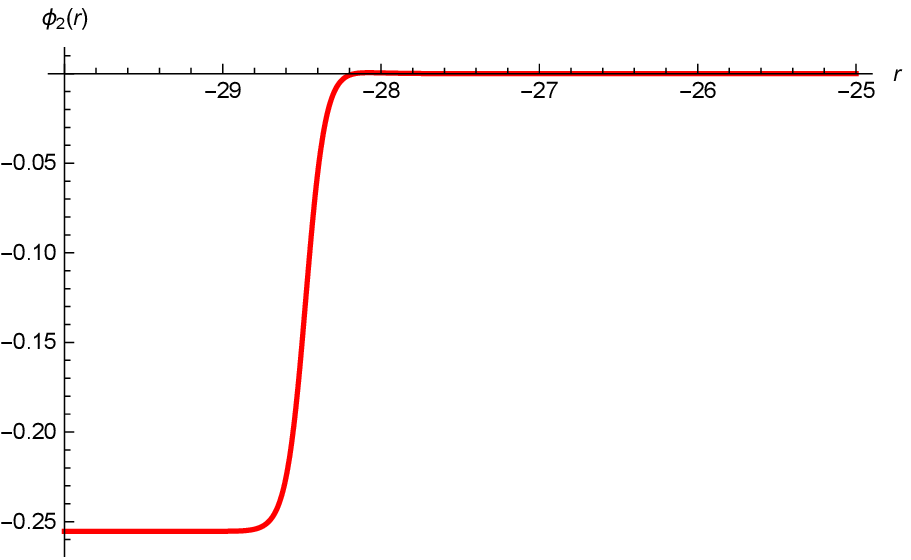}
  \caption{$\phi_2$ solution}
  \end{subfigure}
  \begin{subfigure}[b]{0.32\linewidth}
    \includegraphics[width=\linewidth]{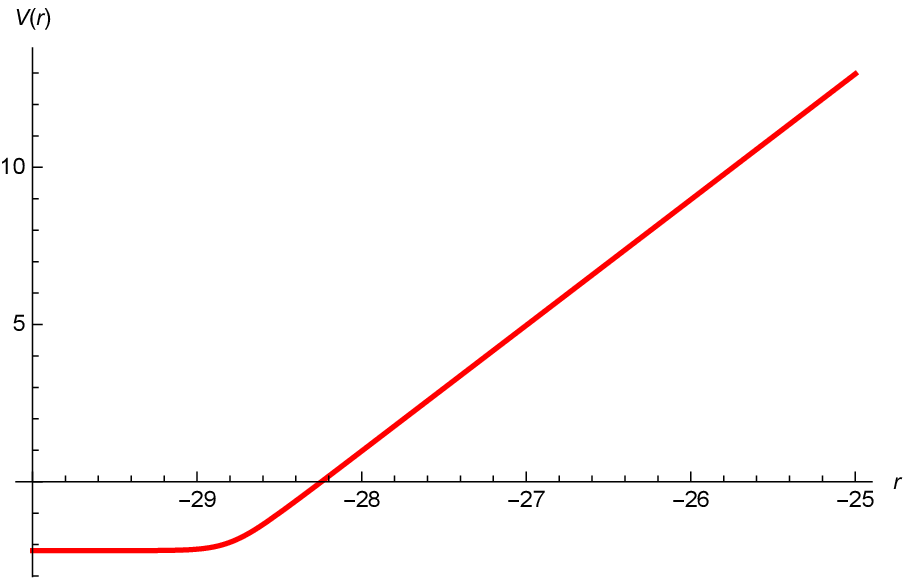}
  \caption{$V$ solution}
  \end{subfigure}
  \begin{subfigure}[b]{0.32\linewidth}
    \includegraphics[width=\linewidth]{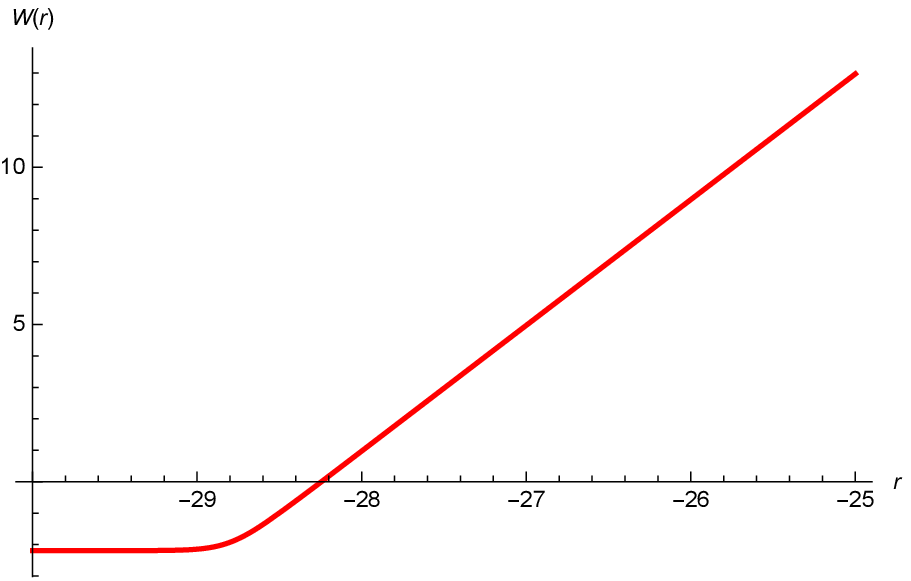}
  \caption{$W$ solution}
  \end{subfigure}
  \caption{An RG flow from $SO(4)$ $N=(1, 0)$ SCFT in six dimensions to two-dimensional $N=(2,0)$ SCFT with $SO(2)_{\text{diag}}$ symmetry dual to $AdS_3\times H^2\times H^2$ solution.}
  \label{SO(2)diagH2xH2flow1}
\end{figure}

\begin{figure}[h!]
  \centering
  \begin{subfigure}[b]{0.32\linewidth}
    \includegraphics[width=\linewidth]{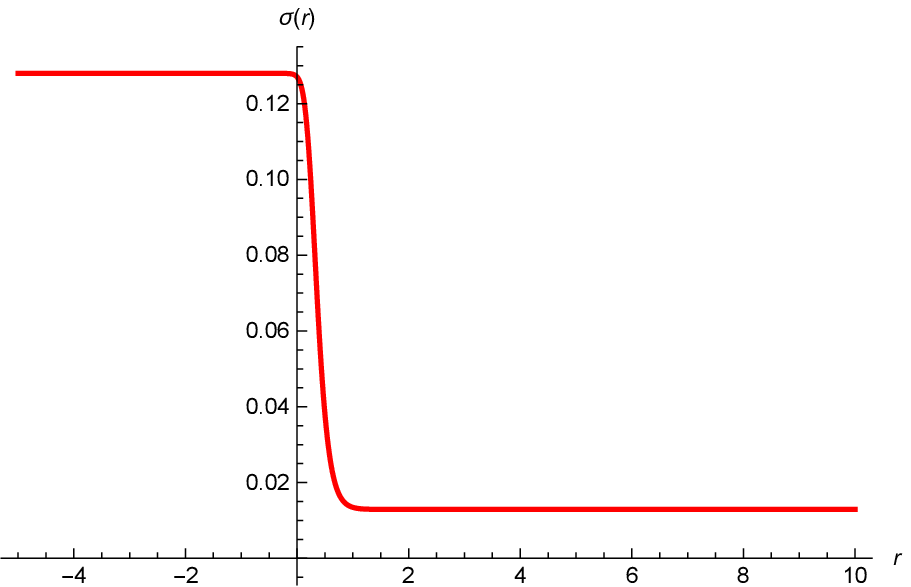}
  \caption{$\sigma$ solution}
  \end{subfigure}
  \begin{subfigure}[b]{0.32\linewidth}
    \includegraphics[width=\linewidth]{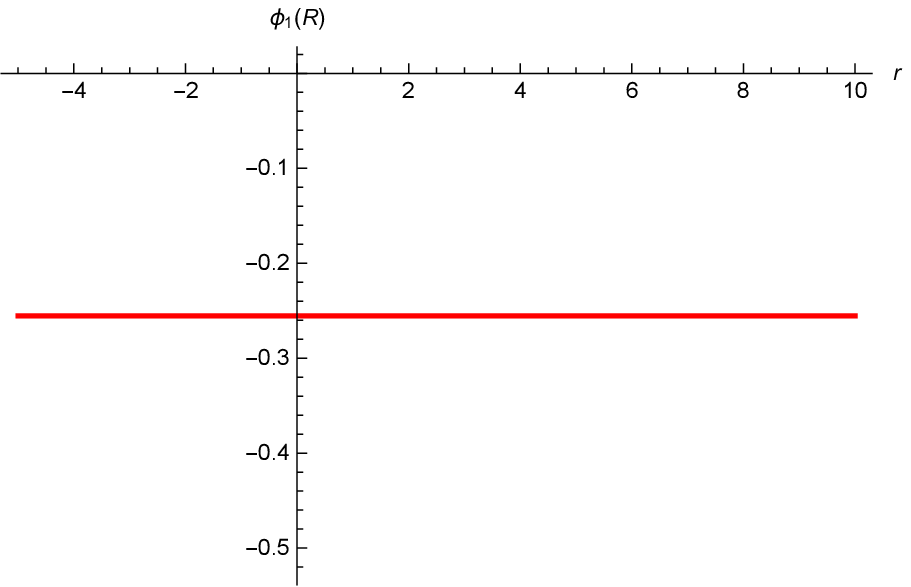}
  \caption{$\phi_1$ solution}
  \end{subfigure}
  \begin{subfigure}[b]{0.32\linewidth}
    \includegraphics[width=\linewidth]{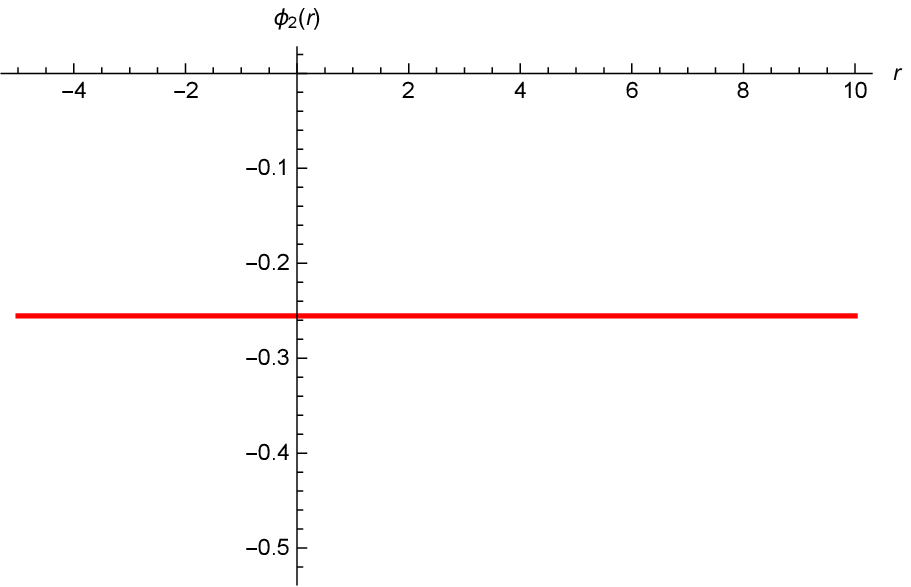}
  \caption{$\phi_2$ solution}
  \end{subfigure}
  \begin{subfigure}[b]{0.32\linewidth}
    \includegraphics[width=\linewidth]{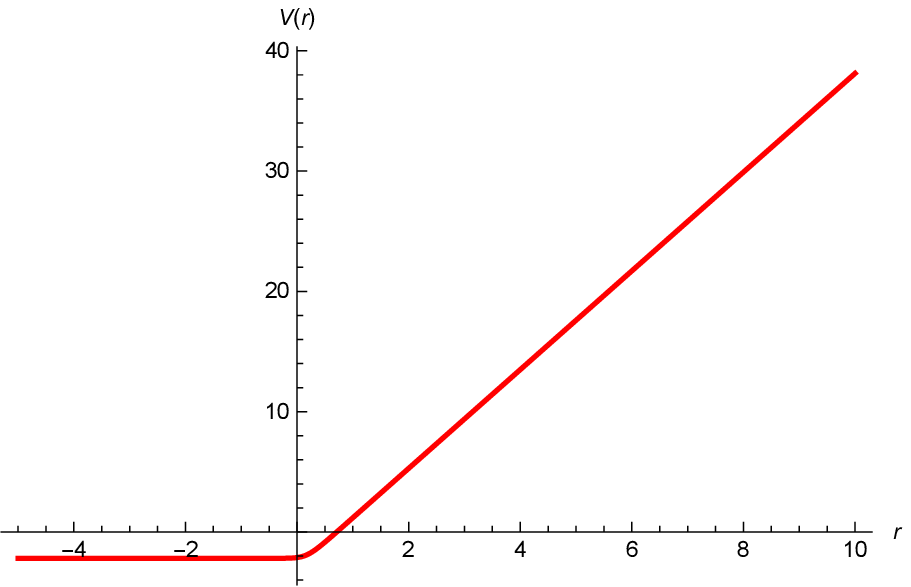}
  \caption{$V$ solution}
  \end{subfigure}
  \begin{subfigure}[b]{0.32\linewidth}
    \includegraphics[width=\linewidth]{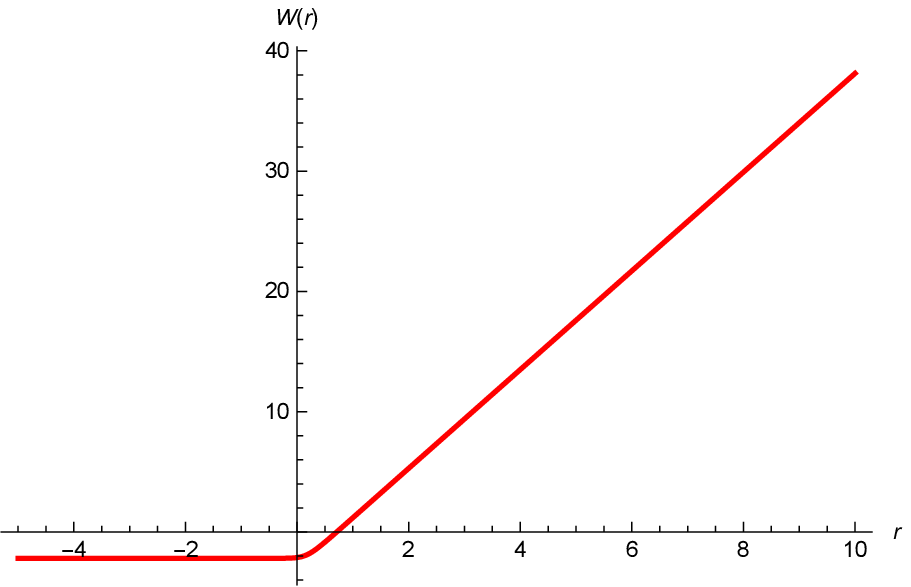}
  \caption{$W$ solution}
  \end{subfigure}
  \caption{An RG flow from $SO(3)$ $N=(1, 0)$ SCFT in six dimensions to two-dimensional $N=(2,0)$ SCFT with $SO(2)_{\text{diag}}$ symmetry dual to $AdS_3\times H^2\times H^2$ solution.}
  \label{SO(2)diagH2xH2flow2}
\end{figure}

Another class of $AdS_3\times H^2\times H^2$ solutions with $\phi_1\neq 0$ is given by
\begin{eqnarray}\label{2SO(2)diagfixedpoint}
\sigma&=&\frac{2}{5}\ln\left[\frac{g_1g_2}{12h\sqrt{(g_2+g_1)(g_2-g_1)}}\right],\nonumber \\
\phi_1&=&\phi_2=\frac{1}{2}\ln\left[\frac{g_2-g_1}{g_2+g_1}\right],\nonumber \\
V&=&W=\frac{1}{10}\ln\left[\frac{27(g_1^2-g_2^2)^4}{16h^2g_1^8g_2^8}\right],\qquad
L_{AdS_3}=\left[\frac{8(g_1^2-g_2^2)^2}{3hg_1^4g_2^4}\right]^{\frac{1}{5}}
\end{eqnarray}
with the condition $g_2>g_1$. Examples of RG flow solutions from the $SO(4)$ and $SO(3)$ $AdS_7$ vacua to these $AdS_3\times H^2\times H^2$ fixed points are respectively shown in figures \ref{SO(2)diagH2xH2flow1} and \ref{SO(2)diagH2xH2flow2} for $g_2=4g_1$ and $h=1$. Note that $\phi_1$ and $\phi_2$ have the same value at both the $SO(3)$ $AdS_7$ and $AdS_3$ fixed points.

\begin{figure}[h!]
  \centering
  \begin{subfigure}[b]{0.32\linewidth}
    \includegraphics[width=\linewidth]{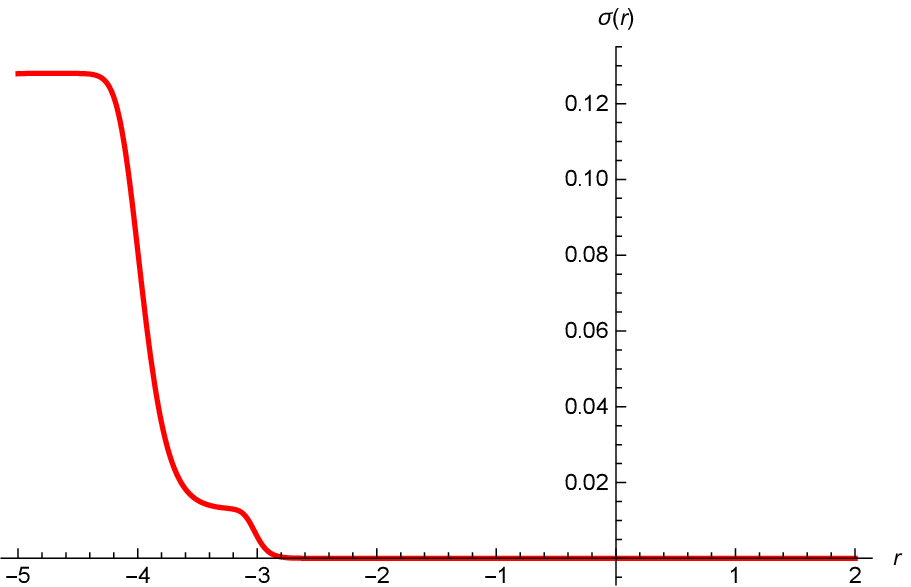}
  \caption{$\sigma$ solution}
  \end{subfigure}
  \begin{subfigure}[b]{0.32\linewidth}
    \includegraphics[width=\linewidth]{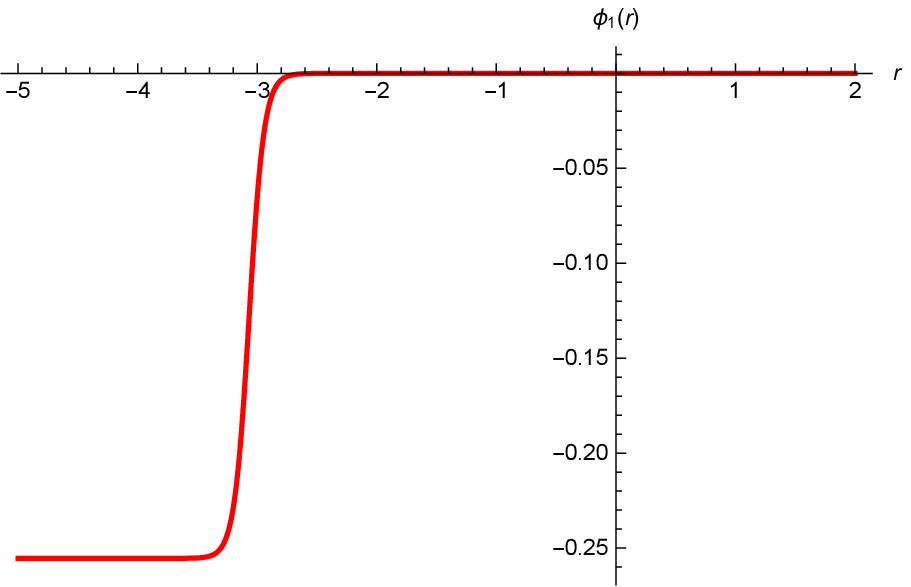}
  \caption{$\phi_1$ solution}
  \end{subfigure}
  \begin{subfigure}[b]{0.32\linewidth}
    \includegraphics[width=\linewidth]{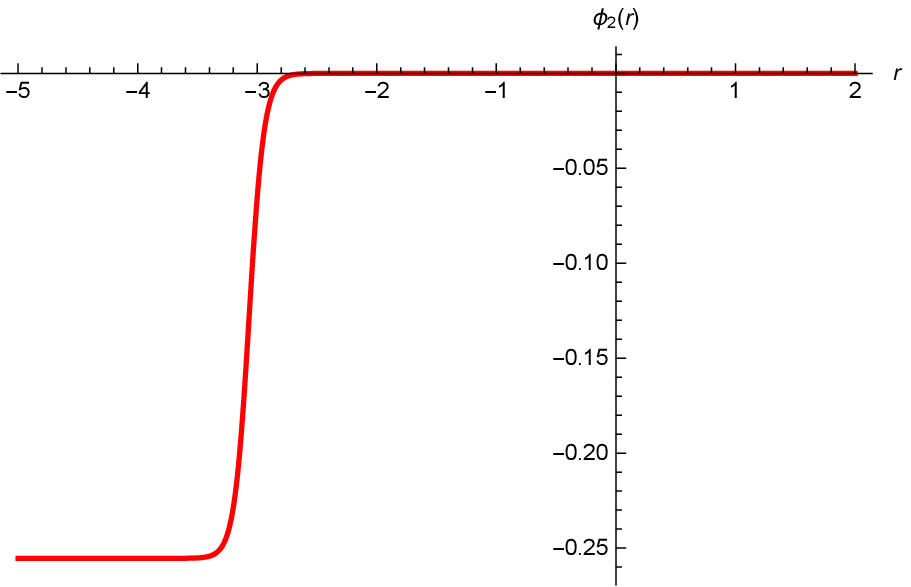}
  \caption{$\phi_2$ solution}
  \end{subfigure}
  \begin{subfigure}[b]{0.32\linewidth}
    \includegraphics[width=\linewidth]{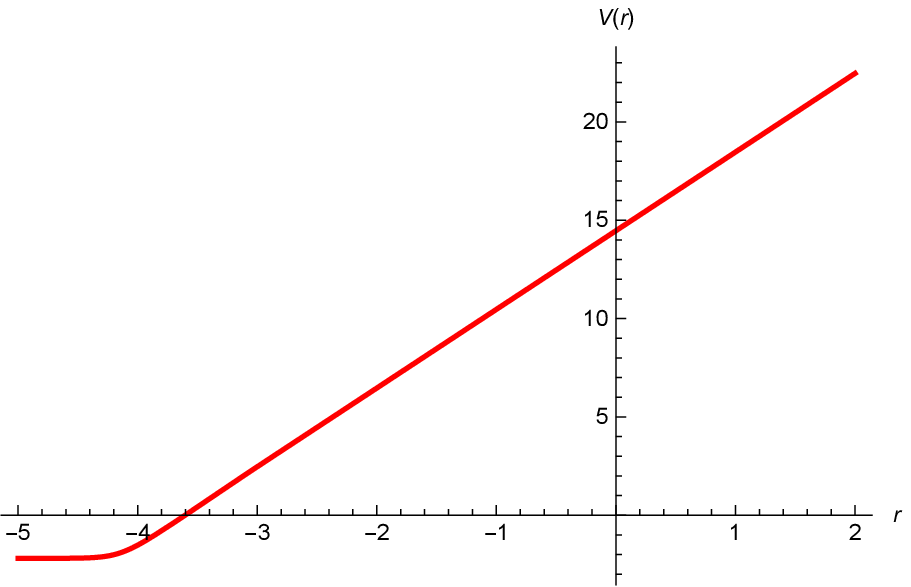}
  \caption{$V$ solution}
  \end{subfigure}
  \begin{subfigure}[b]{0.32\linewidth}
    \includegraphics[width=\linewidth]{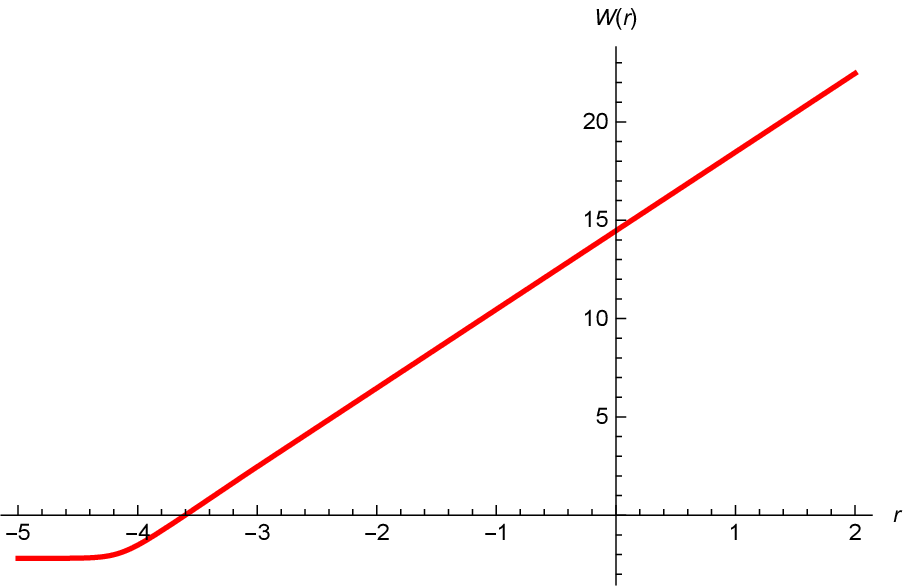}
  \caption{$W$ solution}
  \end{subfigure}
  \caption{An RG flow from $SO(4)$ $N=(1, 0)$ SCFT to $SO(3)$ $N=(1, 0)$ SCFT in six dimensions and then to two-dimensional $N=(2,0)$ SCFT with $SO(2)_{\text{diag}}$ symmetry dual to $AdS_3\times H^2\times H^2$ solution.}
  \label{SO(2)diagH2xH2flow3}
\end{figure}

Moreover, with a suitable set of boundary conditions, there exists an RG flow from $SO(4)$ $AdS_7$ to $SO(3)$ $AdS_7$ fixed points and then to $AdS_3\times H^2\times H^2$ critical point as shown in figure \ref{SO(2)diagH2xH2flow3}. All $AdS_3$ vacua and RG flows in this case cannot be uplifted to eleven dimensions since the existence of these solutions require $g_1\neq g_2$. Therefore, the corresponding holographic interpretation is rather limited.

\subsection{$AdS_3$ vacua with $SO(2)_R$ symmetry}
\begin{figure}[h!]
  \centering
  \begin{subfigure}[b]{0.32\linewidth}
    \includegraphics[width=\linewidth]{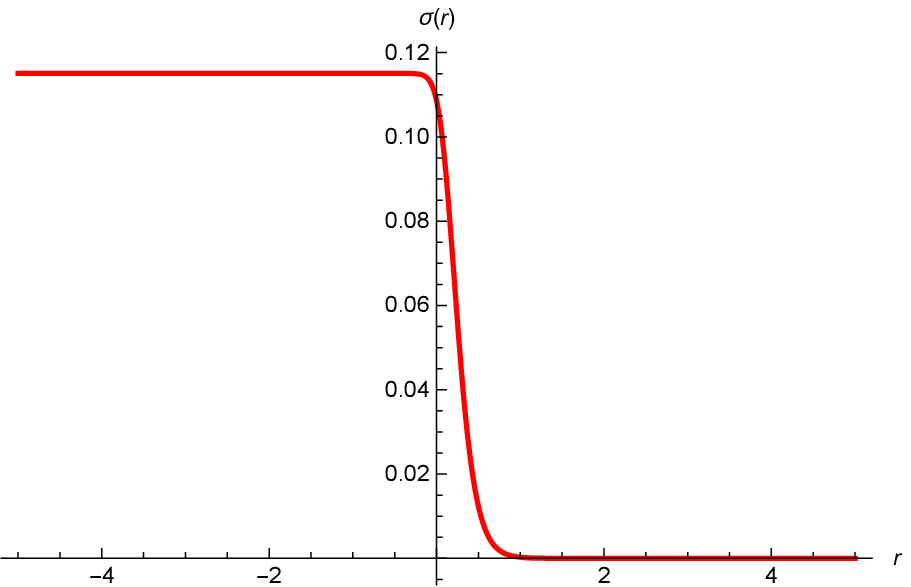}
  \caption{$\sigma$ solution}
  \end{subfigure}
  \begin{subfigure}[b]{0.32\linewidth}
    \includegraphics[width=\linewidth]{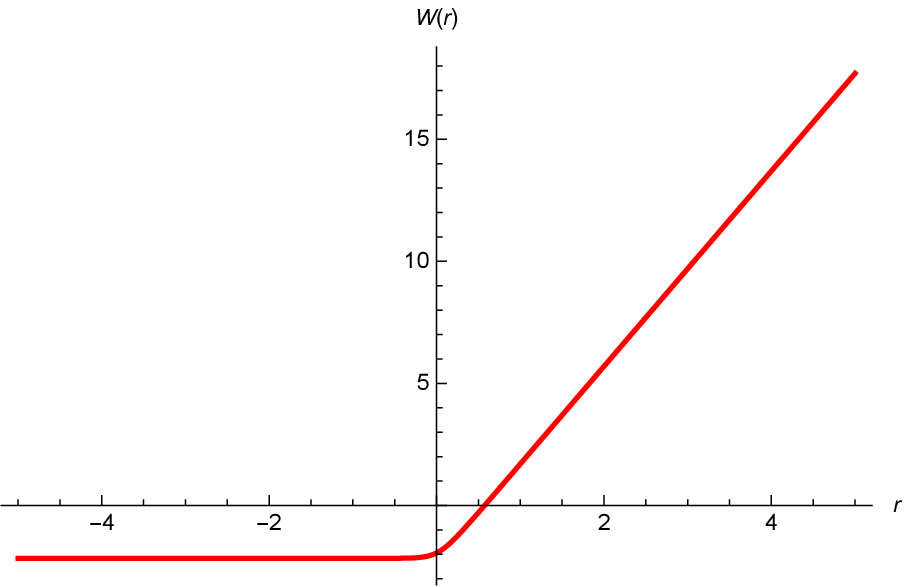}
  \caption{$W$ solution}
  \end{subfigure}
   \begin{subfigure}[b]{0.32\linewidth}
    \includegraphics[width=\linewidth]{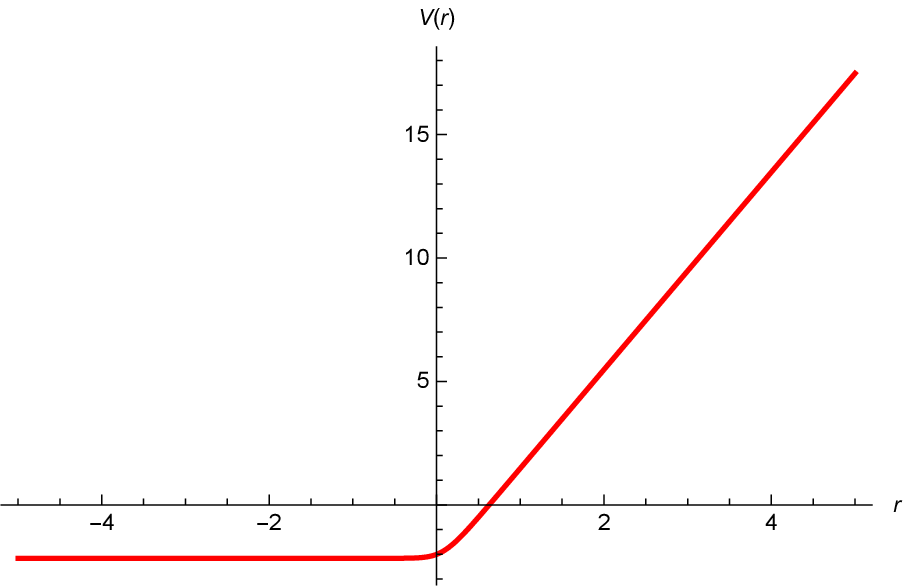}
  \caption{$V$ solution}
  \end{subfigure}
  \caption{An RG flow from $SO(4)$ $N=(1, 0)$ SCFT in six dimensions to two-dimensional $N=(2,0)$ SCFT with $SO(2)_R$ symmetry dual to $AdS_3\times H^2\times H^2$ solution.}
  \label{SO(2)RH2xH2flow}
\end{figure}

We now move on to $AdS_3$ solutions with $SO(2)_R\subset SO(3)_R$ symmetry. There are three $SO(2)_R$ singlet scalars from $SO(3,3)/SO(3)\times SO(3)$ coset. These correspond to non-compact generators $Y_{31}$, $Y_{32}$ and $Y_{33}$. Therefore, the coset representative can be written as
\begin{equation}
L=e^{\phi_1 Y_{31}}e^{\phi_2 Y_{32}}e^{\phi_3 Y_{33}}\, .
\end{equation}
\indent To perform the twist, we take the following ansatz for the $SO(2)_{R}$ gauge field
\begin{equation}
A^3_{(1)}=-\frac{p_{1}}{k_1}e^{-V}\frac{f'_{k_1}(\theta_1)}{f_{k_1}(\theta_1)}e^{\hat{\varphi}_1}-\frac{p_{2}}{k_2}e^{-W}\frac{f'_{k_2}(\theta_2)}{f_{k_2}(\theta_2)}e^{\hat{\varphi}_2}.
\end{equation}
The four-form field strength in this case is given by
\begin{equation}\label{SO(2)RSig2xSig24form}
H_{(4)}=-\frac{1}{8\sqrt{2}h}e^{-2(V+W)}p_{1}p_{2}e^{\hat{\theta}_1}\wedge e^{\hat{\varphi}_1}\wedge e^{\hat{\theta}_2}\wedge e^{\hat{\varphi}_2}\, .
\end{equation}
\indent We can now repeat the same procedure as in the previous two cases to find the corresponding BPS equations. In this case, it turns out that compatibility between the BPS equations and second-order field equations allows only one of the $\phi_i$, $i=1,2,3$, to be non-vanishing. We have verified that any of the $\phi_i$ leads to the same set of BPS equations. We will choose $\phi_1=\phi_2=0$ and $\phi_3\neq 0$ for definiteness. With this choice, the BPS equations are given by
\begin{eqnarray}
U'&=&\frac{1}{5}e^{\frac{\sigma}{2}}\left[g_1e^{-\sigma}+4he^{\frac{3\sigma}{2}}-e^{-2V}p_1-e^{-2W}p_2+\frac{3}{8h}e^{-2(V+W)}p_1p_2\right],\quad\\
V'&=&\frac{1}{5}e^{\frac{\sigma}{2}}\left[g_1e^{-\sigma}+4he^{\frac{3\sigma}{2}}+4e^{-2V}p_1-e^{-2W}p_2-\frac{1}{4h}e^{-2(V+W)}p_1p_2\right], \quad \\
W'&=&\frac{1}{5}e^{\frac{\sigma}{2}}\left[g_1e^{-\sigma}+4he^{\frac{3\sigma}{2}}-e^{-2V}p_1+4e^{-2W}p_2-\frac{1}{4h}e^{-2(V+W)}p_1p_2\right], \quad \\
\sigma'&=&\frac{2}{5}e^{\frac{\sigma}{2}}\left[g_1e^{-\sigma}-16he^{\frac{3\sigma}{2}}-e^{-2V}p_1-e^{-2W}p_2-\frac{1}{4h}e^{-2(V+W)}p_1p_2\right],\quad \\
\phi'_3&=&-e^{-\frac{\sigma}{2}}\left[g_1+e^\sigma(e^{-2V}p_1+e^{-2W}p_2)\right]\sinh{\phi_3}\, .
\end{eqnarray}
For these equations, there exist $AdS_3$ fixed points only for $k_1=k_2=-1$. The resulting $AdS_3\times H^2\times H^2$ solution is given by
\begin{eqnarray}
\phi_3&=&0,\qquad \sigma=\frac{2}{5}\ln\left[\frac{g_1}{12h}\right],\nonumber \\
V&=&W=\frac{1}{10}\ln\left[\frac{27}{16h^2g_1^8}\right],\qquad L_{AdS_3}=\left[\frac{8}{3hg_1^4}\right]^{\frac{1}{5}}\, .
\end{eqnarray}
This solution again preserves four supercharges and corresponds to $N=(2,0)$ SCFT in two dimensions. An example of RG flow solutions from $N=(1,0)$ six-dimensional SCFT to this fixed point for $h=1$ and $\phi_3=0$ is shown in figure \ref{SO(2)RH2xH2flow}. Note that the $AdS_3$ fixed point and the RG flow are also solutions of pure $N=2$ gauged supergravity with $SU(2)$ gauge group.

As in the case of $AdS_3$ solutions with $SO(2)\times SO(2)$ symmetry, the above solutions can be uplifted to eleven dimensions by setting $g_2=g_1$. The eleven-dimensional metric can be obtained from \eqref{11D_metric_SO2_SO2} by setting $\phi=0$ and $A^6_{(1)}=0$, or equivalently $A^{12}=A^{34}\equiv A^3$. The result is given by
\begin{eqnarray}
d\hat{s}^2_{11}&=&\Delta^{\frac{1}{3}}\left[e^{2U}dx^2_{1,1}+dr^2+e^{2V}ds^2_{\Sigma^2_{k_1}}+e^{2W}ds^2_{\Sigma^2_{k_2}}\right]\nonumber \\
& &+\frac{2}{g^2}\Delta^{-\frac{2}{3}}\left(e^{-2\sigma}\cos^2\xi+e^{\frac{\sigma}{2}}\sin^2\xi\right)d\xi^2+\frac{1}{2g^2}\Delta^{-\frac{2}{3}}e^{\frac{\sigma}{2}}\cos^2\xi\left[d\psi^2\right. \nonumber \\
& &\left. +\cos^2\psi(d\alpha-gA^3)^2+\sin^2\psi(d\beta-gA^3)^2\right]\label{11D_metric_SO2R}
\end{eqnarray}
with 
\begin{equation}
\Delta=e^{2\sigma}\sin^2\xi+e^{-\frac{\sigma}{2}}\cos^2\xi\, .
\end{equation}
It should also be pointed out that the seven-dimensional solution in this case has recently been discussed in the context of massive type IIA theory in \cite{universal_6d_7d}. 

\section{Supersymmetric $AdS_3\times M^4_k$ solutions and RG flows}\label{AdS3_M4k}
In this section, we repeat the same analysis for $M^4$ being a Kahler four-cycle and look for solutions of the form $AdS_3\times M^4_k$. For the constant $k=1,0,-1$, the Kahler four-cycle is given by a two-dimensional complex space $CP^2$, a four-dimensional flat space $\mathbb{R}^4$, or a two-dimensional complex hyperbolic space $CH^2$, respectively. The Kahler four-cycle has $U(2)\sim SU(2)\times U(1)$ spin connection. We can perform a twist by using either $SO(2)_R\sim U(1)_R$ or $SO(3)_R\sim SU(2)_R$ gauge fields to cancel the $U(1)$ or $SU(2)$ parts of the spin connection.

\subsection{$AdS_3$ vacua with $SO(2)\times SO(2)$ symmetry}\label{SO(2)xSO(2)KahlerSec}
\begin{figure}[h!]
  \centering
  \begin{subfigure}[b]{0.32\linewidth}
    \includegraphics[width=\linewidth]{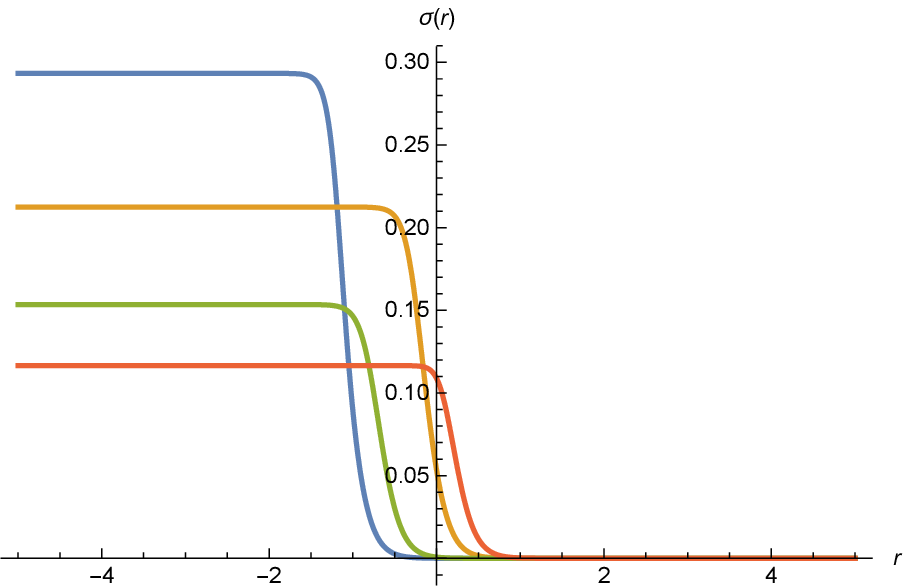}
  \caption{$\sigma$ solution}
  \end{subfigure}
  \begin{subfigure}[b]{0.32\linewidth}
    \includegraphics[width=\linewidth]{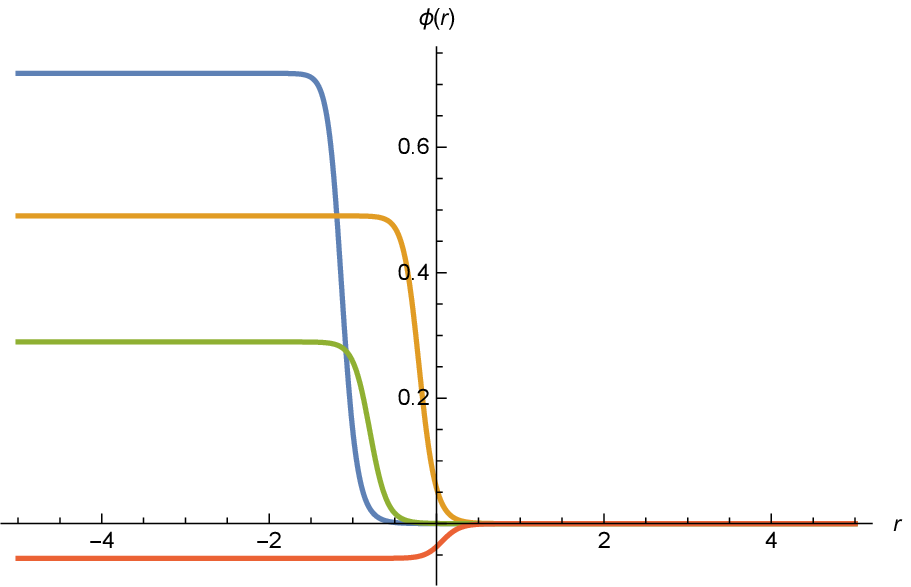}
  \caption{$\phi$ solution}
  \end{subfigure}
  \begin{subfigure}[b]{0.32\linewidth}
    \includegraphics[width=\linewidth]{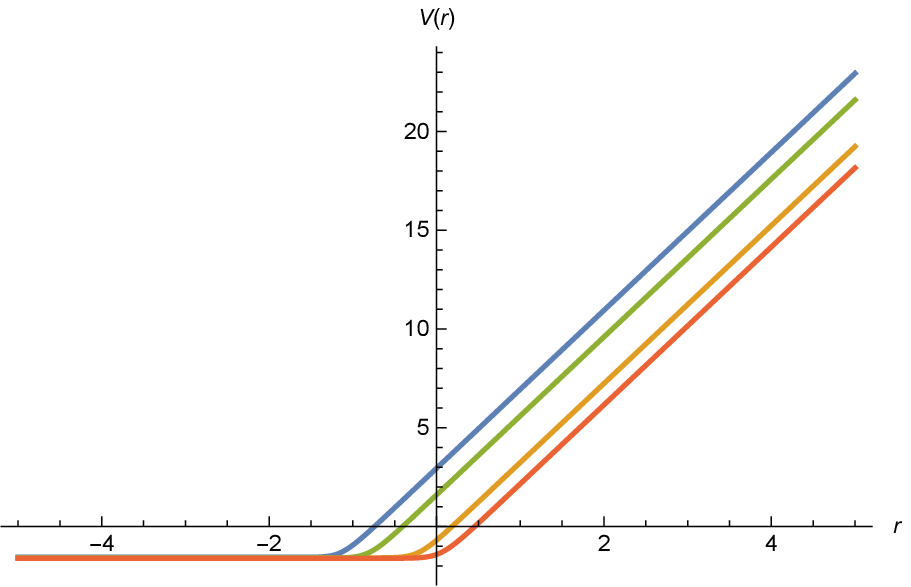}
  \caption{$V$ solution}
  \end{subfigure}
  \caption{RG flows from $SO(4)$ $N=(1, 0)$ SCFT in six dimensions to two-dimensional $N=(2,0)$ SCFT with $SO(2)\times SO(2)$ symmetry dual to $AdS_3\times CH^2$ solution. The blue, orange, green and red curves refer to $p_2=-\frac{1}{64}, -\frac{1}{80}, -\frac{1}{120}, \frac{1}{580}$, respectively.}
  \label{SO(2)xSO(2)Kahlerflow}
\end{figure}

We begin with $AdS_3$ vacua with $SO(2)\times SO(2)$ symmetry and take the following ansatz for the seven-dimensional metric
\begin{equation}\label{AdS3xSO(2)Kahlermetric}
ds_7^2=e^{2U(r)}dx^2_{1,1}+dr^2+e^{2V(r)}ds^2_{M^4_k}\, .
\end{equation}
The metric on the Kahler four-cycle $M^4_k$ is given by
\begin{equation}\label{SO(2)Kahlermetric}
ds^2_{M^4_{k}}=\frac{d\varphi^2}{f^2_k(\varphi)}+\frac{\varphi^2}{f_k(\varphi)}(\tau_1^2+\tau_2^2)+\frac{\varphi^2}{f^2_k(\varphi)}\tau_3^2
\end{equation}
with $\varphi\in[0,\frac{\pi}{2}]$ and the function $f_{k}(\varphi)$ defined by
\begin{equation}
f_{k}(\varphi)=1+k\varphi^2\, .\label{f_M4}
\end{equation}
$\tau_i$, $i=1,2,3$, are $SU(2)$ left-invariant one-forms satisfying $d\tau_i=\frac{1}{2}\varepsilon_{ijk}\tau_j\wedge\tau_k$. Their explicit form is given by
\begin{eqnarray}
\tau_1&=&-\sin{\chi}d\theta+\cos{\chi}\sin{\theta}d\psi,\nonumber \\
\tau_2&=&\cos{\chi}d\theta+\sin{\chi}\sin{\theta}d\psi,\nonumber \\
\tau_3&=&d\chi+\cos{\theta}d\psi\, .\label{SU2_left_1_form}
\end{eqnarray}
The ranges of the coordinates are $\theta\in[0, \pi]$, $\psi\in[0, 2\pi]$, and $\chi\in[0, 4\pi]$.
\\
\indent By choosing the following choice of vielbein
\begin{eqnarray}
e^{\hat{\alpha}}&=& e^Udx^\alpha, \qquad e^{\hat{1}}=e^V\frac{\varphi}{\sqrt{f_k(\varphi)}}\tau_1,\qquad e^{\hat{2}}=e^V\frac{\varphi}{\sqrt{f_k(\varphi)}}\tau_2,\nonumber \\
e^{\hat{r}}&=& dr, \qquad
 e^{\hat{3}}=e^V\frac{\varphi}{f_k(\varphi)}\tau_3,\qquad e^{\hat{4}}=e^V\frac{1}{f_k(\varphi)}d\varphi,\label{AdS3xSO(2)Kahlervielbein}
\end{eqnarray}
we find non-vanishing components of the spin connection
\begin{eqnarray}
{\omega^{\hat{\alpha}}}_{\hat{r}}&=& U'e^{\hat{\alpha}}, \qquad {\omega^{\hat{i}}}_{\hat{r}}=V'e^{\hat{i}}, \, i=1,2,3,\qquad  {\omega^{\hat{4}}}_{\hat{r}}=V'e^{\hat{4}},\nonumber \\
{\omega^{\hat{1}}}_{\hat{4}}&=&\ {\omega^{\hat{2}}}_{\hat{3}}=\frac{1}{\sqrt{f_{k}(\varphi)}}\tau_1, \qquad {\omega^{\hat{1}}}_{\hat{2}}=\frac{(2k\varphi^2+1)}{f_{k}(\varphi)}\tau_3,\nonumber \\
{\omega^{\hat{2}}}_{\hat{4}}&=&\ {\omega^{\hat{3}}}_{\hat{1}}=\frac{1}{\sqrt{f_{k}(\varphi)}}\tau_2, \qquad {\omega^{\hat{4}}}_{\hat{3}}=\frac{(k\varphi^2-1)}{f_{k}(\varphi)}\tau_3\, .\label{AdS3xSO(2)KahlerspinCon}
\end{eqnarray}
\indent We can now perform the twist by turning on $SO(2)\times SO(2)$ gauge fields with the following ansatz
\begin{equation}
A^3_{(1)}=p_1\frac{3\varphi^2}{\sqrt{f_{k}(\varphi)}}\tau_3\qquad \textrm{and}\qquad
A^6_{(1)}=p_2\frac{3\varphi^2}{\sqrt{f_{k}(\varphi)}}\tau_3\, .\label{SO2_SO2_gauge2}
\end{equation}
The associated two-form field strengths are given by
\begin{equation}\label{SO(2)KahlerF}
F^3_{(2)}=3e^{-2V}p_1J_{(2)} \qquad \textrm{and}\qquad F^6_{(2)}=3e^{-2V}p_2J_{(2)}
\end{equation}
where $J_{(2)}$ is the Kahler structure defined by
\begin{equation}\label{KahlerStr}
J_{(2)}=e^{\hat{1}}\wedge e^{\hat{2}}-e^{\hat{3}}\wedge e^{\hat{4}}\, .
\end{equation}
To implement the twist, we impose the following projectors on the Killing spinors
\begin{equation}\label{SO(2)KahlerProjCon}
\gamma_{\hat{1}\hat{2}}\epsilon=-\gamma_{\hat{3}\hat{4}}\epsilon=i\sigma^3\epsilon
\end{equation}
together with the twist condition
\begin{equation}\label{SO(2)KahlerQYM}
g_1p_1=k\, .
\end{equation}
As in the previous cases, we need to turn on the three-form field with the field strength
\begin{equation}\label{SO(2)xSO(2)Kahler4form}
H_{(4)}=\frac{9}{8\sqrt{2}h}e^{-4V}(p_1^2-p_2^2) e^{\hat{1}}\wedge e^{\hat{2}}\wedge e^{\hat{3}}\wedge e^{\hat{4}}.
\end{equation}
\indent With all these and the $\gamma_r$ projector \eqref{gamma_r_projection}, we can derive the following BPS equations
\begin{eqnarray}
U'&=&\frac{1}{5}e^{\frac{\sigma}{2}}\Big[(g_1e^{-\sigma}\cosh{\phi}+4he^{-\frac{5\sigma}{2}})-6e^{-2V}(p_1\cosh{\phi}+p_2\sinh{\phi})\nonumber\\&&+\frac{27}{8h}e^{-\frac{3\sigma}{2}-4V}(p_1^2-p_2^2)\Big],\\
V'&=&\frac{1}{5}e^{\frac{\sigma}{2}}\Big[(g_1e^{-\sigma}\cosh{\phi}+4he^{-\frac{5\sigma}{2}})+9e^{-2V}(p_1\cosh{\phi}+p_2\sinh{\phi})\nonumber\\&&-\frac{9}{4h}e^{-\frac{3\sigma}{2}-4V}(p_1^2-p_2^2)\Big],\\
\sigma'&=&\frac{2}{5}e^{\frac{\sigma}{2}}\Big[(g_1e^{-\sigma}\cosh{\phi}-16he^{-\frac{5\sigma}{2}})-6e^{-2V}(p_1\cosh{\phi}+p_2\sinh{\phi})\nonumber\\&&-\frac{9}{4h}e^{-\frac{3\sigma}{2}-4V}(p_1^2-p_2^2)\Big],\\
\phi'&=&-g_1e^{-\frac{\sigma}{2}}\sinh{\phi}-6e^{\frac{\sigma}{2}-2V}(p_1\sinh{\phi}+p_2\cosh{\phi})
\end{eqnarray}
with $\phi$ being the $SO(2)\times SO(2)$ singlet scalar in \eqref{SO2_SO2_scalar}.
\\
\indent The BPS equations admit an $AdS_3\times CH^2$ fixed point given by
\begin{eqnarray}\label{SO(2)xSO(2)Kahlerfixedpoint}
\sigma&=&\frac{2}{5}\ln\left[\frac{g_1p_1^2}{12h\sqrt{p_1^4-10p_1^2p_2^2+9p_2^4}}\right],\nonumber \\
\phi&=&\frac{1}{2}\ln\left[\frac{p_1^2+2p_1p_2-3p_2^2}{p_1^2-2p_1p_2-3p_2^2}\right],\nonumber \\
V&=&\frac{1}{10}\ln\left[\frac{3^8(p_1^2-p_2^2)^4}{16h^2g_1^3(9p_1p_2^2-p_1^3)}\right],\nonumber \\
L_{AdS_3}&=&\left[\frac{8(p_1^5-10p_1^3p_2^2+9p_1p_2^4)^2}{3hg_1^4(p_1^2-3p_2^2)^5}\right]^{\frac{1}{5}}\, .
\end{eqnarray}
The $AdS_3$ solution preserves four supercharges and exists for
\begin{equation}
-\frac{1}{48h}<p_2<\frac{1}{48h}
\end{equation}
with $g_1=16h$, $k=-1$, and $h>0$. The $AdS_3\times CH^2$ fixed point is dual to an $N=(2,0)$ two-dimensional SCFT.
\\
\indent Examples of RG flows interpolating between this $AdS_3$ fixed point and the $SO(4)$ $AdS_7$ critical point for $h=1$ and different values of $p_2$ are shown in figure \ref{SO(2)xSO(2)Kahlerflow}.

As in the $\Sigma^2\times \Sigma^2$ case, the $AdS_3\times CH^2$ fixed point and the associated RG flows can be uplifted to eleven dimensions by setting $g_2=g_1$. The eleven-dimensional metric can be obtained from \eqref{11D_metric_SO2_SO2} by replacing $e^{2V}ds^2_{\Sigma^2_{k_1}}+e^{2W}ds^2_{\Sigma^2_{k_2}}$ by $e^{2V}ds^2_{M^4_k}$ and using the gauge fields in \eqref{SO2_SO2_gauge2}. We will not repeat it here.

\subsection{$AdS_3$ vacua with $SO(2)_{\text{diag}}$ symmetry}
\begin{figure}[h!]
  \centering
  \begin{subfigure}[b]{0.32\linewidth}
    \includegraphics[width=\linewidth]{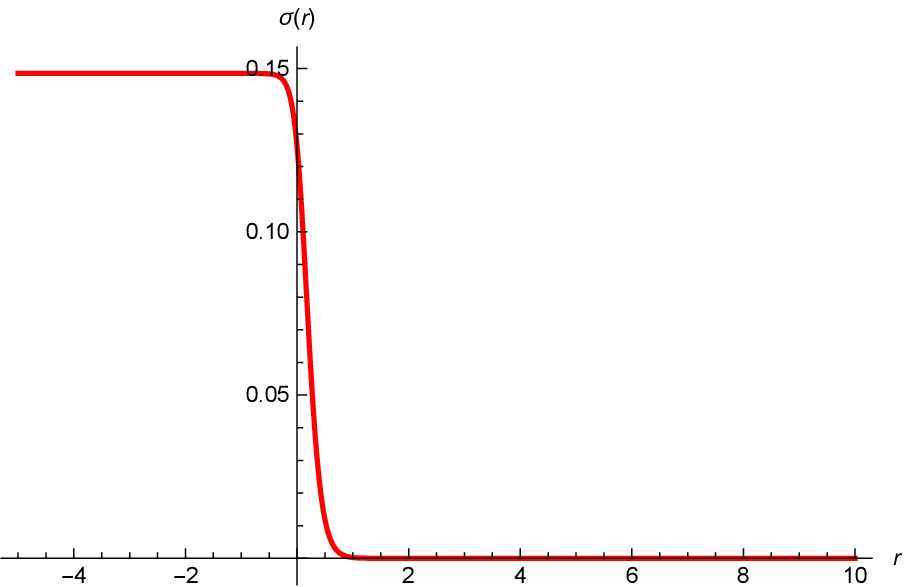}
  \caption{$\sigma$ solution}
  \end{subfigure}
  \begin{subfigure}[b]{0.32\linewidth}
    \includegraphics[width=\linewidth]{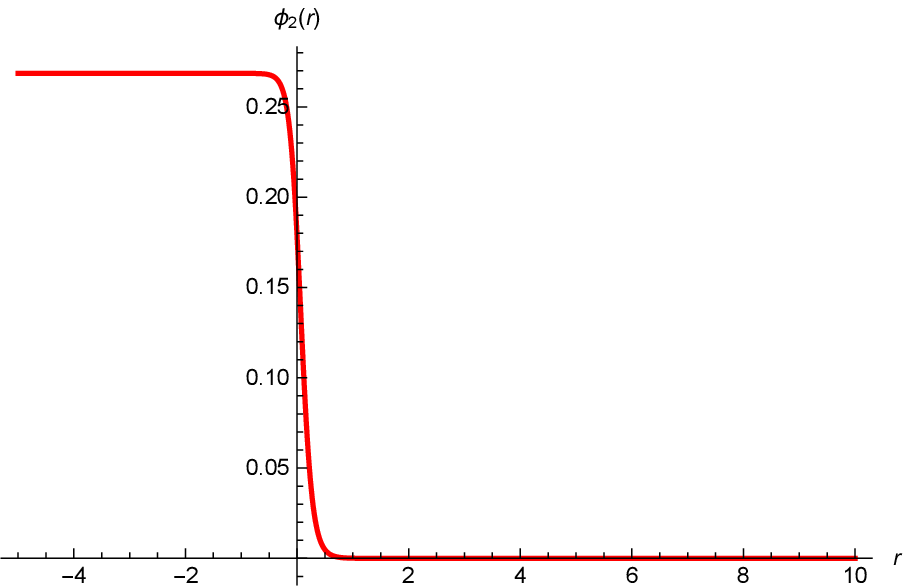}
  \caption{$\phi_2$ solution}
  \end{subfigure}
  \begin{subfigure}[b]{0.32\linewidth}
    \includegraphics[width=\linewidth]{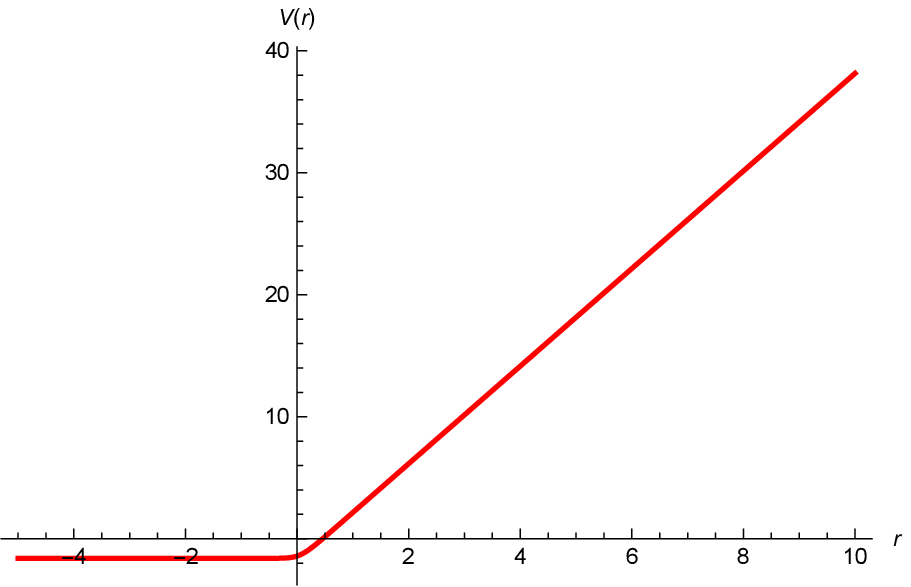}
  \caption{$V$ solution}
  \end{subfigure}
  \caption{An RG flow from $SO(4)$ $N=(1, 0)$ SCFT in six dimensions to two-dimensional $N=(2,0)$ SCFT with $SO(2)_{\text{diag}}$ symmetry dual to $AdS_3\times CH^2$ solution.}
  \label{SO(2)diagKahlerflow0}
\end{figure}

We next consider solutions with smaller residual symmetry $SO(2)_{\text{diag}}\subset SO(2)\times SO(2)$ by imposing the condition $g_2p_2=g_1p_1$. There are three $SO(2)_{\text{diag}}$ singlet scalars with the coset representative given by \eqref{SO2d_scalar}. As in the previous section, compatibility between BPS equations and field equations requires $\phi_1=0$ or $\phi_3=0$, and these two cases are equivalent. We will consider the case of $\phi_3=0$ with the following BPS equations
\begin{eqnarray}
U'&=&\frac{1}{5}e^{\frac{\sigma}{2}}\left[\left(g_1e^{-\sigma}\cosh^2{\phi_1}\cosh{\phi_2}+g_2e^{-\sigma}\sinh^2{\phi_1}\sinh{\phi_2}+4he^{\frac{3\sigma}{2}}\right)\phantom{\frac{g_1}{g_2}}\right. \nonumber \\
& &\left. -6e^{-2V}\left(\cosh{\phi_2}+\frac{g_1}{g_2}\sinh{\phi_2}\right)p_{1}-\frac{27}{8hg_2^2}e^{-\frac{3\sigma}{2}-4V}(g_1^2-g_2^2)p_1^2\right],\quad \\
V'&=&\frac{1}{5}e^{\frac{\sigma}{2}}\left[\left(g_1e^{-\sigma}\cosh^2{\phi_1}\cosh{\phi_2}+g_2e^{-\sigma}\sinh^2{\phi_1}\sinh{\phi_2}+4he^{\frac{3\sigma}{2}}\right)\phantom{\frac{g_1}{g_2}}\right. \nonumber \\
& &\left. +9e^{-2V}\left(\cosh{\phi_2}+\frac{g_1}{g_2}\sinh{\phi_2}\right)p_{1}+\frac{9}{4hg_2^2}e^{-\frac{3\sigma}{2}-4V}(g_1^2-g_2^2)p_1^2\right],\quad\\
\sigma'&=&\frac{2}{5}e^{\frac{\sigma}{2}}\left[\left(g_1e^{-\sigma}\cosh^2{\phi_1}\cosh{\phi_2}+g_2e^{-\sigma}\sinh^2{\phi_1}\sinh{\phi_2}-16he^{\frac{3\sigma}{2}}\right)\phantom{\frac{g_1}{g_2}}\phantom{\frac{g_1}{g_2}}\right. \nonumber \\
& &\left.-6e^{-2V}\left(\cosh{\phi_2}+\frac{g_1}{g_2}\sinh{\phi_2}\right)p_{1}+\frac{9}{4hg_2^2}e^{-\frac{3\sigma}{2}-4V}(g_1^2-g_2^2)p_1^2\right],\\
\phi'_1&=&-e^{-\frac{\sigma}{2}}\cosh{\phi_1}\sinh{\phi_1}(g_1\cosh{\phi_2}+g_2\sinh{\phi_2}),\\
\phi'_2&=&-e^{\frac{\sigma}{2}}\left[\left(g_1e^{-\sigma}\cosh^2{\phi_1}\sinh{\phi_2}+g_2e^{-\sigma}\sinh^2{\phi_1}\cosh{\phi_2}\right)\phantom{\frac{g_1}{g_2}}\right. \nonumber \\
& &\left.+6e^{-2V}\left(\sinh{\phi_2}+\frac{g_1}{g_2}\cosh{\phi_2}\right)p_{1}\right].
\end{eqnarray}
\indent There exist two classes of $AdS_3\times CH^2$ fixed points preserving four supercharges and corresponding to $N=(2,0)$ SCFTs in two dimensions with $SO(2)_{\text{diag}}$ symmetry. With $k=-1$, the first class of $AdS_3\times CH^2$ fixed points is given by
\begin{eqnarray}\label{1SO(2)diagKahlerfixedpoint}
\phi_1&=&0,\qquad\sigma=\frac{2}{5}\phi_2+\frac{2}{5}\ln\left[\frac{g_1g_2^2}{12h(g_2^2+2g_1g_2-3g_1^2)}\right],\nonumber \\
\phi_2&=&\frac{1}{2}\ln\left[\frac{3g_1^2-2g_1g_2-g_2^2}{3g_1^2+2g_1g_2-g_2^2}\right],\nonumber \\
V&=&\frac{1}{10}\ln\left[\frac{3^8(g_1^2-g_2^2)^4}{16h^2g_1^8g_2^6(g_2^2-9g_1^2)}\right],\nonumber \\
L_{AdS_3}&=&\left[\frac{8(9g_1^4 g_2 - 10 g_1^2 g_2^3 + g_2^5)^2}{3hg_1^4(g_2^2-3g_1^2)^5}\right]^{\frac{1}{5}}
\end{eqnarray}
with $g_2>3g_1$ or $g_2<-3g_1$ for $AdS_3$ vacua to exist. An RG flow solution from the $SO(4)$ $AdS_7$ critical point to $AdS_3\times CH^2$ fixed point for $\phi_1=0$, $g_2=4g_1$ and $h=1$ is shown in figure \ref{SO(2)diagKahlerflow0}.

\begin{figure}[h!]
  \centering
  \begin{subfigure}[b]{0.45\linewidth}
    \includegraphics[width=\linewidth]{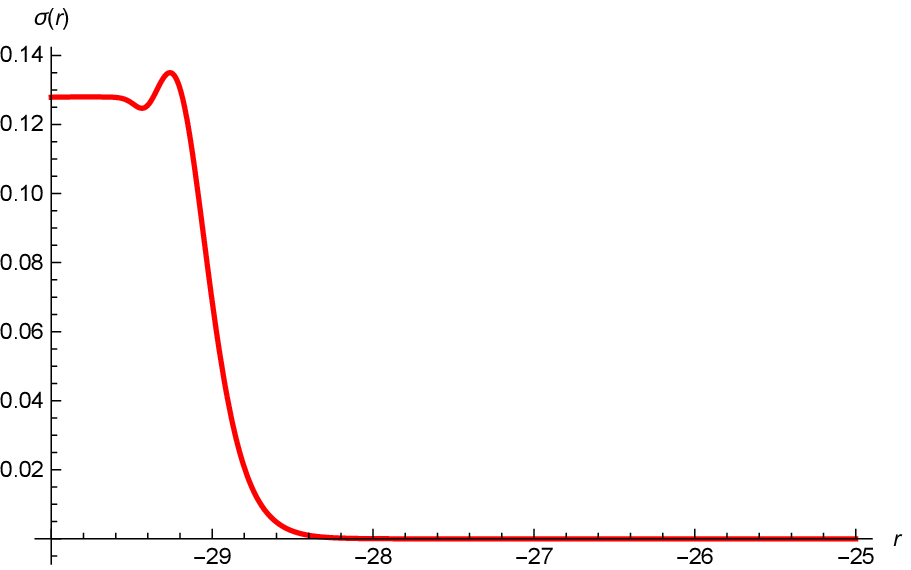}
  \caption{$\sigma$ solution}
  \end{subfigure}
  \begin{subfigure}[b]{0.45\linewidth}
    \includegraphics[width=\linewidth]{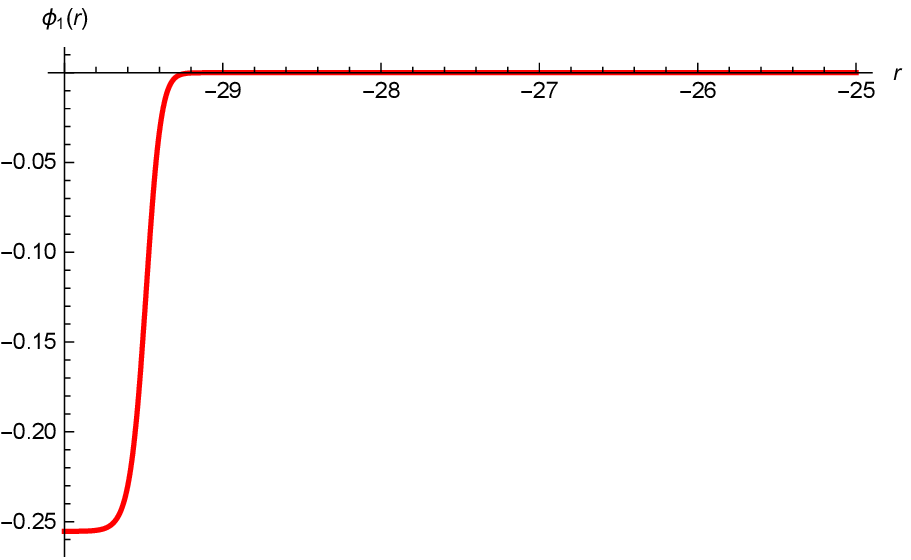}
  \caption{$\phi_1$ solution}
  \end{subfigure}\\
  \begin{subfigure}[b]{0.45\linewidth}
    \includegraphics[width=\linewidth]{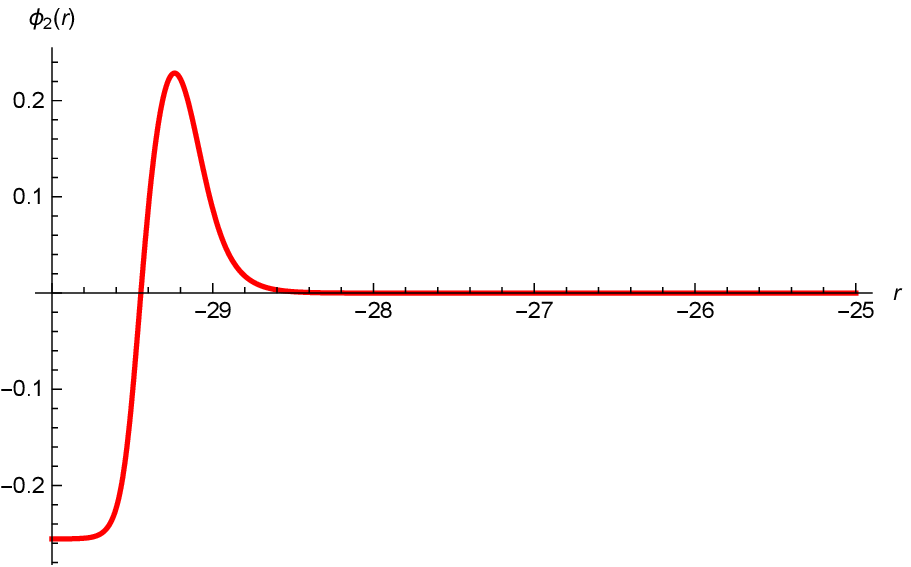}
  \caption{$\phi_2$ solution}
  \end{subfigure}
  \begin{subfigure}[b]{0.45\linewidth}
    \includegraphics[width=\linewidth]{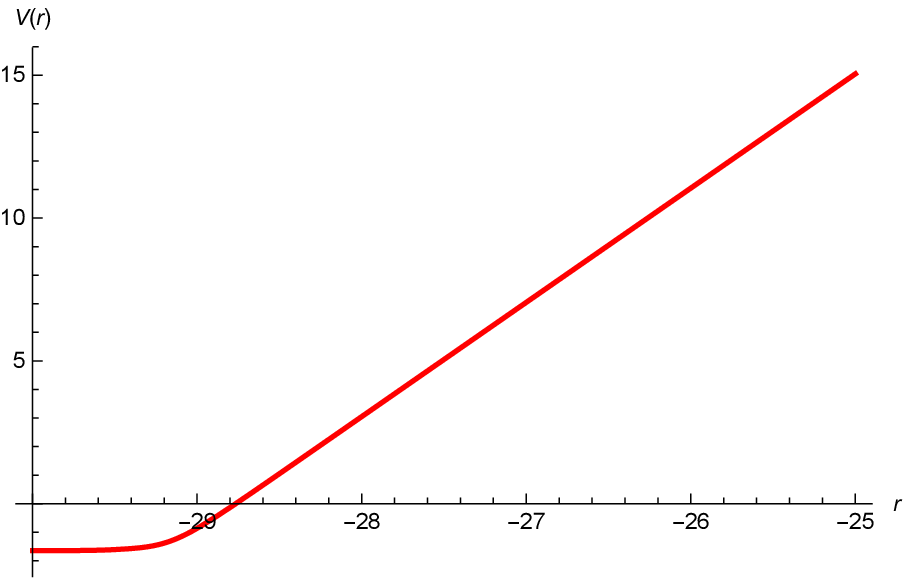}
  \caption{$V$ solution}
  \end{subfigure}
  \caption{An RG flow from $SO(4)$ $N=(1, 0)$ SCFT in six dimensions to two-dimensional $N=(2,0)$ SCFT with $SO(2)_{\text{diag}}$ symmetry dual to $AdS_3\times CH^2$ solution.}
  \label{SO(2)diagKahlerflow1}
\end{figure}

\begin{figure}[h!]
  \centering
  \begin{subfigure}[b]{0.45\linewidth}
    \includegraphics[width=\linewidth]{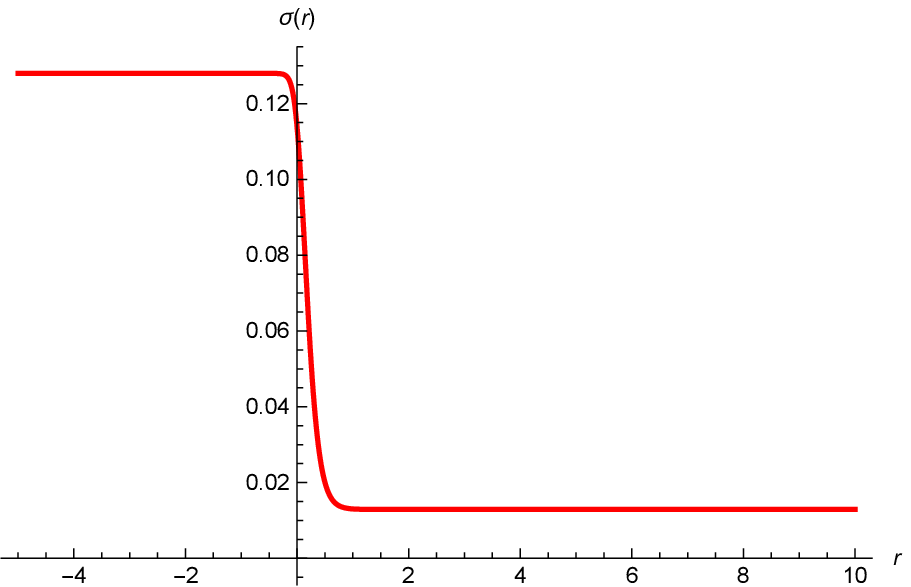}
  \caption{$\sigma$ solution}
  \end{subfigure}
  \begin{subfigure}[b]{0.45\linewidth}
    \includegraphics[width=\linewidth]{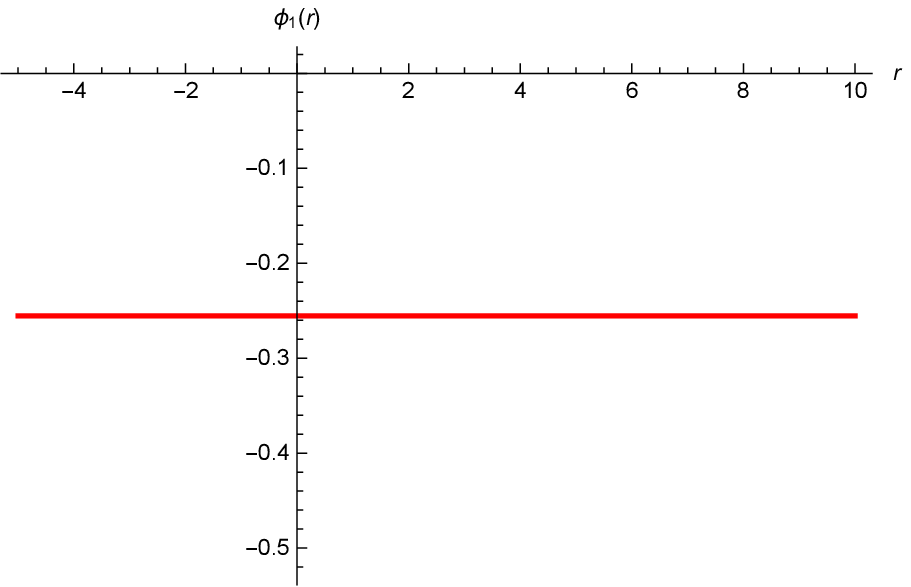}
  \caption{$\phi_1$ solution}
  \end{subfigure}\\
  \begin{subfigure}[b]{0.45\linewidth}
    \includegraphics[width=\linewidth]{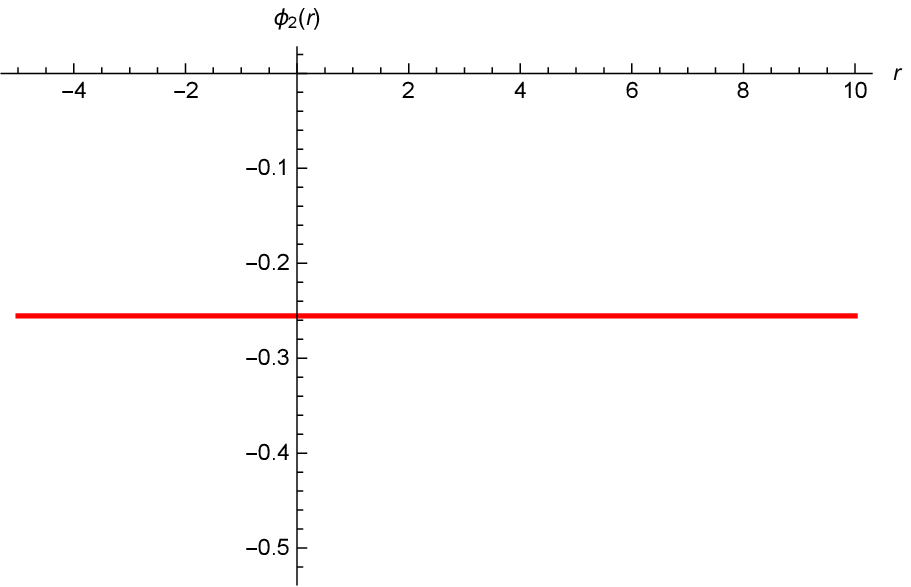}
  \caption{$\phi_2$ solution}
  \end{subfigure}
  \begin{subfigure}[b]{0.45\linewidth}
    \includegraphics[width=\linewidth]{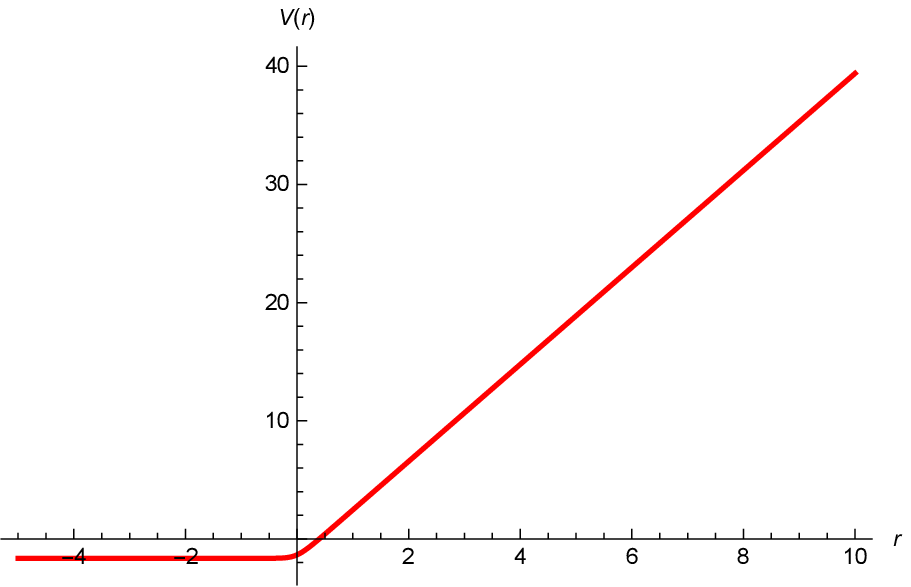}
  \caption{$V$ solution}
  \end{subfigure}
  \caption{An RG flow from $SO(3)$ $N=(1, 0)$ SCFT in six dimensions to two-dimensional $N=(2,0)$ SCFT with $SO(2)_{\text{diag}}$ symmetry dual to $AdS_3\times CH^2$ solution.}
  \label{SO(2)diagKahlerflow2}
\end{figure}

\begin{figure}[h!]
  \centering
  \begin{subfigure}[b]{0.45\linewidth}
    \includegraphics[width=\linewidth]{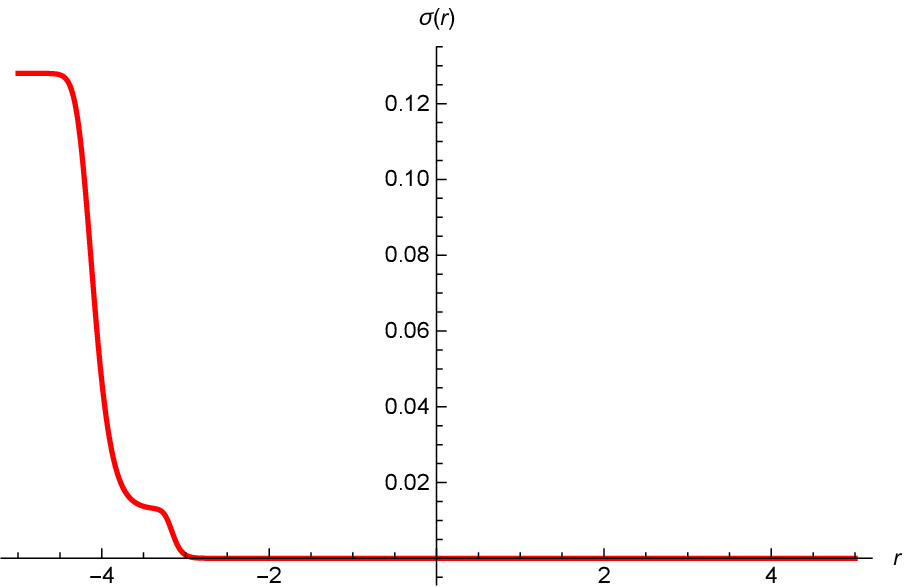}
  \caption{$\sigma$ solution}
  \end{subfigure}
  \begin{subfigure}[b]{0.45\linewidth}
    \includegraphics[width=\linewidth]{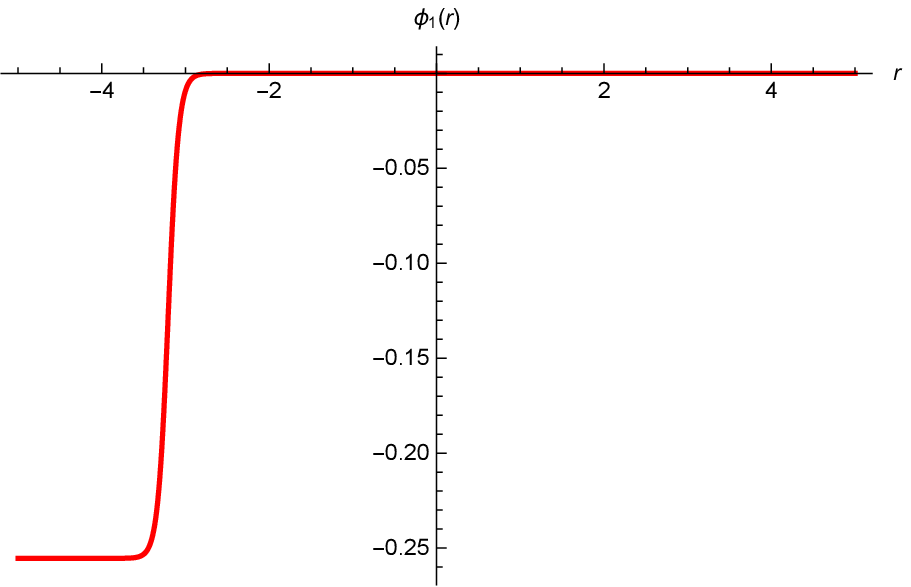}
  \caption{$\phi_1$ solution}
  \end{subfigure}\\
  \begin{subfigure}[b]{0.45\linewidth}
    \includegraphics[width=\linewidth]{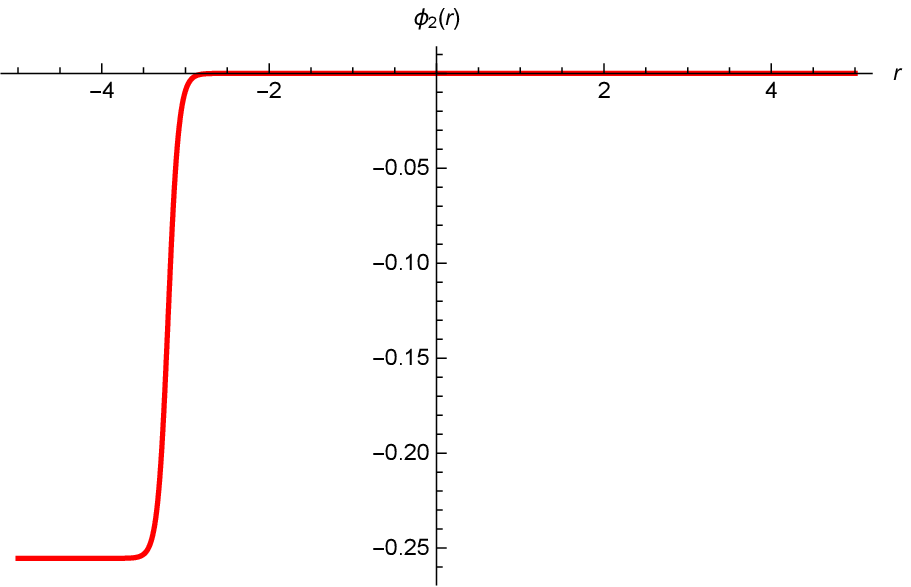}
  \caption{$\phi_2$ solution}
  \end{subfigure}
  \begin{subfigure}[b]{0.45\linewidth}
    \includegraphics[width=\linewidth]{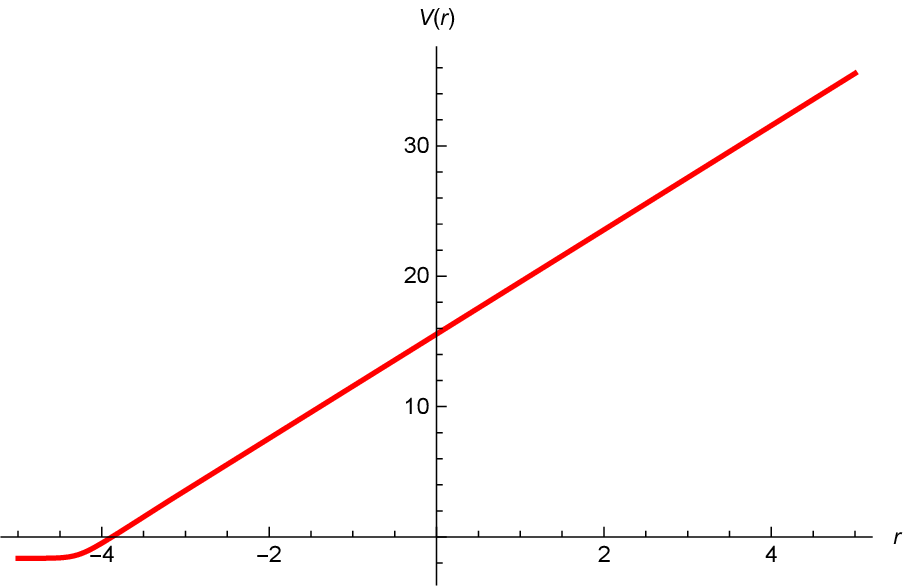}
  \caption{$V$ solution}
  \end{subfigure}
  \caption{An RG flow from $SO(4)$ $N=(1, 0)$ SCFT to $SO(3)$ $N=(1, 0)$ SCFT in six dimensions and eventually to two-dimensional $N=(2,0)$ SCFT with $SO(2)_{\text{diag}}$ symmetry dual to $AdS_3\times CH^2$ solution.}
  \label{SO(2)diagKahlerflow3}
\end{figure}

Another class of $AdS_3\times CH^2$ fixed points is given by
\begin{eqnarray}\label{2SO(2)diagKahlerfixedpoint}
\sigma&=&\frac{2}{5}\ln\left[\frac{g_1g_2}{12h\sqrt{(g_2+g_1)(g_2-g_1)}}\right],\qquad
\phi_1=\phi_2=\frac{1}{2}\ln\left[\frac{g_2-g_1}{g_2+g_1}\right],\nonumber \\
V&=&\frac{1}{5}\ln\left[\frac{3^4(g_1^2-g_2^2)^2}{4hg_1^4g_2^4}\right],\qquad
L_{AdS_3}=\left[\frac{8(g_1^2-g_2^2)^2}{3hg_1^4g_2^4}\right]^{\frac{1}{5}}\, .
\end{eqnarray}
To obtain good $AdS_3$ vacua, we require that $g_2>g_1$. Various RG flows from $N=(1,0)$ six-dimensional SCFTs with $SO(4)$ and $SO(3)$ symmetries to these fixed points for $g_2=4g_1$ and $h=1$ are shown in figures \ref{SO(2)diagKahlerflow1}, \ref{SO(2)diagKahlerflow2} and \ref{SO(2)diagKahlerflow3}.

As in the case of $M^4=\Sigma^2\times \Sigma^2$, all of these $AdS_3$ fixed points and RG flows cannot be uplifted to eleven dimensions using the truncation given in \cite{7D_from_11D}, so we do not have a clear holographic interpretation in this case.

\subsection{$AdS_3$ vacua with $SO(2)_R$ symmetry}
\begin{figure}[h!]
  \centering
  \begin{subfigure}[b]{0.45\linewidth}
    \includegraphics[width=\linewidth]{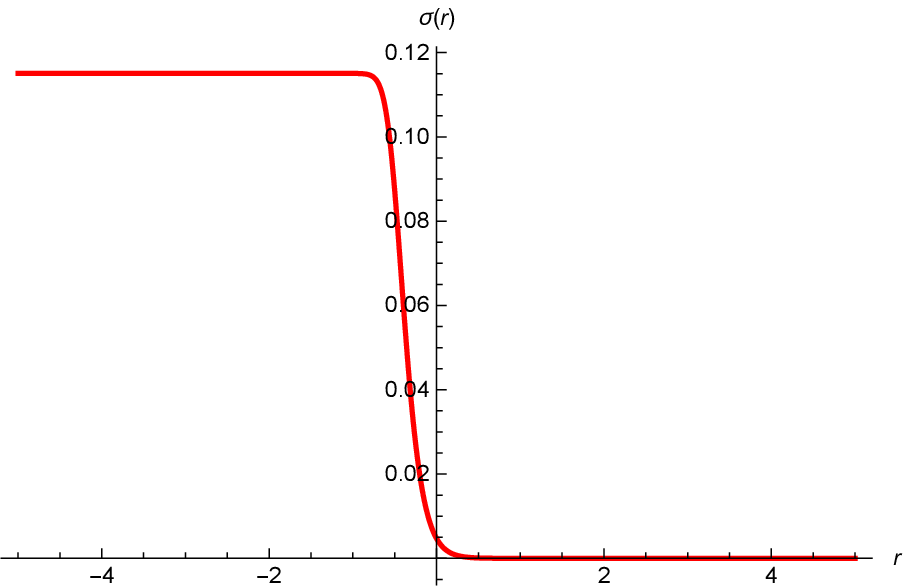}
  \caption{$\sigma$ solution}
  \end{subfigure}
  \begin{subfigure}[b]{0.45\linewidth}
    \includegraphics[width=\linewidth]{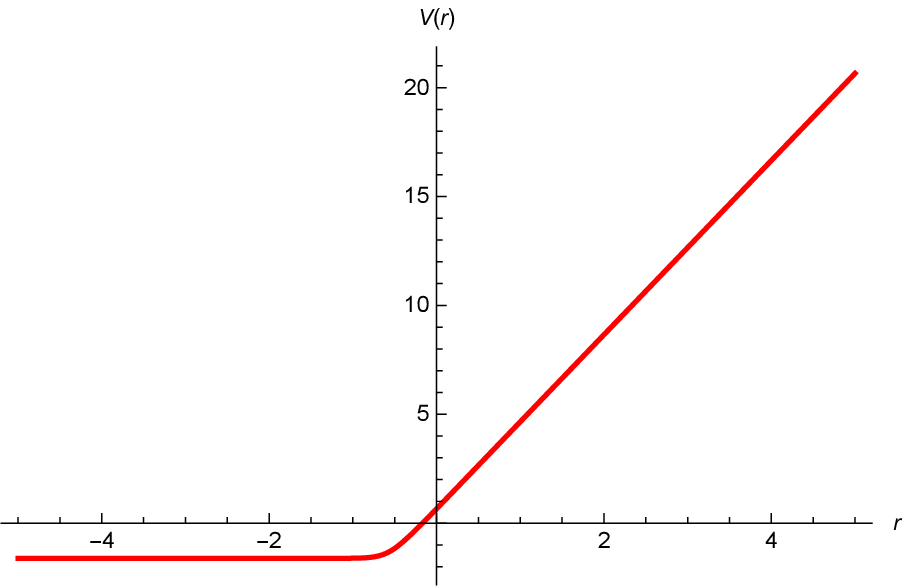}
  \caption{$V$ solution}
  \end{subfigure}
  \caption{An RG flow from $SO(4)$ $N=(1, 0)$ SCFT in six dimensions to two-dimensional $N=(2,0)$ SCFT with $SO(2)_R$ symmetry dual to $AdS_3\times CH^2$ solution.}
  \label{SO(2)RKahlerflow}
\end{figure}

By setting $p_2=0$ in the $SO(2)\times SO(2)$ case, we obtain solutions with $SO(2)_R\subset SO(3)_R$ symmetry. As in the previous case, the three $SO(2)_R$ singlet scalars need to vanish in order for $AdS_3$ fixed points to exist. We will accordingly set all vector multiplet scalars to zero for brevity. The resulting BPS equations are given by
\begin{eqnarray}
U'&=&\frac{1}{5}e^{\frac{\sigma}{2}}\left[g_1e^{-\sigma}+4he^{\frac{3\sigma}{2}}-6e^{-2V}p_1+\frac{27}{8h}e^{-4V}p_1^2\right],\hspace{1cm}\\
V'&=&\frac{1}{5}e^{\frac{\sigma}{2}}\left[g_1e^{-\sigma}+4he^{\frac{3\sigma}{2}}+9e^{-2V}p_1-\frac{9}{4h}e^{-4V}p_1^2\right],\hspace{1cm}\\
\sigma'&=&\frac{2}{5}e^{\frac{\sigma}{2}}\left[g_1e^{-\sigma}-14he^{\frac{3\sigma}{2}}-6e^{-2V}p_1-\frac{9}{4h}e^{-4V}p_1^2\right].\hspace{1cm}
\end{eqnarray}
After imposing the twist condition (\ref{SO(2)KahlerQYM}), we obtain an $AdS_3$ solution for $k=-1$ given by
\begin{equation}\label{SO(2)RKahlerfixedpoint}
\sigma=\frac{2}{5}\ln\left[\frac{g_1}{12h}\right],\qquad V=\frac{1}{10}\ln\left[\frac{3^8}{16h^2g_1^8}\right],\qquad L_{AdS_3}=\left[\frac{8}{3hg_1^4}\right]^{\frac{1}{5}}\, .
\end{equation}
An RG flow from $SO(4)$ $AdS_7$ to this fixed point for $h=1$ is shown in figure \ref{SO(2)RKahlerflow}.

\subsection{$AdS_3$ vacua with $SO(3)_{\text{diag}}$ symmetry}
\begin{figure}[h!]
  \centering
  \begin{subfigure}[b]{0.32\linewidth}
    \includegraphics[width=\linewidth]{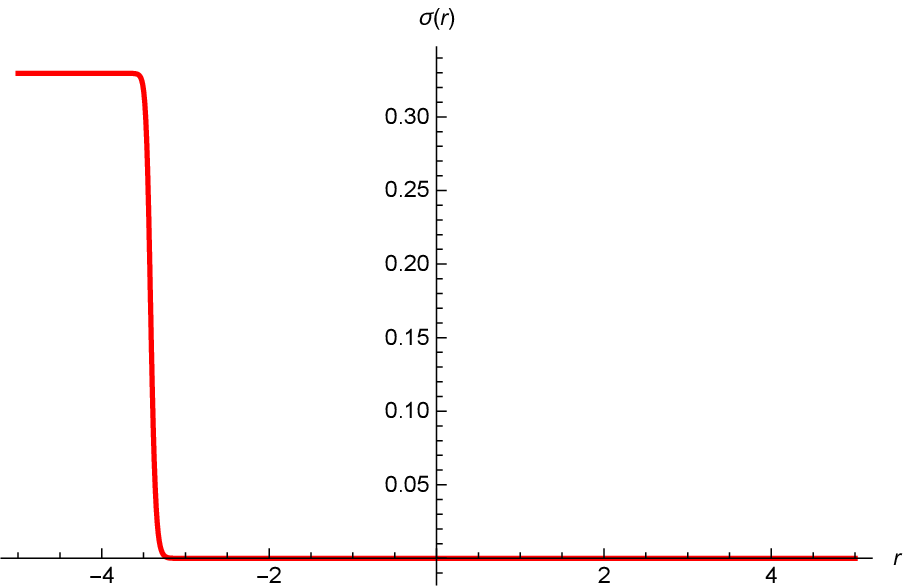}
  \caption{$\sigma$ solution}
  \end{subfigure}
  \begin{subfigure}[b]{0.32\linewidth}
    \includegraphics[width=\linewidth]{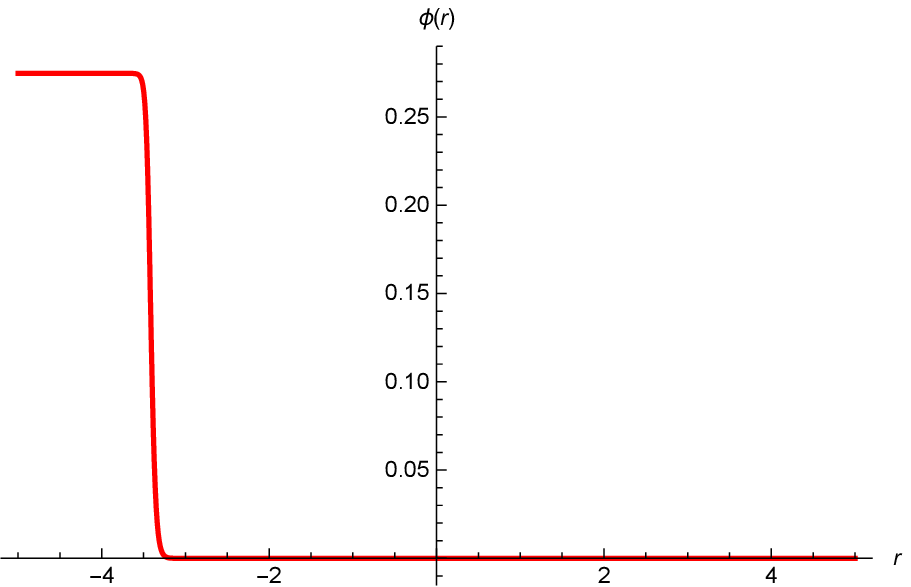}
  \caption{$\phi$ solution}
  \end{subfigure}
  \begin{subfigure}[b]{0.32\linewidth}
    \includegraphics[width=\linewidth]{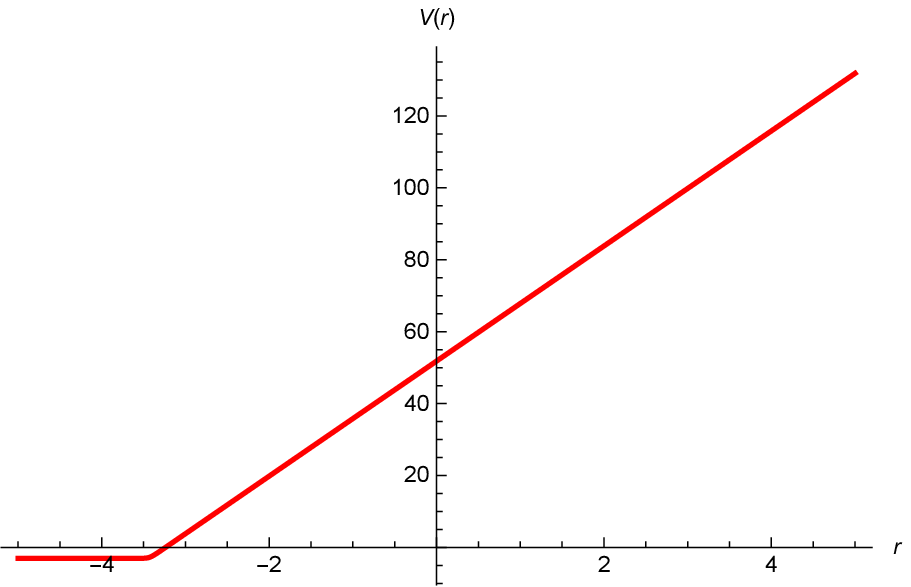}
  \caption{$V$ solution}
  \end{subfigure}
  \caption{An RG flow from $SO(4)$ $N=(1, 0)$ SCFT in six dimensions to two-dimensional $N=(1,0)$ SCFT with $SO(3)_{\text{diag}}$ symmetry dual to $AdS_3\times CH^2$ solution for $g_1=g_2$.}
  \label{samegSO(3)diagflow}
\end{figure}

For Kahler four-cycles with $SU(2)\times U(1)$ spin connection, we can also perform the twist by identifying $SO(3)\sim SU(2)\subset SU(2)\times U(1)$ with the gauge symmetry $SO(3)_{\textrm{diag}}\subset SO(3)\times SO(3)$. In this case, we will use the metric on $M^4_k$ in the form
\begin{equation}
ds_{M^4_k}^2=d\varphi^2+f_k(\varphi)^2(\tau_1^2+\tau_2^2+\tau_3^2)
\end{equation}
with $\tau_i$ being the $SU(2)$ left-invariant one-forms given in \eqref{SU2_left_1_form} and $f_k(\varphi)$ defined in \eqref{fFn}.
\\
\indent With the seven-dimensional vielbein
\begin{eqnarray}
e^{\hat{\alpha}}&=& e^Udx^\alpha,\qquad e^{\hat{r}}= dr, \nonumber \\
e^{\hat{i}}&=&e^Vf_k(\varphi)\tau_i,\qquad i=1,2,3,\qquad e^{\hat{4}}=e^Vd\varphi,
\end{eqnarray}
we can compute the following non-vanishing components of the spin connection
\begin{eqnarray}\label{SO(3)diagspinCon}
{\omega^{\hat{\alpha}}}_{\hat{r}}&=&\ U'e^{\hat{\alpha}}, \qquad {\omega^{\hat{i}}}_{\hat{r}}=V'e^{\hat{i}}, \qquad  {\omega^{\hat{4}}}_{\hat{r}}=V'e^{\hat{4}}, \nonumber \\
{\omega^{\hat{i}}}_{\hat{4}}&=& f'_k(\varphi)\tau_i,\qquad
{\omega^{\hat{i}}}_{\hat{j}}=\epsilon_{ijk} \tau_k\, .
\end{eqnarray}
We then turn on the $SO(3)_{\text{diag}}$ gauge fields as follow
\begin{equation}\label{SO(3)diaggaugefield}
A^i_{(1)}= \frac{g_2}{g_1}A^{i+3}_{(1)}=\frac{p}{k}(f'_{k}(\varphi)+1)\tau_i,\qquad i=1,2,3
\end{equation}
with the two-form field strengths given by
\begin{eqnarray}\label{SO(3)diagKahlerF}
F^1_{(2)}&=& \frac{g_2}{g_1}F^4_{(2)}=e^{-2V}p\ (e^{\hat{1}}\wedge e^{\hat{4}}+e^{\hat{2}}\wedge e^{\hat{3}}),\\
F^2_{(2)}&=& \frac{g_2}{g_1}F^5_{(2)}=e^{-2V}p\ (e^{\hat{1}}\wedge e^{\hat{3}}+e^{\hat{2}}\wedge e^{\hat{4}}),\\
F^3_{(2)}&=& \frac{g_2}{g_1}F^6_{(2)}=e^{-2V}p\ (e^{\hat{1}}\wedge e^{\hat{2}}+e^{\hat{3}}\wedge e^{\hat{4}}).
\end{eqnarray}
As in the previous cases, we also need a non-vanishing four-form field strength
\begin{equation}\label{SO(2)xSO(2)Kahler4form}
H_{(4)}=\frac{3}{8\sqrt{2}hg_2^2}e^{-4V}(g_1^2-g_2^2)p^2 e^{\hat{1}}\wedge e^{\hat{2}}\wedge e^{\hat{3}}\wedge e^{\hat{4}}
\end{equation}
together with the twist condition
\begin{equation}\label{SO(3)diagQYM}
g_1p=k
\end{equation}
and the following projectors
\begin{equation}\label{SO(3)diagProjCon}
\gamma_{r}\epsilon=-\gamma_{\hat{1}\hat{2}\hat{3}\hat{4}}\epsilon=\epsilon \qquad \textrm{and} \qquad
\gamma_{\hat{i}\hat{j}}\epsilon=i \epsilon_{ijk}\sigma^k\epsilon\, .
\end{equation}
It should be noted that the second condition in \eqref{SO(3)diagProjCon} consists of only two independent projectors since $\gamma_{\hat{1}\hat{3}}$ projector can be obtained from the product of those coming from $\gamma_{\hat{1}\hat{2}}$ and $\gamma_{\hat{2}\hat{3}}$. Therefore, the resulting $AdS_3$ fixed points preserve two supercharges corresponding to $N=(1,0)$ superconformal symmetry in two dimensions.
\\
\indent With all these and the coset representative for the $SO(3)_{\textrm{diag}}$ singlet scalar in \eqref{SO3d_coset}, we find the following BPS equations
\begin{eqnarray}\label{SO(3)diagKahlerBPS}
U'&=&\frac{1}{5}e^{\frac{\sigma}{2}}\left[(g_1e^{-\sigma}\cosh^3{\phi}+g_2e^{-\sigma}\sinh^3{\phi}+4he^{\frac{3\sigma}{2}})-\frac{9p^2}{8hg_2^2}e^{-\frac{3\sigma}{2}-4V}(g_1^2-g_2^2)\right. \nonumber \\
& &\left.-6pe^{-2V}\left(\cosh{\phi}+\frac{g_1}{g_2}\sinh{\phi}\right)\right],\\
V'&=&\frac{1}{5}e^{\frac{\sigma}{2}}\left[(g_1e^{-\sigma}\cosh^3{\phi}+g_2e^{-\sigma}\sinh^3{\phi}+4he^{\frac{3\sigma}{2}})+\frac{3p^2}{4hg_2^2}e^{-\frac{3\sigma}{2}-4V}(g_1^2-g_2^2)\right. \nonumber \\
& &\left.+9pe^{-2V}\left(\cosh{\phi}+\frac{g_1}{g_2}\sinh{\phi}\right)\right],\\
\sigma'&=&\frac{2}{5}e^{\frac{\sigma}{2}}\left[(g_1e^{-\sigma}\cosh^3{\phi}+g_2e^{-\sigma}\sinh^3{\phi}-16he^{\frac{3\sigma}{2}})+\frac{3p^2}{4hg_2^2}e^{-\frac{3\sigma}{2}-4V}(g_1^2-g_2^2)\right. \nonumber \\
& &\left.-6pe^{-2V}\left(\cosh{\phi}+\frac{g_1}{g_2}\sinh{\phi}\right)\right],\\
\phi'&=&-\frac{1}{2g_2}e^{-\frac{\sigma}{2}}(g_1\cosh{\phi}+g_2\sinh{\phi})(g_2\sinh{2\phi}+4pe^{\sigma-2V}).\label{lastSO(3)diagKahlerBPS}
\end{eqnarray}
\indent We now look for $AdS_3$ fixed points for the case of $g_2=g_1$ that can be embedded in eleven dimensions. Setting $g_2=g_1$ in the above equations, we find the following $AdS_3\times CH^2$ fixed point
\begin{eqnarray}
\sigma&=&\frac{2}{5}\ln\left[\frac{3^{\frac{3}{4}}g_1}{16h}\right], \qquad \phi=\frac{1}{4}\ln3,\nonumber \\
V&=&\frac{1}{5}\ln\left[\frac{18}{hg_1^4}\right], \qquad L_{AdS_3}=\left[\frac{64}{27hg_1^4}\right]^{\frac{1}{5}}\, .
\end{eqnarray}
An RG flow interpolating between the $SO(4)$ $AdS_7$ vacuum and this $AdS_3\times CH^2$ fixed point is shown in figure \ref{samegSO(3)diagflow}.

\begin{figure}[h!]
  \centering
  \begin{subfigure}[b]{0.32\linewidth}
    \includegraphics[width=\linewidth]{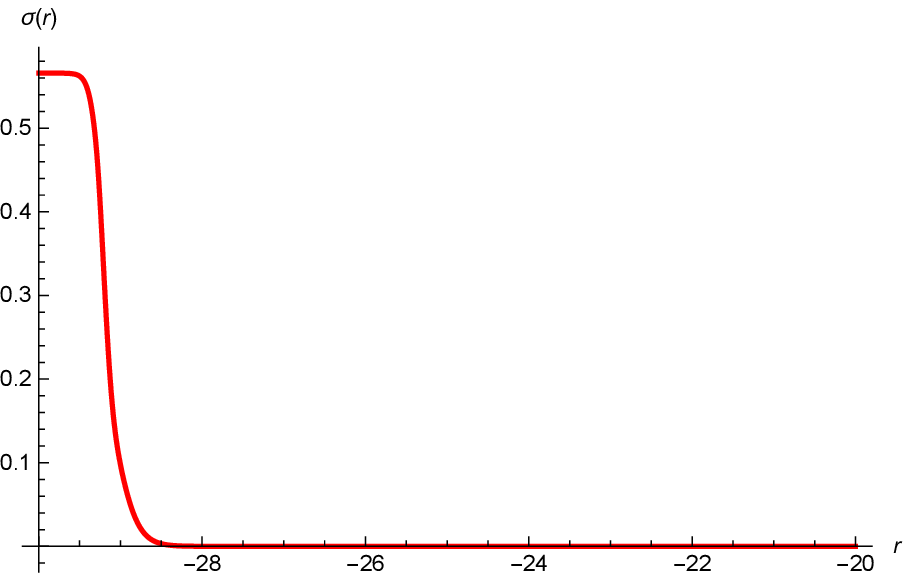}
  \caption{$\sigma$ solution}
  \end{subfigure}
  \begin{subfigure}[b]{0.32\linewidth}
    \includegraphics[width=\linewidth]{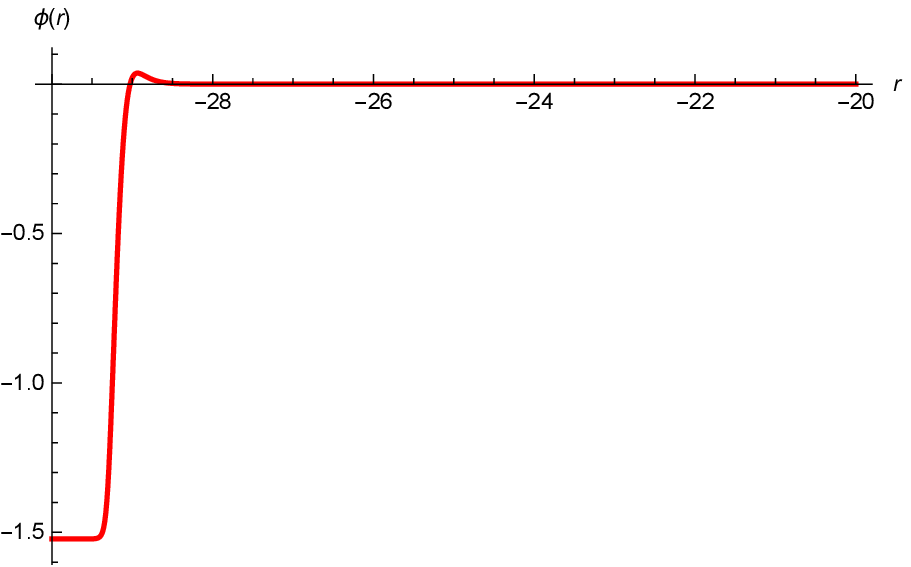}
  \caption{$\phi$ solution}
  \end{subfigure}
  \begin{subfigure}[b]{0.32\linewidth}
    \includegraphics[width=\linewidth]{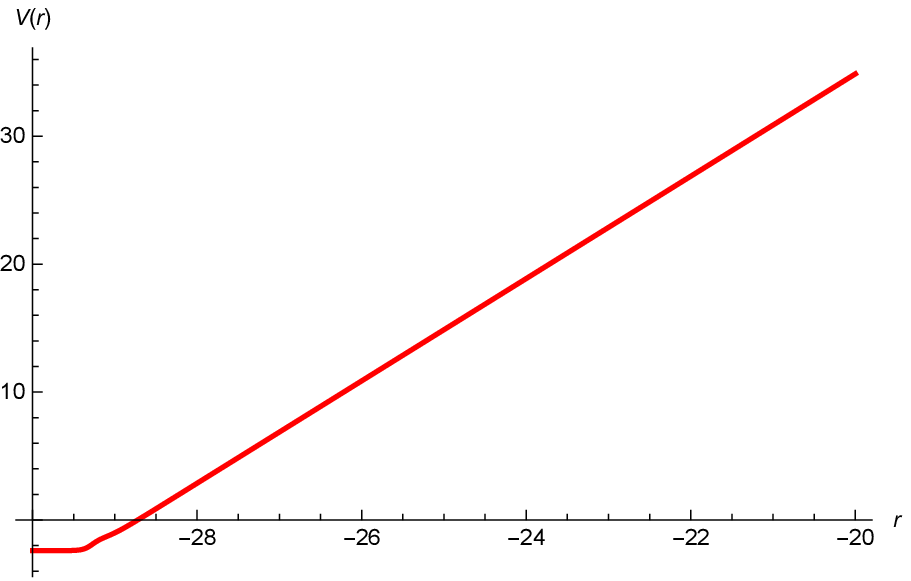}
  \caption{$V$ solution}
  \end{subfigure}
  \caption{An RG flow from $SO(4)$ $N=(1, 0)$ SCFT in six dimensions to two-dimensional $N=(1,0)$ SCFT with $SO(3)_{\text{diag}}$ symmetry dual to $AdS_3\times CH^2$ solution.}
  \label{difgSO(3)diagflow1}
\end{figure}

\begin{figure}[h!]
  \centering
  \begin{subfigure}[b]{0.32\linewidth}
    \includegraphics[width=\linewidth]{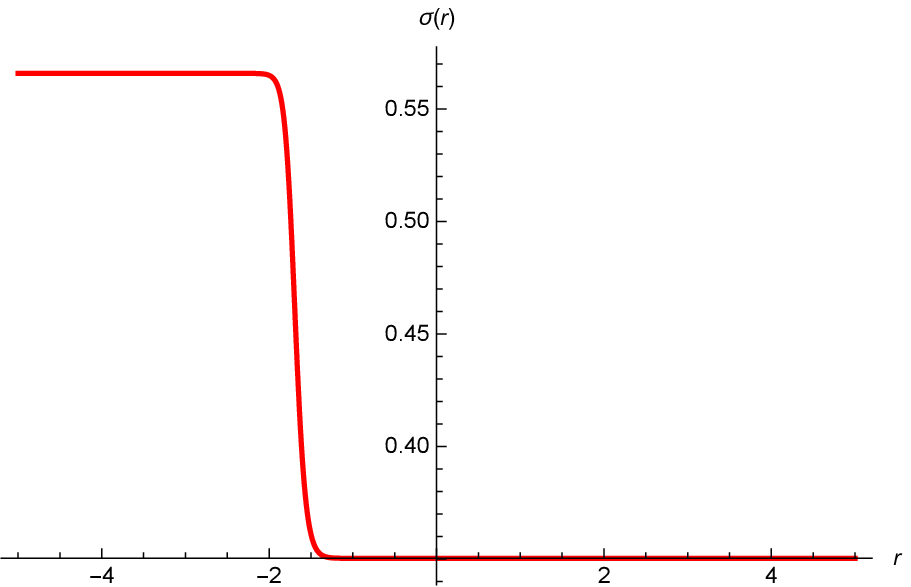}
  \caption{$\sigma$ solution}
  \end{subfigure}
  \begin{subfigure}[b]{0.32\linewidth}
    \includegraphics[width=\linewidth]{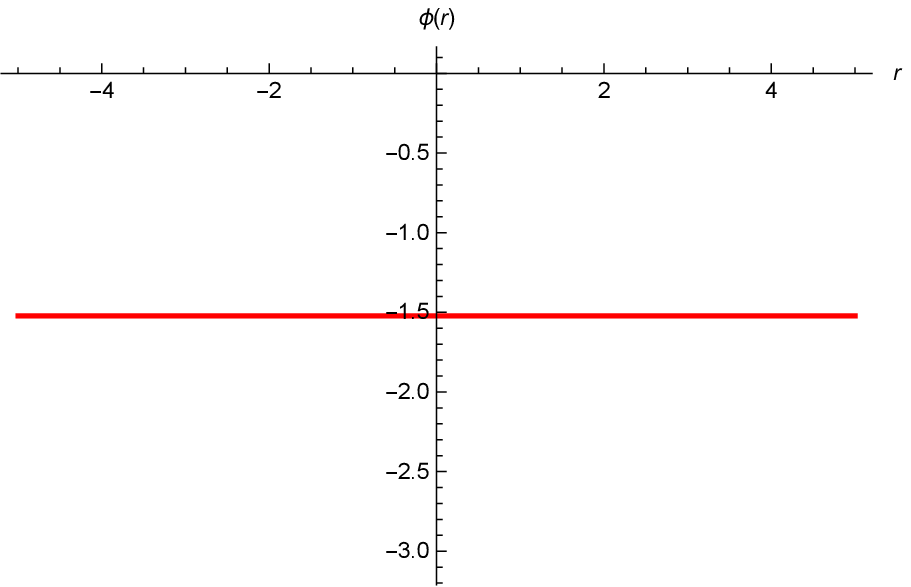}
  \caption{$\phi$ solution}
  \end{subfigure}
  \begin{subfigure}[b]{0.32\linewidth}
    \includegraphics[width=\linewidth]{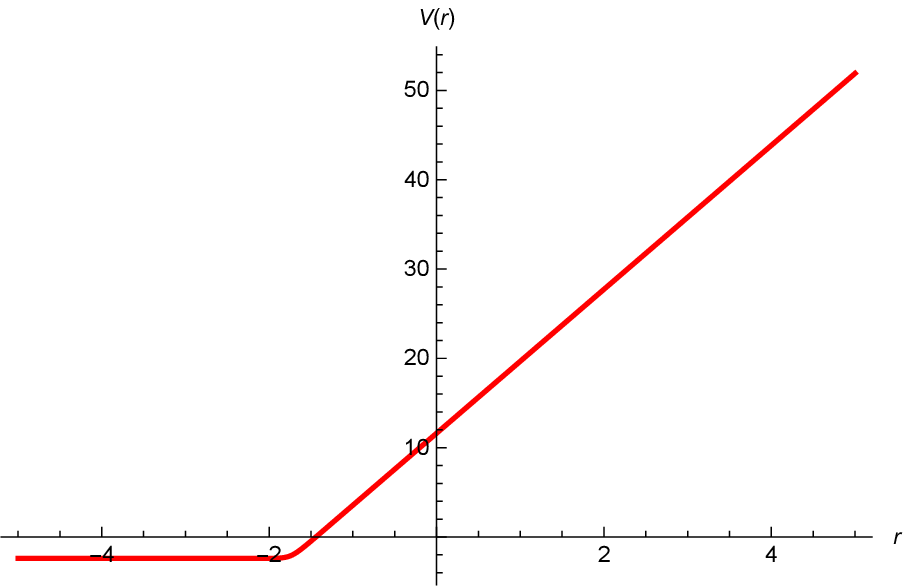}
  \caption{$V$ solution}
  \end{subfigure}
  \caption{An RG flow from $SO(3)$ $N=(1, 0)$ SCFT in six dimensions to two-dimensional $N=(1,0)$ SCFT with $SO(3)_{\text{diag}}$ symmetry dual to $AdS_3\times CH^2$ solution.}
  \label{difgSO(3)diagflow2}
\end{figure}

\begin{figure}[h!]
  \centering
  \begin{subfigure}[b]{0.32\linewidth}
    \includegraphics[width=\linewidth]{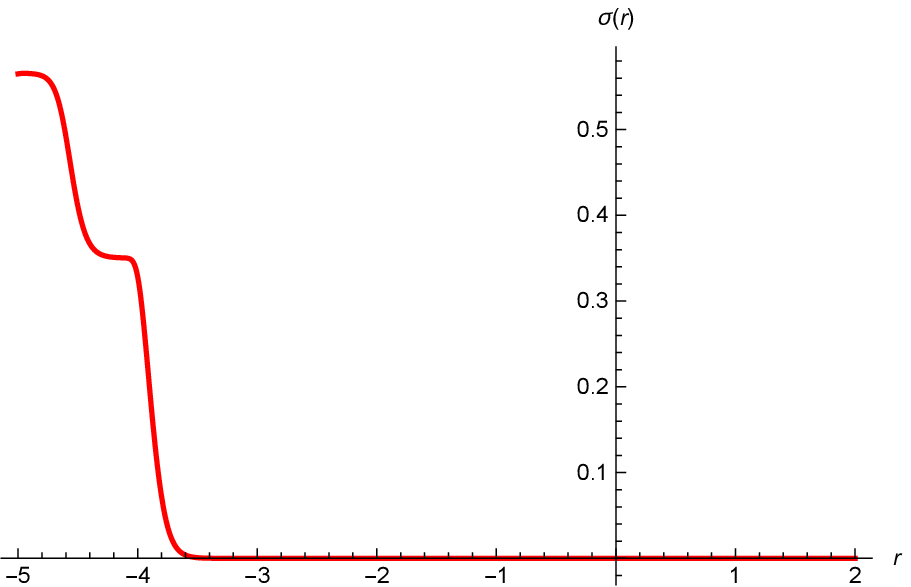}
  \caption{$\sigma$ solution}
  \end{subfigure}
  \begin{subfigure}[b]{0.32\linewidth}
    \includegraphics[width=\linewidth]{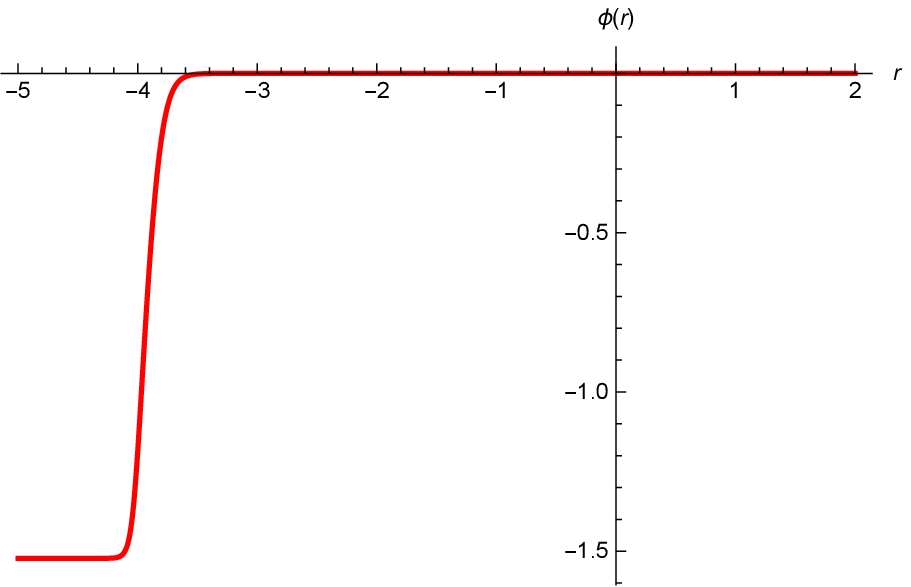}
  \caption{$\phi$ solution}
  \end{subfigure}
  \begin{subfigure}[b]{0.32\linewidth}
    \includegraphics[width=\linewidth]{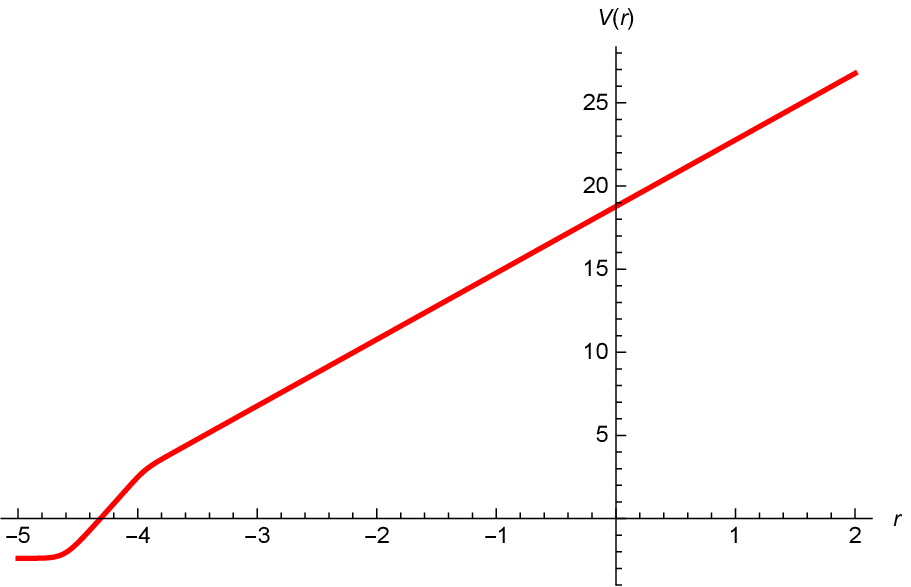}
  \caption{$V$ solution}
  \end{subfigure}
  \caption{An RG flow from $SO(4)$ $N=(1, 0)$ SCFT to $SO(3)$ $N=(1, 0)$ SCFT in six dimensions and to two-dimensional $N=(1,0)$ SCFT with $SO(3)_{\text{diag}}$ symmetry dual to $AdS_3\times CH^2$ solution.}
  \label{difgSO(3)diagflow3}
\end{figure}

We can also uplift this solution to eleven dimensions by first choosing the $S^3$ coordinates
\begin{equation}
\mu^\alpha=(\cos\psi\hat{\mu}^a,\sin\psi),\qquad a,b,\ldots =1,2,3
\end{equation}
with $\hat{\mu}^a$ being coordinates on $S^2$ satisfying $\hat{\mu}^a\hat{\mu}^a=1$. After using the $SL(4,\mathbb{R})/SO(4)$ matrix
\begin{equation}
\tilde{T}^{-1}_{\alpha\beta}=\textrm{diag}(e^\phi,e^\phi,e^\phi,e^{-3\phi})=(\delta_{ab}e^\phi,e^{-3\phi}),
\end{equation}
we find the eleven-dimensional metric
\begin{eqnarray}
d\hat{s}^2_{11}&=&\Delta^{\frac{1}{3}}\left[e^{2U}dx^2_{1,1}+dr^2+e^{2V}[d\varphi^2+f_k(\varphi)^2(\tau_1^2+\tau_2^2+\tau_3^2)]\right]\nonumber \\
& &+\frac{2}{g^2}\Delta^{-\frac{2}{3}}e^{-2\sigma}\left[\cos^2\xi+e^{\frac{5}{2}\sigma}\sin^2\xi (e^{\phi}\cos^2\psi+e^{-3\phi}\sin^2\psi)\right]d\xi^2\nonumber \\
& &+\frac{1}{g^2}\Delta^{-\frac{2}{3}}e^{\frac{\sigma}{2}}\sin\xi\sin\psi\cos\psi(e^{\phi}-e^{-3\phi})d\xi d\psi \nonumber \\
& &+\frac{1}{2g^2}\Delta^{-\frac{2}{3}}e^{\frac{\sigma}{2}}\cos^2\xi \left[(e^{-3\phi}\cos^2\psi+e^{\phi}\sin^2\psi)d\psi^2+e^{\phi}\cos^2\psi D\hat{\mu}^aD\hat{\mu}^a\right]\nonumber \\
& &
\end{eqnarray}
with $\Delta$ given by
\begin{equation}
\Delta=e^{-\frac{\sigma}{2}}\cos^2\xi (e^{-\phi}\cos^2\psi+e^{3\phi}\sin^2\psi)+e^{2\sigma}\sin^2\xi
\end{equation}
and $D\hat{\mu}^a=d\hat{\mu}^a+gA^{ab}\hat{\mu}^b$. The gauge fields $A^{ab}$ are given by
\begin{equation}
 A^{12}=2A^3_{(1)},\qquad A^{13}=-2A^2_{(1)},\qquad A^{23}=-2A^1_{(1)}\, .
\end{equation}
\indent For $g_2\neq g_1$, we find the following $AdS_3$ fixed points
\begin{eqnarray}
\sigma&=&\frac{2}{5}\ln\left[\frac{3g_1g_2}{28h\sqrt{(g_2+g_1)(g_2-g_1)}}\right], \qquad
\phi=\frac{1}{2}\ln\left[\frac{g_2-g_1}{g_2+g_1}\right],\\
V&=&\frac{1}{10}\ln\left[\frac{3087(g_1^2-g_2^2)^4}{16h^2g_1^8g_2^8}\right], \qquad
L_{AdS_3}=\left[\frac{24(g_1^2-g_2^2)^2}{7g_1^4g_2^4h}\right]^{\frac{1}{5}}.
\end{eqnarray}
These are $AdS_3\times CH^2$ solutions with the condition $g_2>g_1$. Finally, we can numerically find RG flow solutions connecting these fixed points to $AdS_7$ vacua with $SO(4)$ and $SO(3)$ symmetries. Examples of these solutions for $g_2=1.1g_1$ and $h=1$ are given in figures \ref{difgSO(3)diagflow1}, \ref{difgSO(3)diagflow2} and \ref{difgSO(3)diagflow3}.

\section{Conclusions}\label{conclusion}
We have studied supersymmetric $AdS_3\times M^4$ solutions of $N=2$ seven-dimensional gauged supergravity with $SO(4)\sim SU(2)\times SU(2)$ gauge group. For $M^4$ being a product of two Riemann surfaces, we have found a large class of $AdS_3\times H^2\times \Sigma^2$ solutions with $SO(2)\times SO(2)$ symmetry for $\Sigma^2=S^2,\mathbb{R}^2,H^2$ similar to the corresponding solutions in maximal $SO(5)$ gauged supergravity studied in \cite{BB}. Furthermore, there exist a number of $AdS_3\times H^2\times H^2$ solutions with $SO(2)_{\text{diag}}$ and $SO(2)_R$ symmetries. In the latter case, all scalars from vector multiplets vanish, so the $AdS_3\times H^2\times H^2$ solution can be interpreted as a solution of pure $N=2$ gauged supergravity with $SU(2)$ gauge group. We have also numerically given various holographic RG flows from supersymmetric $AdS_7$ vacua with $SO(4)$ and $SO(3)$ symmetries to these $AdS_3$ fixed points. The solutions decribe RG flows across dimensions from $N=(1,0)$ SCFTs in six dimensions to two-dimensional $N=(2,0)$ SCFTs in the IR.
\\
\indent For $M^4$ being a Kahler four-cycle, the $AdS_3$ solutions only exist for the Kahler four-cycles with negative curvature. In this case, the spin connection on $M^4$ is a $U(2)\sim SU(2)\times U(1)$ connection. There are two possibilities for performing the twists, along the $U(1)$ and $SU(2)\sim SO(3)$ parts. For a twist by $U(1)\sim SO(2)_R\subset SO(3)_R$, we have found $AdS_3\times CH^2$ fixed points with $SO(2)\times SO(2)$, $SO(2)_{\text{diag}}$ and $SO(2)_R$ symmetries. The solutions preserve four supercharges and correspond to $N=(2,0)$ two-dimensional SCFTs. For a twist along the $SU(2)\sim SO(3)$ part, we have performed the twist by turning on the $SO(3)_{\text{diag}}$ gauge fields. Unlike the previous cases, the $AdS_3$ fixed points in this case preserve only two supercharges. The solutions are accordingly dual to $N=(1,0)$ two-dimensional SCFTs. We have studied RG flows from supersymmetric $AdS_7$ vacua to these geometries as well.
\\
\indent All of these solutions provide a large class of $AdS_3\times M^4$ solutions and RG flows across dimensions from six-dimensional SCFTs to two-dimensional SCFTs. The solutions might be useful in the holographic study of supersymmetric deformations of $N=(1,0)$ SCFTs in six dimensions to two dimensions. For equal $SU(2)$ gauge coupling constants, the $SO(4)$ gauged supergravity can be embedded in eleven-dimensional supergravity. We have also given the uplifted eleven-dimensional metric. These solutions with a clear M-theory origin should be of particular interest in the study of wrapped M5-branes on four-manifolds. 
\\
\indent For solutions with different $SU(2)$ coupling constants, there is no known embedding in string/M theory. Therefore, in this case, the holographic interpretation as RG flows in the dual $N=(1,0)$ SCFTs should be done with some caveats. It would be interesting to look for the embedding of these solutions in ten or eleven dimensions. This could give rise to the full holographic duals of the effective theories on $5$-branes wrapped on four-manifolds. Similar solutions in $N=2$ gauged supergravity with other gauge groups also deserve further study. Finally, it should be noted that the RG flows across dimensions given here can be interpreted as supersymmetric black strings in asymptotically $AdS_7$ space. Our solutions should be useful in the study of black string entropy using twisted indices of $N=(1,0)$ SCFTs along the line of \cite{AdS7_string_Zaffaroni}.
\vspace{0.5cm}\\
{\large{\textbf{Acknowledgement}}} \\
This work is supported by The Thailand Research Fund (TRF) under grant RSA6280022.
\appendix
\section{Truncation ansatz of eleven-dimensional supergravity on $S^4$}
In this appendix, we review relevant formulae for embedding solutions of $N=2$ seven-dimensional gauged supergravity in eleven-dimensional supergravity. Since the $AdS_3\times M^4$ solutions involve all types of seven-dimensional fields namely scalar, vector and three-form fields, the eleven-dimensional four-form field strength is very complicated. Accordingly, we omit an explicit form of the four-form in each case for brevity. It can however be computed by using the formula given in \cite{7D_from_11D} and the mapping between seven- and eleven-dimensional fields given here.
\\
\indent The truncation of eleven-dimensional supergravity on $S^4$ leading to $N=2$ $SO(4)$ seven-dimensional gauged supergravity is described by the metric ansatz 
\begin{eqnarray}
d\hat{s}^2_{11}&=&\Delta^{\frac{1}{3}}ds^2_7+\frac{2}{g^2}\Delta^{-\frac{2}{3}}X^3\left[X\cos^2\xi
+X^{-4}\sin^2\xi \tilde{T}^{-1}_{\alpha\beta}\mu^\alpha\mu^\beta\right]d\xi^2\nonumber \\
& &-\frac{1}{g^2}\Delta^{-\frac{2}{3}}X^{-1}\tilde{T}^{-1}_{\alpha\beta}\sin \xi \mu^\alpha d\xi D\mu^\beta+\frac{1}{2g^2}\Delta^{-\frac{2}{3}}X^{-1}\tilde{T}^{-1}_{\alpha\beta}\cos^2\xi D\mu^\alpha D\mu^\beta\nonumber \\
& &
\end{eqnarray}
with the following definitions
\begin{equation}
D\mu^\alpha =d\mu^\alpha+gA^{\alpha\beta}_{(1)}\mu^\beta\qquad \textrm{and}\qquad \Delta =\cos^2\xi X\tilde{T}_{\alpha\beta}\mu^\alpha\mu^\beta+X^{-4}\sin^2\xi\, .
\end{equation}
$\mu^\alpha$, $\alpha=1,2,3,4$, are coordinates on $S^3$ satisfying $\mu^\alpha\mu^\alpha=1$.
\\
\indent Together with the four-form ansatz given in \cite{7D_from_11D}, the Lagrangian for the resulting $N=2$ gauged supergravity, after multiplied by $\frac{1}{2}$, reads
\begin{eqnarray}
\mc{L}_7&=&\frac{1}{2}R*\mathbf{1}-\frac{1}{8}X^{-2}\tilde{T}^{-1}_{\alpha\gamma}\tilde{T}^{-1}_{\beta\delta}*F^{\alpha\beta}_{(2)}\wedge F^{\gamma\delta}_{(2)}-\frac{1}{8}\tilde{T}^{-1}_{\alpha\beta}*D\tilde{T}_{\beta\gamma}\wedge \tilde{T}^{-1}_{\gamma\delta}D\tilde{T}_{\delta\alpha}\nonumber \\
& &-\frac{1}{4}X^4*F_{(4)}\wedge F_{(4)}+\frac{1}{16}\epsilon_{\alpha\beta\gamma\delta}A_{(3)}\wedge F^{\alpha\beta}_{(2)}\wedge F^{\gamma\delta}_{(2)}-\frac{5}{2}X^{-2}*dX\wedge dX\nonumber \\
& &-\frac{1}{4}gF_{(4)}\wedge A_{(3)}-V*\mathbf{1}
\end{eqnarray}
with the scalar potential given by
\begin{equation}
V=\frac{1}{4}g^2\left[X^{-8}-2X^{-3}\tilde{T}_{\alpha\alpha}+2X^2\left(\tilde{T}_{\alpha\beta}\tilde{T}_{\alpha\beta}-\frac{1}{2}\tilde{T}^2_{\alpha\alpha}\right)\right].
\end{equation}  
A symmetric scalar matrix $\tilde{T}_{\alpha\beta}$, $\alpha,\beta=1,2,3,4$ with unit determinant describes nine scalars in $SL(4,\mathbb{R})/SO(4)$ coset. This is equivalent to $SO(3,3)/SO(3)\times SO(3)$ coset due to the isomorphisms $SO(3,3)\sim SL(4,\mathbb{R})$ and $SO(4)\sim SO(3)\times SO(3)$.
\\
\indent In term of the $SL(4,\mathbb{R})/SO(4)$ coset representative ${\mc{V}_\alpha}^R$ with $SO(4)$ indices $R,S,\ldots=1,2,3,4$, we have the relation 
\begin{equation}
\tilde{T}^{-1}_{\alpha\beta}={\mc{V}_\alpha}^R{\mc{V}_\beta}^S\delta_{RS}\, .
\end{equation} 
The $SO(3,3)/SO(3)\times SO(3)$ coset representative ${L_I}^A$ is related to that of $SL(4,\mathbb{R})/SO(4)$ by the relation
\begin{equation}
{L_I}^A=\frac{1}{4}\Gamma^{\alpha\beta}_I\eta^A_{RS}{\mc{V}_\alpha}^R{\mc{V}_\beta}^S\label{SL4_SO33_coset}
\end{equation}
in which $\Gamma^I$ and $\eta^A$ are chirally projected gamma matrices of $SO(3,3)$ satisfying the relations
\begin{equation}
(\Gamma^I)_{\alpha\beta}(\Gamma^J)^{\alpha\beta}=-4\eta^{IJ}\qquad \textrm{and}\qquad (\Gamma^I)_{\alpha\beta}(\Gamma_I)_{\gamma\delta}=-2\epsilon_{\alpha\beta\gamma\delta}
\end{equation}
and $\Gamma^{I \alpha\beta}=(\Gamma^i_{\alpha\beta},-\Gamma^{i+3}_{\alpha\beta})$, $i=1,2,3$, see more detail in \cite{Eric_N2_7Dmassive}. Note also that $\eta^A_{RS}$ also satisfy similar relations which we will not repeat them here. We use the following choice of $\Gamma^I_{\alpha\beta}$
\begin{eqnarray}
\Gamma^1&=&-i\sigma_2\otimes \sigma_1,\qquad \Gamma^2=-i\sigma_2\otimes \sigma_3,\qquad \Gamma^3=i\mathbf{I}_2\otimes \sigma_2,\nonumber \\
 \Gamma^4&=&i\sigma_1\otimes \sigma_2,\qquad \Gamma^5=-i\sigma_2\otimes \mathbf{I}_2,\qquad \Gamma^6=i\sigma_3\otimes \sigma_2\, .
\end{eqnarray}
\indent All these ingredients lead to the following identification of the fields and parameters in seven and eleven dimensions
\begin{eqnarray}
g_2&=&g_1=16h=2g,\qquad X=e^{-\frac{\sigma}{2}}, \nonumber \\ 
C_{(3)}&=&\frac{1}{\sqrt{2}}A_{(3)},\qquad A^{\alpha\beta}_{(1)}=\Gamma_I^{\alpha\beta}A^{I}_{(1)}\, .
\end{eqnarray}
With this identification, it can also be easily verified that the scalar matrix for the gauge kinetic terms also match
\begin{equation}
a_{IJ}=\frac{1}{4}\tilde{T}^{-1}_{\alpha\gamma}\tilde{T}^{-1}_{\beta\delta}\Gamma_{I}^{\alpha\beta}\Gamma^{\gamma\delta}_J\, .
\end{equation}
\indent For convenience, we explicitly give the $SL(4,\mathbb{R})/SO(4)$ coset representative ${\mc{V}_\alpha}^R$ and $SO(4)$ gauge fields $A^{\alpha\beta}$  as follow. 
\begin{itemize}
\item $SO(3)_{\textrm{diag}}$ singlet scalar:
\begin{eqnarray}
{\mc{V}_\alpha}^R&=&\textrm{diag}(e^{\frac{\phi}{2}},e^{\frac{\phi}{2}},e^{\frac{\phi}{2}},e^{-\frac{3\phi}{2}}),\\
A^{12}&=&A^3+A^6=2A^3,\qquad A^{13}=-A^2-A^5=-2A^2,\nonumber \\
A^{23}&=&-A^1-A^4=-2A^1\, .
\end{eqnarray}
We have used the relation $A^i=\frac{g_2}{g_1}A^{i+3}$ with $g_2=g_1$.
\item $SO(2)\times SO(2)$ singlet scalar:
\begin{eqnarray}
{\mc{V}_\alpha}^R&=&\textrm{diag}(e^{\frac{\phi}{2}},e^{\frac{\phi}{2}},e^{-\frac{\phi}{2}},e^{-\frac{\phi}{2}}),\\
A^{12}&=&A^3+A^6,\qquad A^{34}=A^3-A^6\, .
\end{eqnarray}
\item $SO(2)_{\textrm{diag}}$ singlet scalars:
\begin{eqnarray}
{\mc{V}_\alpha}^R&=&\begin{pmatrix}
e^{\frac{\phi_2}{2}}0 & 0 &  0& 0\\
0& e^{\frac{\phi_2}{2}} &0 &0\\
0&0 & e^{\phi_1-\frac{\phi_2}{2}}\cosh\phi_3 & e^{\phi_1-\frac{\phi_2}{2}}\sinh\phi_3 \\
0& 0 & e^{-\phi_1-\frac{\phi_2}{2}}\sinh\phi_3 & e^{-\phi_1-\frac{\phi_2}{2}}\cosh\phi_3
\end{pmatrix},\\
A^{12}&=&2A^3\, .
\end{eqnarray}
\end{itemize}
In all cases, it can be verified using the relation \eqref{SL4_SO33_coset} that the above ${\mc{V}_\alpha}^R$ give precisely ${L_I}^A$ in the main text.


\end{document}